\documentclass[11pt,fleqn]{article} 
\usepackage{amsmath,amsfonts, amssymb, braket, bbold}
\usepackage{empheq}
\usepackage{eufrak, mathtools}
\usepackage{tabularx}
\usepackage{boldline}
\usepackage{graphicx,color,multicol,multirow}
\usepackage{tikz}
\usetikzlibrary{trees,er,snakes,shapes,mindmap}
\usetikzlibrary{decorations.pathmorphing}
\usetikzlibrary{decorations.markings}
\usepackage{hyperref}
\usepackage{titlesec}
\titleformat{\section}
 {\normalfont\fontfamily{bch}\fontsize{14pt}{16pt}\bfseries\color{black}}
{\thesection}{1em}{}
\titleformat{\subsection}
 {\normalfont\fontfamily{bch}\fontsize{12pt}{16pt}\bfseries\color{black}}
{\thesubsection}{1em}{}
\def \beq  {\begin{equation}}
\def \eeq  {\end{equation}}
\def \beqar {\begin{eqnarray}}
\def \eeqar {\end{eqnarray}}
\allowdisplaybreaks
\tikzset{
    photon/.style={decorate, decoration={snake, amplitude =1.5pt}, draw=red, line width = 1pt},
    electron/.style={draw=blue, line width = 1pt, postaction={decorate},
        decoration={markings,mark=at position .65 with {\arrow[draw=black]{>}}}},
        proton/.style={draw=black, line width=1.5pt, postaction={decorate},
        decoration={markings,mark=at position .55 with {\arrow[draw=black]{>}}}},
             neutron/.style={draw=darkblue, line width=1.5pt, postaction={decorate},
        decoration={markings,mark=at position .55 with {\arrow[draw=darkblue]{>}}}},
          quark/.style={draw=brown!70, line width = 1pt, postaction={decorate},
        decoration={markings,mark=at position .55 with {\arrow[draw=brown]{>}}}},
    gluon/.style={decorate, draw= blue!60,
        decoration={coil,amplitude=3.2 pt, segment length= 4.0pt}} 
}
\def\sqr#1#2{{\vcenter{\vbox{\hrule height.#2pt
\hbox{\vrule width.#2pt height#1pt \kern#1pt
\vrule width.#2pt}\hrule height.#2pt}}}}

\def\S {{\cal S}}
\def\la {{\langle}}
\def\ra {{\rangle}}
\def\vx {{\vec x}}
\def\vy {{\vec y}}
\def\vk {{\vec k}}
\def\vf {{\varphi}}

\def\bvf {{\bar \varphi}}
\def\dag {{\dagger}}

\def\vq {{\vec q}}

\def\Tr {{\rm Tr}}
\def \tr {{\rm tr}}

\def\bp {\bar p}
\def\ba {\bar{a}}
\def\bA {\bar{A}}
\def\bD {\bar{D}}
\def\bE {\bar{E}}
\def\bG {\bar{G}}
\def\bx {\bar{x}}
\def\by {\bar{y}}
\def\bu {\bar{u}}
\def\bv {\bar{v}}
\def\bw {\bar{w}}
\def\vk {\vec{k}}
\def\vx {{\vec x}}
\def\vz {\vec{z}}
\def\vy{\vec{y}}
\def\vv {\vec{v}}
\def\vu {\vec{u}}
\def\vw {\vec{w}}

\def\dag {\dagger}
\def\del {\partial}
\def\bdel{\bar{\partial}}
\def\a {\alpha}

\def\e {\epsilon}
\def\d {\delta}
\def\s {\sigma}

\def\bz {{\bar{z}}}

\def\A {{\cal A}}

\def\C {{\cal C}}
\def\D {{\cal D}}
\def\E {{\cal E}}
\def\F {{\cal F}}
\def\G {{\cal G}}
\def\H {{\cal H}}

\def\O {{\cal O}}
\def\P {{\cal P}}
\def\T {{\cal T}}

\def\vf {{\varphi}}
\def \bvf {{\bar \varphi}}

\def\half{\textstyle{1\over 2}}

\mathchardef\mhyphen="2D
\begin{document}
\fontfamily{bch}\fontsize{12pt}{16pt}\selectfont
\def \CMP {{Commun. Math. Phys.}}
\def \PRL {{Phys. Rev. Lett.}}
\def \PL {{Phys. Lett.}}
\def \NPBProc {{Nucl. Phys. B (Proc. Suppl.)}}
\def \NP {{Nucl. Phys.}}
\def \RMP {{Rev. Mod. Phys.}}
\def \JGP {{J. Geom. Phys.}}
\def \CQG {{Class. Quant. Grav.}}
\def \MPL {{Mod. Phys. Lett.}}
\def \IJMP {{ Int. J. Mod. Phys.}}
\def \JHEP {{JHEP}}
\def \PR {{Phys. Rev.}}
\def \JMP {{J. Math. Phys.}}
\def \GRG{{Gen. Rel. Grav.}}
\begin{titlepage}
\null\vspace{-62pt} \pagestyle{empty}
\begin{center}
\vspace{1.3truein} {\Large\bfseries
The Schr\"odinger Representation and 3d Gauge Theories}
\\
{\Large\bfseries ~}\\
\vskip .2in
{\Large\bfseries~ }\\
{\large\sc V.P. Nair}\\
\vskip .2in
{\sl Physics Department\\
City College of the CUNY\\
New York, NY 10031}\\
 \vskip .1in
\begin{tabular}{r l}
{\sl E-mail}:&\!\!\!{\fontfamily{cmtt}\fontsize{11pt}{15pt}\selectfont vpnair@ccny.cuny.edu}\\
\end{tabular}
\vskip .5in

\centerline{\large\bf Abstract}
\end{center}
In this review we consider the Hamiltonian analysis of Yang-Mills theory and some variants of it in three spacetime dimensions using the Schr\"odinger representation. This representation, although technically more involved than the usual covariant formulation, may be better suited for some nonperturbative issues. Specifically for the Yang-Mills theory, 
we explain how to set up the Hamiltonian
formulation in terms of manifestly gauge-invariant variables
and set up an expansion scheme for solving the Schr\"odinger equation.
We review the calculation of the string tension, the Casimir energy
and the propagator mass and compare with the results from lattice simulations.
The computation of the first set of corrections to the string tension, 
string breaking effects, extensions to
the Yang-Mills-Chern-Simons theory and to the supersymmetric cases
are also discussed. We also comment on how entanglement for the vacuum state can be formulated in terms of the BFK gluing formula.
The review concludes with a discussion of the status and prospects
of this approach.

\vskip .2in
\noindent This is an expanded version of the lectures given at
{\it Understanding Confinement: Prospects in Theoretical Physics}
Summer School at the Institute for Advanced Study, Princeton, July 2023.

\end{titlepage}

\fontfamily{bch}\fontsize{12pt}{16pt}\selectfont
\pagestyle{empty} 

\tableofcontents 



\newpage
\fontfamily{bch}\fontsize{12pt}{16pt}\selectfont
\pagestyle{plain} 
\def\theequation{\thesection.\arabic{equation}}
\section{Introduction}\label{Intro}
Gauge theories have a foundational role in physics since they are the basic paradigm for the formulation of the Standard Model (SM) of fundamental particles and their interactions. The great success of the SM therefore makes it imperative that we understand the structure of gauge theories in different environments and kinematic regimes.
Covariant perturbation theory for gauge theories is by now a well-developed and powerful technique and it is adequate for the analysis of the electroweak sector of the SM for most questions of interest.
The situation for the strong nuclear forces, described by Quantum Chromodynamics (QCD), is very different. The high energy regime of QCD
(energies $\gtrsim 10\, {\rm GeV}$) can be analyzed using perturbation theory by virtue of asymptotic freedom.
But the low energy regime, where the interaction strength is large and 
where perturbation theory is no longer applicable, remains a real challenge.
Decades of work have led to a fairly good qualitative understanding
of the low energy regime of nonabelian gauge theories, but quantitative
analysis of important questions such as how quarks bind together to form hadrons, what the nucleonic and nuclear matrix elements for
the electroweak transitions of hadrons are, etc., is difficult.
Lattice gauge theory, combined with large scale numerical simulations, has been the reliable workhorse for most questions of a nonpertrubative nature and, indeed, it has produced a number of useful results. However, it is important to correlate these results with an analytical approach to arrive at a complete or more comprehensive understanding of the physics of gauge theories.

In this review, we will describe an approach which is very different from covariant perturbation theory, namely, the Schr\"odinger representation in field theory where we use Hamiltonians and seek wave functions (actually functionals) which are solutions of the Schr\"odinger equation.
Although this representation goes back to the early days of field theory, and has the conceptual simplicity of elementary quantum mechanics, it has rarely been used because of many perceived difficulties. Nevertheless, it may be more suitable for certain types of questions in field theory. To cite an elementary example, recall that a spacetime approach in terms of path integrals can be used to work out the bound state energy levels and transition matrix elements for the Hydrogen atom, but, as anyone who tries to do so will realize immediately, it is much simpler to use the Hamiltonian and the Schr\"odinger equation. 

We will be considering the application of this method mainly to the three (or 2+1)
dimensional Yang-Mills (YM) theory. However, it is useful to start with a few
general observations. 
Consider a simple scalar field theory with a classical action of the form\footnote{We use $d{\rm V}$ to denote the volume element for the spatial manifold.}
\beq
S = \int dt d{\rm V}\,\left[ {1\over 2} \del_\mu \phi \, \del^\mu \phi - {m^2 \over 2}
\phi^2  - \lambda \phi^4\right]
\label{Intro1}
\eeq
In the canonical quantization of this theory, we start with the equal-time commutation rules, say at time $t = 0$,
\beqar
[ \phi (\vx , 0),  \phi (\vy , 0)]&=& 0\nonumber\\
{}[ \phi (\vx , 0),  \pi (\vy , 0)]&=& i \,\delta(\vx - \vy)\label{Intro2}\\
{}[ \pi (\vx , 0),  \pi (\vy , 0)] &=& 0\nonumber
\eeqar
where $\pi (\vx, 0) = {\dot \phi} (\vx, 0)$.
This suggests that we can define a set of $\phi$-diagonal states
$\ket{\vf}$ obeying $\phi(\vx, 0) \ket{\vf} = \vf (\vx) \ket{\vf}$, where
$\vf(\vx)$ is a c-number function. A Schr\"odinger wave function for a
state $\ket{\alpha}$ will take the form
\beq
\Psi_\alpha [\vf ] = \braket{\vf| \alpha}
\label{Intro3}
\eeq
It is a functional of $\vf$. The commutation rules (\ref{Intro2}) can then be
represented as
\beqar
\bra{\vf} \phi (\vx, 0) \ket{\alpha} &=& \vf(\vx) \, \Psi_\alpha [\vf]\nonumber\\
\bra{\vf} \pi(\vx, 0 )  \ket{\alpha}&=& - i {\delta \over \delta \vf (\vx)}\, \Psi_\alpha [\vf]
\label{Intro4}
\eeqar
This is the Schr\"odinger representation of the commutation rules.

The Hamiltonian corresponding to the action (\ref{Intro1}) has the form
\beq
\H = \int d{\rm V} \left[ {1\over2 } \pi^2  + {1\over 2} \phi (-\nabla^2 + m^2) \phi + \lambda \phi^4
\right]
\label{Intro5}
\eeq
The idea is that we can use this to write down and solve the Schr\"odinger equation.
The vacuum state of the theory, represented by the wave function
$\Psi_0[\vf]$, would thus satisfy
\beqar
 \H \, \Psi_0[\vf] &=&\int_{\rm V} \left[ - {1\over2 } \left( {\delta^2 \over \delta \vf(\vx) \delta \vf(\vx)}\right)  + {1\over 2} \vf(x) \omega^2 (x,y) \vf(y)+ \lambda \vf^4(x)
\right] \Psi_0[\vf]\nonumber\\
&=& 0\label{Intro6}\\
\int_{x,y} \vf(x) \omega^2(x,y) \vf(y) &\equiv& \int_x \vf(x) (-\nabla^2 + m^2) \vf(x)
\nonumber
\eeqar
where we have used the Schr\"odinger representation
to write $\H$ as a functional differential operator
which can act on $\Psi_0[\vf]$.

A number of potential problems are evident at this stage.
As with any field theory, we need regularization and renormalization.
In covariant perturbation theory, the regularized action has the form
\beq
S = \int \left[ Z_3 \left[ {1\over 2} \del_\mu \phi \, \del^\mu \phi - {m^2 \over 2}
\phi^2 + {\delta m^2 \over 2} \phi^2\right] - Z_1\lambda \phi^4\right]
\label{Intro7}
\eeq
where $Z_1$, $Z_3$ and $\delta m^2$ will depend on the regularization parameter $\Lambda$ (upper cutoff on momenta) and are chosen so as to render all correlation functions finite as $\Lambda \rightarrow \infty$.
The situation in the Schr\"odinger representation is more complicated.
We have functional derivatives at the same point $\vx$ in the 
$\delta^2/\delta \vf^2$-term, so it needs regularization and
a $Z$-factor. A similar statement applies to the $\vf (-\nabla^2 \vf)$-term.
The mass term will need an additive renormalization as well, so we need
a term ${\half} \delta m^2 \vf^2$. And finally we need regularization and a
$Z$-factor for the interaction term.
At this stage, we could envisage independent regularizations for the
terms $\delta^2 /\delta \vf^2$ and $\vf (-\nabla^2 \vf)$, since we have a separation of space and time and Lorentz invariance is not manifest.
The requirement of Lorentz invariance will relate the $Z$-factors for
these two terms. The regularization must be so chosen as to ensure this,
Lorentz invariance is not automatic as in covariant perturbation theory.
This is one of the complications of the Schr\"odinger representation for
field theories.

There is one other issue associated with Poincar\'e invariance.
One of the commutation rules for the Poincar\'e algebra is
\beq
[K_i, P_j ] = i \,\delta_{ij} \H
\label{Intro8}
\eeq
where $P_j$ is the total momentum and $K_i$ is the generator of Lorentz boosts. Taking the expectation value of this with the vacuum state shows that if the vacuum is to be Lorentz invariant, we must have
$\bra{0} \H \ket{0} = 0$. So, for maintaining Poincar\'e invariance, $\H$ must be redefined by subtracting a certain $c$-number term to ensure this; this is the version of the familiar
normal ordering in the present context.

In addition to the Hamiltonian, we must ensure that the wave functions
(which are functionals of the field)
are well-defined. In general, this will require additional counterterms.
One way to understand the genesis of such counterterms is to think of the wave function at time $t_1$ as
defined by a path integral over the region $t < t_1$ as in
\beq
\Psi[\vf, t_1] = \int [\D {\tilde \vf}]\,e^{i S [{\tilde \vf}, t_1, t_0] } \, \Psi[\vf', t_0],
\hskip .2in {\tilde \vf} (\vx, t_1) = \vf(\vx ), ~{\tilde \vf} (\vx, t_0) = \vf'(\vx )
\label{Intro9}
\eeq
The functional integration is over all paths ${\tilde \vf} (\vx, t )$ with
the boundary values shown.
In the course of carrying out calculations using this form, we will be renormalizing an action defined on a spacetime region with boundaries (the time-slices at $t_0$ and $t_1$) and it will require counterterms on the boundaries.
These take the form \cite{Symanzik}
\beq
\Psi \rightarrow 
\exp\left[  i \int d{\rm V}\,\left( Z_5 \varphi \partial_0 \varphi 
~+ \Delta\, \varphi^2 \right) \right]\, \Psi
\label{Intro10}
\eeq
The Hamiltonian itself takes the form
\beqar
\H &=& \int d{\rm V}\, {1\over2 }\left[ - {1\over Z_3 Z_0} \left( {\delta^2 \over \delta \vf(\vx) \delta \vf(\vx)}\right)_{\rm reg}   + Z_3 Z_0 \Bigl((\nabla \vf )^2 +
(m^2 - \delta m^2) \vf^2 \Bigr)_{\rm reg} \right]\nonumber\\
&& \hskip .2in +\int d{\rm V}\, Z_1 \Bigl(\lambda \vf(x)^4
\Bigr)_{\rm reg} - E_0
\label{Intro11}
\eeqar
The factor $Z_0$ is proportional to $Z_5$ introduced in (\ref{Intro10}).\footnote{The one-loop calculation of $Z_0$ is outlined in \cite{Min-N}.}
The boundary counterterms are another complication, in general, for the Schr\"odinger representation.

With the formalism as outlined above, and using a point-splitting regularization, Symanzik was able to prove the renormalizability of the $\phi^4$-theory in the Schr\"odinger representation \cite{Symanzik}. (While renormalizability of this theory in the covariant formalism was relatively straightforward and known for many years, there was even a general feeling, before Symanzik's work, that the theory was not renormalizable in the Schr\"odinger representation.
There is some new physics which emerges in this formalism as well.
Symanzik used the Schr\"odinger representation to analyze Casimir energies. Further, the additional $Z$-factor ($Z_5$) introduced by Symanzik can also be related to a new critical exponent, see \cite{Luscher}.) 

A useful observation worth mentioning at this stage is that the vacuum wave function for the free theory (with $\lambda = 0$) is given by
\beq
\Psi_0 [\vf] = \left[ \det \left( {\sqrt{k^2+m^2}\over \pi}\right)_{\rm reg}\right]^{\half}  \, \exp\left( - {1\over 2} \int_{x,y} \vf(x) \bigl[\sqrt{k^2+ m^2}\bigr]_{x,y, {\rm reg}} \vf(y)\right)
\label{Intro12}
\eeq
with $E_0 = {\half} \bigl[\sqrt{k^2+ m^2}\bigr]_{x,x, {\rm reg}}$.

Given the additional complications with regard to regularization and renormalization, compared to covariant perturbation theory, one might wonder whether it is worth the trouble to pursue the Schr\"odinger representation in field theory. For certain questions of a nonperturbative nature, the answer seems to be a qualified yes. 
The kinetic operator in the Hamiltonian may be viewed as the 
Laplace operator on the infinite-dimensional space of field configurations and if we have some knowledge of the geometry and topology of this space, it can shed light on the spectrum of the Hamiltonian.
A key inspirational paper in this context was by Feynman, who
analyzed Yang-Mills theory in 2+1 dimensions \cite{Feyn}.
These theories are rather optimal candidates for the Schr\"odinger representation since there is no renormalization of the coupling constant, so some of the aforementioned problems can be avoided.
Feynman tried to argue that the space of gauge-invariant 
configurations (gauge potentials modulo gauge transformations)
is compact and hence can lead to a discrete spectrum for
the Laplacian and ultimately a mass gap for the theory.
This is not quite true, the configuration space is not compact, as shown by 
Singer, who however argued that the sectional curvature of the space is positive
\cite{Singer}.\footnote{Feynman's analysis was modeled on his earlier very successful analysis of superfluid Helium. The comparison of the two cases and
some of the nuances of the gauge theory are outlined in
\cite{Nair-He}.}
This is suggestive in view of the Lichnerowicz bound on the
lowest eigenvalue of the Laplacian.
Explicitly, if the Ricci tensor $R_{\alpha\beta}$ of a
compact Riemannian manifold of dimension $n$ has a positive
lower bound
\beq
R_{\alpha \beta} \geq \mu^2 (n -1) \, g_{\alpha \beta} ,
\label{Intro13}
\eeq
where $\mu^2$ is a constant parameter of dimension $({\rm length})^{-2}$,
then the lowest eigenvalue of the ($-$Laplacian) satisfies the bound
\beq
\lambda \geq \mu^2 \, n.
\label{Intro14}
\eeq
In the present case, a simple extension of this argument is not possible since we are dealing with an infinite dimensional manifold. So
regularizations are needed to define the Laplacian, Ricci tensor, etc.
before we can even consider the proper formulation of
a similar bound.

Feynman's arguments and Singer's analysis were carried out
before we had an exact expression for the 
volume element for the configuration space.
What we shall do here is to revisit this problem in the light of later developments.
The basic analysis and results are from
\cite{KKN1}-\cite{KKN4}, \cite{KN5}.
We will see that, modulo certain approximations and caveats as explained in detail below, there are a few key quantitative (and encouraging) results which emerge from our analysis:
\begin{enumerate}
\item There is an analytic formula for the string tension which compares very favorably with numerical estimates from lattice simulations. 
\item One can calculate the Casimir energy for a parallel plate arrangement; this too compares very favorably with the lattice simulations.
\end{enumerate}
There are also some additional insights obtained regarding supersymmetric theories, entanglement, etc., which we will comment on later.

Unlike the 1+1 dimensional cases, the 2+1 dimensional Yang-Mills theories have propagating degrees of freedom, so one might consider them to be closer to the 4d Yang-Mills theories; this is an added motivation for analyzing these theories. But they are also relevant for the high temperature
(${\mathsf{T}}$) limit of 4d Yang-Mills theories. In this limit, the 4d (or 3+1 dimensional) theory reduces to a (Euclidean) 3d Yang-Mills theory with  coupling constant $e^2 = g^2 {\mathsf{T}}$, where $g$ is the coupling constant of the 4d-theory.
Electric fields and time-dependent processes become irrelevant.
The mass gap of the 3d theory, from the point of view of the 4d theory, 
becomes the magnetic mass since it controls
the screening of magnetic fields in the gluon plasma \cite{Gross-PY}.
So the identification  of the propagator mass in the 3d theory (either analytically or via the lattice simulation of the Casimir effect, as explained
in section \ref{Res}) 
will be important
for the 4d theory at high temperatures.

Three dimensional space is also famous as the home ground of the 
3d Chern-Simons (CS) theory, with all its ramifications including
knot theory, conformal field theory, etc. 
For the CS theory also, a beautiful analysis
can be carried out in the Schr\"odinger representation. For a review, see
\cite{Jackiw}. Some facets of the Hamiltonian analysis of the
CS theory will also be discussed in section \ref{CStheory}.

Since there are diverse concepts involved as well as a number
of comments and digressions, it may be useful to give a layout of 
what is to follow before getting to the technical details.
Just for the (2+1)-dimensional YM theory, the analysis and the main results are given in sections \ref{Par}, \ref{Vol}, \ref{Ham}, \ref{SchE} and
\ref{Res}. In section \ref{Par}, we will give various arguments to
show that the complex components of the gauge potentials, for an $SU(N)$ gauge theory, can be parametrized as
\beq
A = -\del M\, M^{-1}, \hskip .2in \bA = M^{\dagger -1} \bdel M^\dagger
\label{Intro15}
\eeq
where $M$ is an $SL(N, \mathbb{C})$ matrix.
As the next step in setting up the Hamiltonian formulation, the
gauge-invariant volume element $d\mu (\C )$ on the space of the
fields ($A$, $\bA$) modulo gauge transformations is calculated
in section \ref{Vol} using the parametrization
(\ref{Intro15}).  This serves to define the inner product for
wave functions as
\beq
\braket{1|2} = \int d\mu (\C )\, \Psi_1^* \, \Psi_2
\label{Intro16}
\eeq
As the next logical step in setting up the Schr\"odinger formulation,
we will work out the form of the Hamiltonian $\H$ in terms of a set of
gauge-invariant variables.
The relevant variables will turn out to be a current of the form
$J\sim \del H \, H^{-1}$, $H = M^\dagger M$.
Once we have the Hamiltonian in a form where the redundant
gauge degrees of freedom have been eliminated, one can 
proceed to the Schr\"odinger equation.
In section \ref{SchE}, we will explain how a systematic expansion scheme
for solving the Schr\"odinger equation can be set up and we will
work out the solution 
to the lowest two orders.
The resulting vacuum wave function will then be used to
calculate the string tension and the Casimir energy 
for the nonabelian gauge theory in section \ref{Res}; we will also compare these results
with numerical estimates based on lattice simulations.
These sections,
namely, \ref{Par}, \ref{Vol}, \ref{Ham}, \ref{SchE} and
\ref{Res},
will constitute the main thread of arguments regarding the use of the
Schr\"odinger representation for the Yang-Mills theory in 2+1 dimensions.

Sections \ref{gaugeprinciple} and \ref{Conf} discuss
key ideas from
 the general formulation of gauge theories useful for the analysis
 of YM(2+1).
 The discussion of the propagator mass in section
 \ref{promass} is meant primarily for the analysis of the Casimir energy in
 section \ref{Res}, but also serves to formulate the alternate argument for the wave function given in section \ref{SchE}.
 We have also commented on the Casimir scaling versus the sine-law
 for string tensions in section \ref{Res}.
Section \ref{s-break} is about string-breaking and 
screenable representations. 

The main thrust of the remainder of the text is about extensions of the
Yang-Mills theory, with a Chern-Simons term added to the action as well as
supersymmetric theories. The analysis is presented
mainly in section \ref{Ext}. The integration measure for the inner product plays a crucial role in our approach to these theories. We present an indirect way to 
calculate this measure in terms of the Knizhnik-Zamolodchikov equation and the finite renormalization of the level number of the Chern-Simons term.
In the spirit of staying within the Schr\"odinger representation, 
and for a sense of completeness,
we have added a short section (\ref{CStheory}) on the Chern-Simons theory
where we show how this renormalization arises within the Hamiltonian 
approach.

There have been suggestions about the form of the vacuum wave function
other than our solution to the Schr\"odinger equation.
Some of these are reviewed and commented on {\it vis-a-vis} our
solution in section \ref{Alter}.
Entanglement is the one concept in the quantum theory
which is presented most directly in terms of states or wave functions
and hence the Schr\"odinger representation is the most natural framework
for understanding this feature.
We discuss entanglement in the Yang-Mills theory
in section \ref{Ent}; the focus here
is on the so-called contact term and how it can be related to the
BFK gluing formula.
The last section (\ref{Concl}) is on prospects as well as a status report.

There are four Appendices. Appendix \ref{AppA} just outlines
some conventions.
 Appendix \ref{TopC} is on the geometry and topology of the configuration space and
 is not essential for a first reading.
 It does however, touch upon the issue of the Gribov problem
 \cite{Grib}.
Appendix \ref{AppC} is on
regularization of the operators.
While regularization is rather technical and hence relegated to an Appendix,
it is important for the results derived in the main text. 
Appendix \ref{AppD} is
on the calculation of corrections to the vacuum wave function and string tension. It 
shows that the first set of corrections (within the expansion
scheme of section \ref{SchE}) to the string tension are small, a result 
which is crucial for the eventual justification
of the expansion procedure.

\section{The gauge principle}\label{gaugeprinciple}
\setcounter{equation}{0}
The quintessential example of a gauge theory is quantum electrodynamics
describing the interaction of electrons and positrons with the electromagnetic field. The starting point for this theory is the Lagrangian
\beq
L (\psi, {\bar \psi}, A) = {\bar \psi} \left[ i \gamma^\mu (\del_\mu - i A_\mu ) - m \right] \psi
- {1\over 4 e^2} F_{\mu\nu}F^{\mu\nu}
\label{G1}
\eeq
where $\psi$ is a 4-component spinor field in four dimensions representing the electron-positron field, ${\bar \psi} = \psi^\dagger \gamma^0$,
and $A_\mu$ is the vector potential for the electromagnetic field.
Also $F_{\mu\nu}$ is the field strength tensor defined as
$F_{\mu\nu} = \del_\mu A_\nu - \del_\nu A_\mu$.
The components of $F_{\mu\nu}$ are related to the electric ($E_i$) and
magnetic ($B_i$) fields as $F_{0i} = E_i$, $F_{ij} = \epsilon_{ijk} B_k$.
The charge of the electron is $e$ and its mass is $m$. Also, $\gamma^\mu$ are the Dirac $\gamma$-matrices obeying\footnote{Our conventions and specific realizations are discussed in Appendix \ref{AppA}.}
\beq
\gamma^\mu \, \gamma^\nu + \gamma^\nu \, \gamma^\mu = 
2\, \eta^{\mu\nu} \, {\mathbb{1}}
\label{G2}
\eeq

The key property of the Lagrangian (\ref{G1}) for our analysis
is gauge invariance. If we make a change of variables
as
\beqar
\psi \rightarrow \psi^g = g \, \psi, \hskip .1in&&\hskip .1in
{\bar \psi} \rightarrow {\bar \psi}^g = {\bar \psi} \, g\nonumber\\
A_\mu \rightarrow A^g_\mu &=& g\, A_\mu \, g^{-1} - i \del_\mu g \, g^{-1}
= A_\mu + \del_\mu \theta ,
\label{G3}
\eeqar
where $g = e^{i \theta}$, we find
\beq
L (\psi^g, {\bar \psi}^g, A^g ) = L(\psi, {\bar \psi}, A)
\label{G4}
\eeq
Notice that $F_{\mu\nu}$ (i.e., $E_i$, $B_i$) is unchanged 
by the transformation (\ref{G3}). Classically the motion of a charged particle is governed by the Lorentz force law which involves only $E_i$, $B_i$.
Hence, classically the entire dynamics is insensitive to the transformation
(\ref{G3}). Therefore, the gauge degree of freedom, namely $\theta (x)$,
represents a redundancy in the dynamical variables used to describe the
theory. Going to the quantum theory, notice that we can set $A^g$ to zero
along a line by defining $\theta (x)$ as 
\beq
\theta(x) = \int_{x_0, C}^x dx^\mu A_\mu = \int_{x_0, C}^x
A
\label{G5}
\eeq
where $C$ denotes a path connecting the point
$x_0^\mu$ to $x^\mu$. In this case, $\psi$ acquires a phase factor
$e^{i \theta} = e^{i\int_C A}$. But the value
of $\theta (x)$ depends on the path $C$ and not just the end-point
$x^\mu$ of the path. So, in general, we cannot eliminate
$A_\mu$; we would need $A_\mu$ to set up the theory rather than just $F_{\mu\nu}$. 

The function $g = e^{i \theta}$ used in (\ref{G3}) is an element of the group
$U(1)$, so the gauge symmetry in (\ref{G3}), (\ref{G4}) is 
a $U(1)$ gauge symmetry. The generalization of this to an arbitrary 
Lie group $G$
is as follows. Consider a set of fields $\psi^i$, $i = 1, 2, \cdots, N$
which transform as an $N$-dimensional representation
$R$ of the group $G$; i.e.,
\beq
(\psi^i)^g = g^{ij} \, \psi^j
\label{G7}
\eeq
We define a covariant derivative $D_\mu \psi$ as
\beq
(D_\mu \psi)^i = \del_\mu \psi^i + (A_\mu)^{ij} \, \psi^j
\label{G8}
\eeq
where $A_\mu$ is an element of the Lie algebra of
$G$, with $(A_\mu)$ as its matrix representative in the chosen representation
$R$. Thus, if $\{ T^a\}$ denote a basis for the Lie algebra of $G$,
with $a = 1, 2, \cdots, {\text{dim}}\,G$,
realized as matrices in the representation $R$,
\beq
(A_\mu)^{ij}  = - i A^a_\mu \, (T^a)^{ij}
\label{G9}
\eeq
We also define the gauge transform of $A$ as
\beq
A^g_\mu = g \, A_\mu \, g^{-1} - \del_\mu g \, g^{-1}
\label{G10}
\eeq
This is also in the matrix notation. The derivative $D_\mu \psi$ is covariant in the sense that
\beqar
(D^g_\mu \psi^g )^i &=&\left[ \del_\mu (g\psi) + ( g \, A_\mu \, g^{-1} - \del_\mu g \, g^{-1} ) (g \psi)\right]^i\nonumber\\
&=& g^{ij} (\del_\mu \psi + A_\mu \psi)^j = g^{ij} (D_\mu \psi)^j
\label{G11}
\eeqar
A particular case of interest would be for
fields transforming according to the adjoint
representation of the group. In this case, $(T^a)^{ij} = -i f^{aij}$, where
$f^{aij}$ are the structure constants of the Lie algebra of $G$
in the chosen basis. Thus they are given by
$[T^a , T^b ] = i f^{abc} T^c$. In this case
$(D_\mu \psi)^a = \del_\mu\psi^a  + f^{a bc} A^b_\mu \psi^c$.

The commutator of covariant derivatives defines the field strength tensor as
\beqar
[D_\mu, D_\nu] &=& F_{\mu\nu} =  (-i T^a)\, F^a_{\mu\nu}\nonumber\\
F^a_{\mu\nu}&=& \del_\mu A^a_\nu - \del_\nu A^a_\mu 
+ f^{abc} A^b_\mu A^c_\nu
\label{G12}
\eeqar
By construction, $F_{\mu\nu}$ transforms homogeneously
under gauge transformations as
\beq
F^g_{\mu\nu} = g\, F_{\mu\nu} \,g^{-1}
\label{G13}
\eeq
If we have a unitary representation of the group on the fields
$\psi$, we have ${\bar \psi}^g = {\bar \psi} g^\dagger
= {\bar \psi} g^{-1}$, so that a Lagrangian consistent with
 gauge invariance is
 \beq
 L (\psi, {\bar\psi}, A) = {\bar \psi} \left[
 i \gamma^\mu D_\mu - m\right] \psi - {1\over 4 e^2} F^a_{\mu\nu} 
 F^{a \mu\nu}
 \label{G14}
 \eeq
 This is the kind of Lagrangian we use for coupling of quarks
 to the gluons (particles corresponding to $A^a_\mu$) in quantum chromodynamics (QCD). The left and right chiral components of the fermion
 field couple to the gauge field in an identical fashion, so the coupling is vectorial in nature.
 The Standard Model also involves chiral or axial
 couplings of the quarks and leptons to various gauge fields.
 Most of of our analysis will be for the pure gauge theory,
 and when we discuss gauge fields in interaction
 with matter, we will mostly consider
 vectorial couplings. The action for the gauge field part of the Lagrangian
 (\ref{G14}) is the Yang-Mills action
 \beqar
 S_{\text{YM}} &=&  - {1\over 4 e^2} \int  dt d{\rm V}\, F^a_{\mu\nu} 
 F^{a \mu\nu}\nonumber\\
 &=& {1\over 2 e^2} \int dt d{\rm V}\, (E^a_i E^a_i - B^a_i B^a_i )
 \label{G15}
 \eeqar
 For the special case of a $U(1)$ gauge theory where 
 $\text{dim}\,G = 1$, this action agrees with
 the action for the electric and magnetic fields in electrodynamics.
 The nonabelian analogs of these fields
 can be written out as
 \beqar
 E^a_i &=& F^a_{0i} = {\del A^a_i \over \del t} - \del_i A_0^a + f^{abc} A^b_0 A^c_i
 \nonumber\\
 &=& {\del A^a_i \over \del t}  - ( D_iA_0)^a\label{G16}\\
 B^a_i&=& {1\over 2}  \e_{ijk} F^a_{jk} =  {1\over 2}  \e_{ijk} (\del_j A_k^a  - \del_k A_j^a + f^{abc} A^b_j A^c_k )\hskip .2in {\text{(3+1 dim)}}
 \nonumber\\
B^a&=& {1\over 2}  \e_{jk} F^a_{jk} =  {1\over 2}  \e_{jk} (\del_j A_k^a  - \del_k A_j^a + f^{abc} A^b_j A^c_k )~~~\hskip .2in {\text{(2+1 dim)}}
 \nonumber
 \eeqar
 The equations of motion for the Yang-Mills theory are
 \beqar
 (D_i E_i)^a &=& 0\nonumber\\
 {\del E^a_i \over \del t} &=&\begin{cases}
 - \e_{ijk} (D_j B_k)^a\hskip .2in&{\text{3+1 dim}}\\
  - \e_{ij} (D_j B)^a\hskip .2in&{\text{2+1 dim}}\\
  \end{cases}
 \label{G16a}
 \eeqar
 The first of these is the Gauss law familiar from electrodynamics, now generalized to the nonabelian case. The second is an equation of motion,
 in the sense of defining time-evolution, for the field
 $E^a_i$.
 
 Our aim is to consider the Hamiltonian formulation
 of the YM theory using the functional
 Schr\"odinger formulation.
 From now on, unless specifically indicated, we
 will consider $2+1$ dimensions.
 As a first step, by extending the action in (\ref{G15})
 to a general curved manifold with metric $g_{\mu\nu}$ as 
 $F^a_{\mu\nu}F^{a \mu\nu} \rightarrow \sqrt{-g}\,g^{\mu\alpha} g^{\nu\beta}F^a_{\mu\nu}F^{a}_{\alpha\beta} $ and taking the variation with respect
 $g^{\mu \nu}$, we find the energy-momentum tensor for the theory
 as
 \beq
 T_{\mu\nu} = {1\over e^2} \left[ - \eta^{\alpha\beta} F^a_{\mu\alpha} F^{a}_{\nu\beta}
 + {1\over 4} \eta_{\mu\nu} F^a_{\alpha\beta} F^{a\alpha \beta}
 \right]
 \label{G17}
 \eeq
 This identifies the Hamiltonian as
 \beq
 \H = \int d{\rm V} \, T_{00} = {1\over 2 e^2} \int d {\rm V} \,  ( E^a_i E^a_i + B^a B^a )
 \label{G18}
 \eeq
 To obtain the Poisson brackets, or the commutation rules for the fields in the quantum theory, we need the canonical or symplectic structure for the fields.
 From the term involving time-derivatives of the fields
 in (\ref{G15}), we can identify this as\footnote{
This can be obtained as follows. Consider the action, which depends on a set of fields which we denote generically as $\phi^a$, and which is
defined over the time-interval $[ t, t_0]$.
A general variation of $S$ will have the form
\[
\delta S = \int d{\rm V}  \alpha_a \delta \phi^a\Bigr]^t_{t_0} + \int dt d{\rm V}\,
\E_a (\phi) \delta \phi^a\nonumber
\]
The first term is the surface term on the time-slices at $t$ and at $t_0$.
The second term is an integral over the spacetime region.
The equation of motion is then given by $\E_a (\phi) = 0$.
The quantity $\int d{\rm V} \alpha_a \delta \phi^a$ is the canonical or symplectic
one-form on the space of fields. Its exterior derivative on the space of fields
is the canonical structure $\omega_{\rm symp}$.
In the present case, from the action, with $A_0^a =0$, we get
$\alpha = (1/e^2) \int d{\rm V} E^a_i \delta A^a_i$ which leads to 
\ref{G19}).
}
 \beq
 \omega_{\rm symp} = {1\over e^2} \int d{\rm V} \, \delta E^a_i \, \delta A^a_i
 = \int d{\rm V} \, \delta \Pi^a_i \, \delta A^a_i, \hskip .2in
 \Pi^a_i = {E^a_i \over e^2}
 \label{G19}
 \eeq
 This is to be interpreted as a differential two-form in the space of field
 configurations $(E^a_i, A^a_i )$; we use $\delta$ to denote the exterior
 derivative on the space of fields.
  On the spatial manifold at a fixed time, $E^a_i$ is to be treated as
 an independent variable since it involves the time-derivative of
 $A^a_i$. It is proportional to the canonical momentum
 $\Pi^a_i$ conjugate to $A^a_i$.
 The equal-time commutation rules defined by (\ref{G19}) are
 \beqar
 [A^a_i (x), A^b_j (y) ] &=& 0\nonumber\\
{} [ E^a_i (x), E^b_j (y) ] &=& 0\nonumber\\
{} [ E^a_i (x) , A^b_j (y) ] &=& -i \,e^2 \delta_{ij} \delta^{ab} \delta^{(2)}(x-y)
 \label{G20}
 \eeqar
The commutation rules (\ref{G20}) show that $\Pi^a_i$ is the variable canonically conjugate to
$A^a_i$. There is no variable conjugate to
$A_0^a$. Put another way, the canonical momentum for $A^a_0$ is zero. 
If we augment $\omega_{\rm symp}$ by the addition of a term
$\int \delta \Pi^a_0 \, \delta A^a_0$, then we must carry out a reduction of the phase space
by setting $\Pi^a_0$ to zero as a constraint, $\Pi^a_0 \approx 0$
(in the sense of Dirac's theory of constraints).
As a conjugate constraint, we can use $A^a_0 \approx 0$. Thus the pair
$(\Pi_0^a, A^a_0 )$ will be eliminated from the theory.

The Hamiltonian equations of motion which follow from the canonical brackets are obtained as
 \beqar
 {\del A^a_i \over \del t} &=& E^a_i\nonumber\\
 {\del E^a_i \over \del t} &=& - \e_{ij} (D_j B)^a
 \label{G21}
 \eeqar
 Notice that the first of these equations requires $A^a_0 = 0$ for consistency with the definition in (\ref{G16}). If we did not set $A^a_0 $ to zero, we would need to add a term to the Hamiltonian to obtain the result
 (\ref{G16}). The canonical Hamiltonian and the Hamiltonian defined by
 $T_{00}$ would differ by terms proportional to the constraint.
 With $A^a_0 = 0$, the first of the equations in
 (\ref{G21}) reproduces the definition of $E^a_i$. The second equation agrees with the second of the Lagrangian equations of motion
 in (\ref{G16a}).
 
 In terms of the canonical momentum, the first of the Lagrangian equations
 in (\ref{G16a})
 reads $(D_i \Pi_i)^a = 0$, so it does not involve time-derivatives.
 Therefore  it cannot be reproduced as a Hamiltonian
 equation of motion. For equivalence of the Hamiltonian formulation to the Lagrangian given as (\ref{G16a}), we have to impose 
  $(D_i \Pi_i)^a = 0$ as an additional condition.
  It should be viewed as a constraint on the 
 phase space variables or on the initial data.
 
 We have restricted the field variables (by use of the freedom of gauge transformations) to some extent by setting
 $A^a_0 = 0$. But the theory would still allow for gauge transformations $g$
 which do not depend on time, so that they preserve the condition
 $A^a_0 = 0$. The constraint $D_i E_i = 0$  may be viewed as the statement of this residual gauge freedom. We can then choose a constraint conjugate to
 $D_i E_i$, say $\nabla \cdot A \approx 0$ for example,
 and carry out a further canonical reduction to obtain $\omega_{\rm symp}$
 on the reduced phase space (where $D_i E_i= 0$ and $\nabla_i A_i = 0$).
 We can then formulate Poisson brackets and commutators in terms of this reduced
$\omega_{\rm symp}$. This is the approach of gauge-fixing, $\nabla_i A_i = 0$ being the gauge-fixing condition. Alternatively, in the quantum theory
 we can impose $D_i E_i = 0$
 not as an operator condition but
 as a condition on states or wave functions. This is the approach we will be pursuing.
 
 As is well-known, conditions imposed in terms of operators should be understood as valid with suitable smearing using test functions.
 The nature of the test functions is crucial to determining the physical
 consequences of the theory.
 We consider the smeared operator
 \beq
 G_0 (\theta ) = \int d{\rm V} \,\theta^a (D_i \Pi_i )^a
 \label{G23}
 \eeq
 If we impose the condition 
 \beq
 G_0 (\theta ) \Psi = 0 
 \label{G23a}
 \eeq
 on the wave functions $\Psi$ in the theory,
 for consistency, we will also need the commutator
 $[G_0 (\theta ), G_0 (\theta')] $ to vanish on $\Psi$.
 From the canonical commutation rules (\ref{G20}) it is easy to check that
 \beqar
 [G_0(\theta), G_0 (\theta' ) ] &=& i G_0 (\theta \times \theta') 
 + i \oint_{\del {\rm V}}  (\theta \times \theta')^a \Pi_i^a \, dS_i
 \label{G23b}\\
  (\theta \times \theta')^a &\equiv& f^{abc} \theta^b \theta'^c
  \nonumber
 \eeqar
We see that we cannot consistently impose (\ref{G23a}) unless
$\Pi^a_i$ vanishes fast enough as we approach $\del {\rm V}$ or
at spatial  infinity.
This would in turn amount to requiring all charges to vanish (this will be clearer soon), which is not something we can impose {\it a priori} in the theory. The only other option is to require
the test functions to vanish on $\del {\rm V}$. In this case, the surface term in
(\ref{G23b}) will vanish and we have a closed algebra
for the $G_0(\theta)$'s and the condition
(\ref{G23a}) can be consistently imposed.
 In terms of its action on fields, we find
 \beqar
 e^{i G_0 (\theta )} \,\left[ \int d{\rm V}\, A^a_i v_i \right] \,
 e^{- i G_0 (\theta )} &=&\int d{\rm V}\, A^a_i v_i
 + i [ G_0 (\theta ), \int d{\rm V}\, A^a_i v_i] + \cdots\nonumber\\
 &=& \int d{\rm V} \,A^a_i v_i + \int d{\rm V}\, \theta^a (\nabla_i v_i) -
 f^{abc} \int d{\rm V}\, A^b_i \theta^c v_i + \cdots\nonumber\\
 &=& \int d{\rm V}\, A^a_i v_i - \int d{\rm V}\,  (D_i \theta)^a v_i  + \cdots
 + \oint_{\del {\rm V}} \theta^a v_i dS_i\nonumber\\
 &=& \int d{\rm V}\, (A_i - D_i \theta )^a v_i , \hskip .2in {\text{if}}~~
 \theta^a \rightarrow 0 ~~{\text{on}} ~ \del V
 \label{G24}
 \eeqar
 For the electric field we find
 \beq
  e^{i G_0 (\theta )} \,\left[ \int d{\rm V}\, E^a_i w_i \right] \,
 e^{- i G_0 (\theta )} = \int d{\rm V}\, E^a_i w_i + f^{abc}\int d{\rm V}\, \theta^b E^c_i
w_i + \cdots
 \label{G25}
 \eeq
  (In (\ref{G24}) and (\ref{G25}),  $v_i$ and $w_i$ are test functions for $A_i^a$ and $E^a_i$.)
 The right hand sides of these equations
  are of the form of infinitesimal gauge transformations
 (\ref{G10}), (\ref{G13}) with $g= e^{ -i t^a \theta^a} 
 \approx 1 - i t^a \theta^a$.
This means that the operator $G_0 (\theta)$ will generate infinitesimal
 gauge transformations of $(A^a_i, E^a_i )$ provided
 $\theta^a$ vanishes at spatial infinity (or on the boundary of the spatial volume under consideration).
 Since the Hamiltonian is invariant under gauge transformations,
 $[ G_0 (\theta ), \H ] = 0$, and hence the requirement $G_0 (\theta )  \Psi
 = 0$ will be preserved under time-evolution as well.
 The closed algebra (\ref{G23b}) is a statement of the
group property that a sequence of infinitesimal transformations of the form (\ref{G24}), (\ref{G25}) can be used to generate a
 finite transformation. 
 We can now define an infinite dimensional group
 $\G_*$ as follows:
 \beq
 \G_* = \{ {\text{Set of}}~ g(x): {\text{Space}}~ \rightarrow G ~{\text{such that}}~
 g(x) \rightarrow 1~  {\text{on}}~ \del V\}
 \label{G25a}
 \eeq
 If we consider all of $\mathbb{R}^2$, we may define
 $\G_*$ as
 \beq
 \G_* = \{ {\text{Set of}}~ g(x):  \mathbb{R}^2 \rightarrow G ~{\text{such that}}~
 g(x) \rightarrow 1~  {\text{as}}~ \vert \vx \vert \rightarrow \infty\}
 \label{G25b}
 \eeq
 The condition (\ref{G23a}) is the statement that
 all wave functions in the theory are invariant under 
 gauge transformations $g \in \G_*$.
 In this sense, $\G_*$ is the true gauge group of the theory.
 To distinguish wave functions or states which are more general
 and do not necessarily obey (\ref{G23a}), we refer to states satisfying
 (\ref{G23a}) as ``physical states".
 
Given that states or wave functions
obey (\ref{G23a}), for the matrix element of an operator ${\cal O}$
we can write
\beqar
\bra{\Psi_1} {\cal O} \ket{\Psi_2}
&=& \bra{\Psi_1} e^{i G_0 (\theta) } {\cal O} e^{-i G_0(\theta) }  \ket{\Psi_2} 
\nonumber\\
&=& \bra{\Psi_1} {\cal O} \ket{\Psi_2} + i \bra{\Psi_1}[G_0(\theta), {\cal O} ]\ket{\Psi_2} + \cdots
\label{G25c}
\eeqar
This will give an inconsistent result unless
we have $[G_0(\theta), {\cal O} ] = 0$. 
Therefore, we can say that an operator
 ${\cal O} $ is an observable and can have well-defined matrix
 elements only if it weakly commutes with $G_0 (\theta )$,
 i.e., if $\bra{\Psi_1} [ G_0 (\theta ), {\cal O} ] \ket{\Psi_2} = 0$,
 for all physical states $\Psi_1$, $\Psi_2$.

 We now turn to another set of transformations of interest.
 Towards this, we first consider transformations of the type
 (\ref{G10}), (\ref{G13}) where $g \in G$ is a constant not necessarily
 equal to one
 on the spatial manifold; i.e., the transformations are
 \beq
 A_i \rightarrow g\, A_i g^{-1}, \hskip .2in E_i \rightarrow g\, E_i g^{-1}
 \label{G25d}
 \eeq
 The Hamiltonian (\ref{G18}) is clearly invariant under these.
 Further, this is not a gauge transformation and cannot be removed
 by the choice of a suitable element of $\G_*$ since elements of
 $\G_*$ must become the identity on $\del {\rm V}$ or at spatial infinity.
  So these transformations (\ref{G25d}) generate a 
 Noether-type symmetry and
 the states of the system can be classified by representations of the
 group $G$. In fact the transformations (\ref{G25d}) with constant
 $g$'s correspond to charge rotations and the states
 in the various irreducible representations of $G$ correspond to states with
 different possible charges.
 Notice that, by
choice of the action of $G_0 (\theta)$, we can go from
 $g$ to $g'(x) g$ where $g'(x) \rightarrow 1$ on the boundary.
 This will allow us to change the value of $g$ in the bulk, 
but the combined transformation
 $g'(x) g$ still has the value $g$ (which is not necessarily the identity)
 on the boundary.
 Thus it is really the asymptotic value of the group element that
 defines the Noether symmetry. With this in mind, it
is also useful to consider another set of operators
 \beq
 G(\theta ) = - \int d{\rm V}\, (D_i \theta)^a \Pi^a_i
 \label{G26}
 \eeq
 These coincide with $G_0 (\theta )$ for those test functions 
 $\theta^a$ which vanish on $\del {\rm V}$, but, in general, we can consider
 $G(\theta )$ even for those functions
 $\theta^a$ which do not vanish on $\del {\rm V}$.
 (Notationally, we distinguish the two by using the subscript on
 $G_0 (\theta )$ to indicate that it is for the case when
 $\theta^a \rightarrow 0$ on $\del {\rm V}$.)
 It is easy enough to check that
 \beqar
 i [ G(\theta ), \int d{\rm V}\, A^a_i v_i ] &=& \int d{\rm V}\, (A_i - D_i \theta )^a
 v_i\nonumber\\
  {}[G (\theta ), G (\theta') ] &=& i \, G (\theta \times \theta')
  \label{G27}
  \eeqar
 So $G(\theta )$ does generate gauge transformations 
 (as in (\ref{G10}), (\ref{G13})) even for
 $\theta^a \neq 0$ on $\del {\rm V}$. (But recall that these are not true gauge transformations as they are not elements of $\G_*$.)
 We can use the freedom of gauge transformations by
 $G_0 (\theta')$ to change the value of $\theta$ everywhere
 except on the boundary. Thus $G(\theta)$ is characterized by the boundary value
 $\theta$ (modulo the action of $G_0 (\theta' )$).
The commutation rules also give
\beqar
G_0 (\theta' ) \, G(\theta ) \,\Psi &=& G(\theta )\, G_0(\theta') \,\Psi
+i G_0(\theta'\times \theta )\,\Psi\nonumber\\
&=& 0
\label{G28}
\eeqar
 so that $G(\theta )\Psi$ are also states compatible with the 
 requirement of (\ref{G23a}).
 In other words, the action of $G(\theta )$ on $\Psi$'s will generate
 physical states in the theory.
 Among the  operators  $G(\theta)$ there are the ones
 mentioned earlier where
 $\theta^a$ on $\del {\rm V}$ or spatial infinity is a constant
 (that is, independent of
 angular directions), but not necessarily the identity.
 These generate charge rotations and hence they lead to
 the charged states of the theory. 
  More generally, the operators
 $G(\theta)$ for those $\theta$ which may have nontrivial
 angular dependence or is a nonconstant function on $\del {\rm V}$
 generate observable dynamical degrees of freedom localized
 on the boundary. They are usually referred to as edge states.
 Notice that $[G_0(\theta' ), G(\theta )] = 0$.
 
 The fact that the wave functions corresponding to physical states are gauge invariant means that their normalization has to be defined with a gauge-invariant volume element. Since $A_i^a $ at different spatial points commute, we can consider $A$-diagonal wave functions
 $\Psi (A)$. (We can equally well consider $E$-diagonal ones, but for the
 moment we stay with $\Psi (A)$.)
 Thus $\Psi_1^* \Psi_2$ for physical states will be gauge-invariant and integration
 over all configurations $A$ will clearly diverge. To define the proper volume element, we start by defining 
 \beq
 \A = \{ \text{ Set of all  gauge potentials such that } F_{ij} \rightarrow 0
~ \text{as}~ \vert \vx \vert \rightarrow 0\}
\label{G29}
 \eeq
 We impose a mild condition on the gauge potentials. Also, here, by gauge potential we mean a Lie-algebra valued one-form on the spatial manifold,
 $A = (-i t^a) A^a_i dx^i$. This space is actually an affine space,
 i.e., any two points on $\A$ can be connected by a straight line
 as
 \beq
 A(\tau )= A^{(1)} \, (1- \tau ) + A^{(2)} \, \tau
 \label{G30}
 \eeq
 where $\tau$ is a real parameter $0\leq \tau \leq 1$.
The straight line (\ref{G30}) connects $A^{(1)} $ at $\tau = 0$ to
 $A^{(2)}$ at $\tau = 1$. The key point here is that, for any
 value of $\tau$, $A (\tau)$ transforms as a gauge potential,
 \beq
 A^g(\tau ) = g \, A(\tau ) \, g^{-1} - d g \, g^{-1}
 \label{G31}
 \eeq
 Hence the entire straight line (\ref{G30})  is in $\A$.
Because of this
property, the topology of the $\A$ is trivial, it is a flat contractible space.
We can then consider the space $\C = \A / \G_*$ which is the space of
all gauge potentials modulo gauge transformations.
The configurations of the form $A^g = g A g^{-1} - dg g^{-1}$, for $g \in \G_*$,
give the orbit of $A$ under the action of $\G_*$. So $\C$ will also be
referred to as the space of $\G_*$-orbits in $\A$, or the gauge-orbit
space, for short.
This is the space of physical configurations. (If we consider the
phase space, it will also
have the momentum conjugate to the variables
in $\C$.)
The wave functions are defined as functions on $\C$.
Therefore the inner product for states should be defined with an integration measure (or the volume element) for the space
$\C$. Expressed mathematically,
\beq
\braket{1|2} = \int d\mu (\C) \, \Psi_1^* \, \Psi_2
\label{G32}
\eeq
A similar statement can be made if we choose to represent states
by wave functions which are functions of $E$ as well.

The Hamiltonian (\ref{G18}) in terms of its action on $\Psi$ can be written
as
\beq
\H\, \Psi = {1\over 2 } \int dV \,\left[ - e^2 {\delta^2 \over \delta A^a_i \delta A^a_i } 
+ {B^a B^a\over e^2} \right] \, \Psi
\label{G33}
\eeq
where we have used the functional Schr\"odinger representation of
$E^a_i$,
\beq
E^a_i = -i e^2 {\delta \over \delta A^a_i} 
\label{G34}
\eeq
The functional differential operator, or the kinetic energy term in
(\ref{G33}) is the functional Laplace operator on the space $\A$.
But since it acts on $\Psi$'s which are gauge-invariant,
it can be viewed as the Laplace operator on the space $\C$.

We are now in a position to assemble the ingredients needed for the
Hamiltonian formulation of the theory.
First of all, the Hamiltonian has the form
\beq
\H  = {1\over 2 } \int d{\rm V} \,\left[ - e^2 \Delta_\C
+ {B^2\over e^2} \right] \, \Psi
\label{G41}
\eeq
where $\Delta_\C$ is the Laplace operator on the configuration space
$\C$. The wave functions themselves are gauge-invariant, i.e., defined
as functions on $\C$. Their inner product for states
$\ket{1}$ and $\ket{2}$ is given by
(\ref{G32}), where $d\mu (\C)$ is the volume element 
on $\C$.\footnote{While the wave functions are gauge-invariant, if we consider different coordinate patches on $\C$, they
 may require nontrivial transition functions as we move from one patch to another. Thus more
accurately the wave functions are sections of a line bundle on
$\C$. Since we will not be considering different coordinate
patches on $\C$ for most of our discussion,
this qualification is not important at this point.
See however, the discussion in Appendix \ref{TopC}.}

Thus the key ingredients we need to calculate are
the Laplacian $\Delta_\C$ and the volume element
$d\mu (\C)$. Both of these have to be defined with
suitable regularizations, as would be the case for any field theory.
Further, as mentioned in section \ref{Intro},
since we are using the Hamiltonian approach, we do not have
manifest Lorentz invariance. So we do have to verify that the regularizations are compatible with Lorentz invariance.
The final ingredient to getting physical predictions would be a method to solve the Schr\"odinger equation, once we have the
inner product and the regularized Hamiltonian operator.

\section{Confinement}\label{Conf}
\setcounter{equation}{0}

One of the key features of a nonabelian gauge theory is 
the possibility of confinement of particles or fields in nontrivial representations of the gauge group. As indicated in the last section, {\it a priori}
we should allow for charged states which are generated by
$G(\theta)$ which was defined in (\ref{G26}). Confinement refers to the 
statement that, in the nonabelian Yang-Mills theory,
the dynamics is such that
there are no charged states in the physical spectrum. 
Put another way, such states have infinite energy and 
therefore cannot be dynamically excited.
Although this is not a proven fact, there are strong indications to
support the idea of confinement.
However, a direct analysis of the spectrum of the Hamiltonian, with a view to elucidating confinement, has not yet been successful.
A possible strategy would then be to look for observables which can serve as useful diagnostics of confinement and to try to calculate them in some way.
 The most important among these is the Wilson loop
operator defined by
\beq
W_R(C) = \Tr \left[\P \exp\left( -\oint_{C} A_\mu dx^\mu
\right)\right]
\label{Conf1}
\eeq
Here $A_\mu = -i T_a A_\mu^a$ and $T_a$, which are the generators of the Lie algebra,
are in the representation $R$. This is indicated by the subscript on
$W(C)$. The integral is over a closed curve $C$. 
The matrices $A_\mu$ at different 
points along the curve do not commute in general, likewise
$A_\mu$ and $A_\nu$ do not commute in general.
So there has to be an ordering prescription in how the line integral
is evaluated. This is taken to be path-ordering, by which we mean the following. Let us parametrize the curve as $x^\mu (\tau)$,
$0\leq \tau\leq 1$, 
and divide the interval of $\tau$ into a sequence of
infinitesimal segments, say $n$ of them, 
each of extent $\e$. Thus we have a set of points
$0$, $\tau_1$, $\cdots$, $\tau_n$, $\tau_{i+1} - \tau_i = \e$,
with $\tau_n = 1$. 
We take $n \rightarrow \infty$ and $\e \rightarrow 0$ in the end as usual.
Then the path-ordered integral from $x^\mu = x^\mu (0)$
to $y^\mu = x^\mu (\tau_n) = x^\mu (1)$ is given by
\beqar
W(y, x, C) &=&
\P \exp\left( - \int_{x}^y A_\mu dx^\mu \right) \nonumber\\
&=& \exp\left( - \int_{\tau_{n-1}}^{\tau_n} A_\mu {dx^\mu\over d\tau} d\tau
\right)  \exp\left( - \int_{\tau_{n-2}}^{\tau_{n-1}} A_\mu {dx^\mu\over d\tau} d\tau
\right)\cdots\nonumber\\
&&\cdots  \exp\left( - \int_{\tau_{0}}^{\tau_{1}} A_\mu {dx^\mu\over d\tau} d\tau
\right)\nonumber\\
 &=&\exp\left( - \int_{x(\tau_{n-1})}^{y} A_\mu dx^\mu
\right)  \exp\left( - \int_{x(\tau_{n-2})}^{x(\tau_{n-1})} A_\mu dx^\mu
\right)\cdots
\label{Conf2}
\eeqar
For an open interval, we have the gauge transformation property
\beq
\left[\P \exp\left( - \int_{x}^y A^g_\mu
dx^\mu
\right) \right]^{ij} = \left[g(y)~\P \exp\left( - \int_{x}^y A_\mu
dx^\mu
\right)~g^{-1}(x)\right]^{ij}
\label{Conf3}
\eeq
This follows from the fact that $W(y, x, C)$ obeys
\beq
\left[ {\del \over \del y^\mu} + A_\mu (y) \right]\, W(y,x, C) = 0
\label{Conf3a}
\eeq
For the closed curve, we have $y^\mu = x^\mu$ and we take the trace of
the resulting expression to define $W(C)$ as in (\ref{Conf1}).
The transformation property (\ref{Conf3})
 shows that once we close the curve and take the trace, we get a gauge-invariant quantity.
The Wilson loop operators are thus observables of the theory.
In fact, by choosing all possible closed curves, we get an
over-complete set of observables. All other observables can be
constructed from $W_R(C)$.

The Wilson loop operator is important for another reason as well.
The expectation value of $W_R(C)$ is related to the interaction energy
of a heavy particle-antiparticle pair belonging to the representation $R$ and
its conjugate. Such a pair can be used as a probe into the dynamics of 
the gauge theory. They are taken to be heavy so that their own dynamics is trivial and does not complicate the interpretation of the result, since the focus is on the gauge theory.

In order to relate $W_R(C)$ to the energy of a particle-antiparticle pair, we
will start by considering the process where we start with
a heavy static particle-antiparticle pair separated by a spatial distance
$L$ at a certain time $x^0$. 
We will use $\phi$ and $\phi^\dagger$ as the annihilation
and creation operators for the particle;
$\chi$ and $\chi^\dagger$ will play a similar role for the antiparticle.
Since these are taken to be heavy, the action for these fields is
just the usual nonrelativistic action, but we can even omit the
$(\nabla^2/ 2 {\rm M})$-part. Thus
\beq
{S} (\phi ,\chi )= \int dt d{\rm V}~ \left[ i \phi^{\dagger} D_0 \phi
+ i \chi^\dagger D_0 \chi  \right] \label{Conf4}
\eeq
where $D_0 \phi = \del_t \phi +A_0 \phi$
and $ D_0 \chi = \del_t +A^*_0 \chi$ are the covariant derivatives of
$\phi$ and $\chi$, respectively. This is in accordance with the fact that
the fields transform under gauge transformations as
$ \phi \rightarrow  g \phi$, $\chi \rightarrow g^*\chi$.
We start with a gauge-invariant state corresponding to the particle-antiparticle pair 
separated by a spatial distance $L$. This state can be represented as
\beq
F^\dagger (x^0, x^1, x^1+L)\ket{0} 
= \phi^{\dagger i}(x)\, W^{ij}(x,y) ~\chi^{\dagger j}(y)\, \ket{0}
\label{Conf5}
\eeq
where $x= (x^0 , x^1)$, $y =(x^0, x^1+L)$ and $W^{ij}(x,y)$ is as
in (\ref{Conf2}) over, say, a straight line segment.  We have taken the
separation of the pair to be along the $x^1$-direction, for simplicity.

Let $\H$ be the Hamiltonian for the Yang-Mills theory coupled to
these matter fields $\phi$, $\chi$.
As usual, we can set
$A_0$ to zero; the $A_0$-dependent terms in (\ref{Conf4})
are then zero but will contribute to $\H$ via the Gauss law, which now
takes the form
\beq
{(D_i E_i)^a\over e^2} + \phi^\dagger T^a \phi + \chi^\dagger(- {\tilde T}^a )\chi = 0
\label{Conf6}
\eeq
Here ${\tilde T}^a $ is the transpose of $t^a$, corresponding to the conjugate representation. ($T^a \rightarrow - {\tilde T}^a$ is the conjugation operation in the Lie algebra.)

We now consider the time-evolution of the state (\ref{Conf5}) by an imaginary amount
$-iT$ and then take its overlap with (\ref{Conf5}). The amplitude for this is
given by
\beq
\la 0\vert F ~e^{-HT} F^\dagger \vert 0\ra \approx
{\cal N} ~e^{-E(L)T} \label{Conf7}
\eeq
where ${\cal N}$ is some prefactor related to the normalization of $F$,
and $E(L)$ is the energy of the pair.
We are interested in taking $T$ to be large, so that
$E(L)$ will be the energy
of the lowest energy state which can be created by $F^\dagger$. 
Since the particles are heavy and static, $E(L)$ is basically just 
the interaction energy of
the pair due to the gauge field. 

By the usual technique of the slicing of the time-interval,
we can represent this amplitude as a
Euclidean functional integral
\beqar
\la 0\vert F ~e^{-HT} F^\dagger \vert 0\ra 
&=&
\int [dA\, d\phi\, d\chi ]\,
\exp \left[-{S}_E(A, \phi ,\chi )\right]~\nonumber\\
&&\hskip .2in\chi^i (y') W^{*j i} (y', x') \phi^j (x')~\phi^{\dagger r} (x)
W^{rs} (x,y)
\chi^{\dagger s}(y)
\label{Conf8}
\eeqar
where $x' =( x^0 +T , x^1)$,
$y' =(x^0 +T , x^1+L)$.
The $(\phi, \chi)$-part of the Euclidean action which appears in this
functional integral is given by
\beq
{S}_E (\phi ,\chi ) = \int d\tau d{\rm V}~ \left[ \phi^{\dagger}
{\del \phi\over \del \tau} +  \chi^\dagger {\del \chi \over
\del \tau}  \right] \label{Conf9}
\eeq
This leads to the propagators
\beqar
\la \phi^i(x) \phi^{\dagger j} (x') \ra &=& \delta^{ij}\,
\theta (\tau -\tau')\, \delta (\vx- \vy)\nonumber\\
\la \chi^i(x) \chi^{\dagger j} (x') \ra &=& \delta^{ij}\,
\theta (\tau -\tau')\, \delta (\vx- \vy)\label{Conf10}
\eeqar
where $\theta (\tau - \tau')$ is the step function and $\tau$ denotes the Euclidean time-coordinate.
The amplitude in (\ref{Conf8}) then reduces to
\beqar
\la 0\vert F ~e^{-HT} F^\dagger \vert 0\ra 
&=&\int [dA] ~e^{-{S}_{\rm YM} } ~ W^{ij} (y', x') ~W^{ji}(x,y) \nonumber\\
&=& \int [dA] ~e^{-{S}_{\rm YM} }~ W_R (C)\nonumber\\
&=&\la W_R (C) \ra
\label{Conf11}
\eeqar
where $C$ is the rectangle with vertices $x, y, x', y'$. 
Since $A_0 =0$, we can put in the two time-like segments
for free to complete the loop.
Comparing this expression with (\ref{Conf7}), we see that
\beq
\la W_R (C) \ra \approx {\cal N} 
e^{-E(L) T}\label{Conf12}
\eeq
This shows that the Euclidean expectation value of a large Wilson loop
can be used to identify
 the interaction energy of a heavy static particle-antiparticle
pair. 
Even though we used the $A_0 =0$ gauge, $W_R(C)$ is 
gauge-invariant, and so are energies of gauge-invariant states.
Thus the result holds true in general. 

If the interaction energy $E(L)$ increases with the separation $L$,
say, $E(L) \rightarrow \infty$ as $L \rightarrow \infty$,
then it will cost arbitrarily large energy to remove a charged particle from its
conjugate to an arbitrarily far away point,
 if the pair is created by any process. This is what 
 we expect if there is confinement.
 In the case of nonabelian gauge theories, the expectation is that
 the interaction energy will grow linearly with $L$,
 i.e.,  $E(L) =\sigma L$. The coefficient $\sigma$ is known as the
 string tension.
 In terms of the Wilson loop, this statement is expressed as
\beqar
\la W_R (C)\ra &\approx& {\cal N} \exp (-\sigma ~L \,T )\nonumber\\
&\approx& {\cal N} \exp (- \sigma ~A_C)
\label{Conf13}
\eeqar
where $A_C$ is the area of the minimal surface whose boundary is
$C$. 

The use of the term ``string tension" is related to the following
qualitative picture of confinement.
If we consider a heavy particle-antiparticle pair, the 
expectation is that
the chromoelectric flux lines connecting the particle 
and the antiparticle are collimated to a thin tube
of flux, which we refer to as the string, by the properties of 
the vacuum. Since the energy of a string would increase linearly with
the length, the proportionality factor being the tension of the string,
this picture
would explain the linear rise of the potential.

Equation (\ref{Conf13}) shows that the area-law behavior of the expectation value $\la W_R (C)\ra$ 
can be used as a test of confinement. This works for all
representations which cannot be screened.
Since the average in $\la W_R (C)\ra$ is done with the Yang-Mills
action, the theory allows for the dynamical generation of gluons,
which belong to the adjoint representation of the group $G$.
Thus when we impart energy to a particle-antiparticle pair, separating
the constituents, $E(L)$ can grow to a point where
it becomes possible to create a number of gluons spontaneously.
If the representation $R$ is such that
$R \otimes ({\rm Adjoint})$ (or $R \otimes {\rm Adjoint} \otimes {\rm Adjoint} \cdots$) contains the trivial representation\footnote{Sometimes one needs the tensor product of $R$ with several adjoint representations to get a trivial representation upon reduction. An example is the group $G_2$,
for which the fundamental representation $\underline{7}$
can be screened by three gluons, 
i.e., $\underline{7} \otimes {\rm Adjoint} \otimes {\rm Adjoint} \otimes {\rm Adjoint}
\supset 1 \oplus \cdots$. The product with
several {\rm Adjoint}s in parentheses is included to take care of such possibilities.}
(these are called screenable representations),
then the pair-configuration can decompose into a particle-gluon(s)
state (of zero charge) and an antiparticle-gluon(s) state (also of
zero charge).
The interaction energy between these composites is no longer $E(L)$,
since each has zero charge,
so they can be separated far from each other. Correspondingly, 
$\la W_R (C)\ra$ will not exhibit an area law. 
Thus, while confinement
continues to be true (since the particle-gluon(s) state and the
antiparticle-gluon(s) state each has zero charge), 
the expectation value of the Wilson loop is no longer a good
diagnostic tool. 

The picture in terms of the string of flux connecting the particle-antiparticle pair is that the string breaks by the spontaneous production of gluons, which leads to
new composites of zero charge and hence there is no longer any string of
flux connecting these states.

From the argument given above, we see that,
strictly speaking, $\la W_R (C)\ra$  is useful only for nonscreenable
representations, namely, those for which
$R \otimes {\rm Adjoint} \otimes {\rm Adjoint} \cdots$ does not contain the trivial representation.
(While confinement is obtained for screenable representations as well,
$\la W_R (C)\ra$ is not a good diagnostic for it.)
Nevertheless, our argument with $E(L)$ shows that we should expect
the area law to hold until $E(L)$ becomes large enough to create a pair (or more in some cases) of gluons. So for a limited range of $L$, the area law 
for $\la W_R (C)\ra$  can still be obtained and can still be useful.

\section{Parametrization of gauge fields}\label{Par}
\setcounter{equation}{0}
\begin{quotation}
\fontfamily{bch}\fontsize{10pt}{16pt}\selectfont
\noindent 
In this section we show that the complex combinations of the spatial components of the gauge potentials can be parametrized as
$A= - \del M M^{-1}$, $\bA = M^{\dagger -1} \bdel M^\dagger$,
where $M$ is a complex invertible matrix.
This is done by using the Hodge decomposition of a vector and noting that it has the form of an infinitesimal pure gauge transformation with complex
parameters.
A similar parametrization is also shown for the sphere $S^2$ using
group theoretic arguments.
\end{quotation}
\fontfamily{bch}\fontsize{12pt}{16pt}\selectfont
We will now consider a special parametrization for the gauge fields which will facilitate working out the Hamiltonian and the volume element $d\mu (\C )$
in terms of manifestly gauge-invariant variables \cite{Ati-B}, see
also \cite{Bos-N}.
We are primarily interested in Yang-Mills theories on flat
$(2+1)$-dimensional space, so the spatial manifold is
$\mathbb{R}^2$. 
The two spatial coordinates $x^1$, $x^2$ can be combined
into the complex combinations
$z = x^1- i x^2$, $\bz = x^1 +i x^2$, with the corresponding derivatives
\beq
\del \equiv \del_z = {1\over 2} (\del_1 + i \del_2), \hskip .2in
\bdel \equiv\del_\bz = {1\over 2} (\del_1 - i \del_2 )
\label{par1}
\eeq
As explained before, we can take
$A_0 = 0$.
For the Abelian gauge theory, for the spatial components of $A$, we can use the Hodge decomposition
\beq
A_i = -i (\del_i \theta + \e_{ij} \del_j \vf )
\label{par2}
\eeq
for real functions $\theta$ and $\vf$ on $\mathbb{R}^2$.
Here we use antihermitian $A_i$ so that the covariant derivative is
$(\del_i + A_i )$, similar in form to what is usually used for the nonabelian case.
For the complex components, we can write
\beq
A \equiv A_z = {1\over 2} (A_1 + i A_2) = 
-\del_z \Theta, \hskip .2in \Theta = \vf + i \theta
\label{par3}
\eeq
with $\bA \equiv A_\bz = - (A_z)^\dagger$. 

The gauge potentials for the nonabelian case are of the form
$A_i = (-i t_a) A^a_i$. We will consider the gauge group $SU(N)$ for simplicity, so that $t_a$ may be taken as $N\times N$ hermitian traceless matrices. For a small neighborhood around  $A= 0$, 
the fields may be considered as Abelian and
we expect  a result similar to (\ref{par3}).
We may thus write
\beq
A \equiv A_z = -\del_z \Theta + {\cal O} (\Theta^2)
\label{par4}
\eeq
where $\Theta$ is also an $N \times N$ traceless matrix.
Because it is complex, we may regard it as the group
parameter of an element of $SL(N, \mathbb{C})$ (represented as
an $N \times N$ matrix). The expression (\ref{par4}) is then of the
form of a pure gauge near the identity in $SL(N, \mathbb{C})$, i.e.,
for an element $M = e^\Theta \approx 1 + \Theta$.
We can then ``integrate" (\ref{par4}) (i.e., compose it with a series of
infinitesimal group translations in $SL(N, \mathbb{C})$) and write it in the form
\beq
A = - \del M \, M^{-1}, \hskip .2in M \in SL(N, \mathbb{C})
\label{par5}
\eeq
With $A_\bz = - (A_z)^\dagger$, the full parametrization is thus
\beq
\boxed{
A = -\del M\, M^{-1}, \hskip .2in
\bA = M^{\dagger -1} \bdel M^\dagger}
\label{par6}
\eeq 
While we have obtained this result for the group $SU(N)$, it is easy to see how it will generalize. For a Lie group $G$, $\Theta$ is combination of the generators of the group with complex coefficients, so the parametrization
(\ref{par6}) will hold in general with $M$ as an element of
the complexification $G^\mathbb{C}$ of the group $G$.
(It may be worth emphasizing that while $A$ has the form of a pure gauge
for $G^{\mathbb{C}}$, it is not a pure gauge when the allowed
gauge transformations
are in $G$.)

In (\ref{par2}), the term $\del_i \theta$ denotes the gauge transformation for 
the group $U(1)$. More generally, for the nonabelian case,
 gauge transformations take the form\footnote{There are other ways to parametrize $A$'s. One could even use $A_z = -\del_z \Theta$, without any further terms of order $\Theta^2$. In this case, $\Theta$ will transform in a rather complicated way under gauge transformations. The 
 simple transformation law (\ref{par7}) is the real advantage of using
 the $SL(N, \mathbb{C})$ version.}
 \beq
 M \rightarrow g\, M, \hskip .2in  g \in SU(N) ~({\text{or more generally~} \in G})
 \label{par7}
 \eeq
 The gauge invariant degrees of freedom are thus given by
 \beq
 H = M^\dagger M
 \label{par8}
 \eeq
  The factors of $g$ and $g^\dagger$
 in the transformation of $M$, $M^\dagger$ 
 cancel out and $H$ is invariant. Since $M$ modulo the
 $SU(N)$ transformations $g$ define $SL(N, \mathbb{C})/SU(N)$, 
 the gauge-invariant degrees of freedom can be taken as
 the set of mappings from $\mathbb{R}^2$ to this
 coset space $SL(N, \mathbb{C})/SU(N)$ 
 (or more generally to $G^\mathbb{C}/G$).
 The hermitian matrix $H$ parametrizes the coset 
 $SL(N, \mathbb{C})/SU(N)$.
 
 The advantage of the parametrization (\ref{par6}) is precisely that
 the gauge transformations take the homogeneous form
 in (\ref{par7}), as left translations by $G$ on the matrix $M$, so that
 we can easily identify all gauge-invariant degrees of freedom.

There is another way to argue for the parametrization
(\ref{par6}). We can obtain a similar parametrization on
$S^2$, viewed as the complex projective space $\mathbb{CP}^1$, and then take a large radius limit to get the result
(\ref{par6}) for $\mathbb{R}^2$.
(The parametrization of gauge fields for this case has been worked out
in \cite{Aga-N}.)
The space $\mathbb{CP}^1 \sim S^2$ is equivalent to the
coset space
$SU(2)/ U(1)$. We can thus use an element $u$ of $SU(2)$ as coordinates
for $\mathbb{CP}^1$, with the identification
$u \sim u \, h$, $h \in U(1) \subset SU(2)$. Local coordinates
$z$, $\bz$ can be related to this using the parametrization
\beq
u = {1 \over \sqrt{1 + \bz z}} \left( \begin{matrix}
1&z\\ -\bz & 1\\ \end{matrix} \right) \left( \begin{matrix}
e^{i \alpha /2} & 0\\ 0& e^{-i \alpha/2} \\ \end{matrix} \right)
\label{par9}
\eeq
The $U(1)$ angle $\alpha$ can be eliminated via the identification
$u \sim u h$.
We can define three coordinates $x^a$
by $u \sigma^3 u^{-1} = - \sigma^a\, x^a$.
Clearly $x^a$ are invariant under
$u \rightarrow u h$, so they can be viewed as coordinates
on the coset space $SU(2)/ U(1)$.
Further, $\sigma\cdot x \, \sigma\cdot x = 
u \sigma^3 u^{-1} u \sigma^3 u^{-1}  = 1$, so that
$x^a x^a =1$.
Explicitly, for the parametrization (\ref{par9}),
\begin{align}
x^1 = {z + \bz \over 1+ \bz z}, \hskip .2in
x^2 = i {z-\bz \over 1+ \bz z}, \hskip .2in
x^3 &= { \bz z - 1\over  1+ \bz z }
\label{par10}\\
(x^1)^2 + (x^2)^2 + (x^3)^2 &= 1\nonumber
\end{align}
These correspond to the
embedding of $S^2$ in $\mathbb{R}^3$, with
a stereographic projection onto the complex plane,
with the south pole mapped to $z = 0$
and the north pole mapped to $\vert z\vert \rightarrow \infty$.
These coordinates cover $S^2$ except for a small region around
the north pole. (A second coordinate patch can be used around the north pole, by choosing $e^{i \alpha /2} = \sqrt{z/\bz}$ (away from the south pole, so $z \neq 0$). Effectively this amounts to an inversion of
$z$. The two coordinate patches will give full coverage of the sphere.)
The metric on the coset space $SU(2)/U(1)$ is the
Fubini-Study metric for $\mathbb{CP}^1$ given by
\beq
ds^2 = {dz\, d\bz \over (1+ \bz z)^2}
\label{par11}
\eeq

We now consider unitary irreducible representations (UIR) of $SU(2)$.
A basis for the Lie algebra of $SU(2)$ in the defining
$2\times 2$ matrix representation is given by
$\sigma_a/2$, so that we may write $u$ as
\beq
u = e^{i \sigma_a \xi^a/2}
\label{par12}
\eeq
where the parameters $\xi^a$ can be taken as functions of 
$z$, $\bz$, $\alpha$ or vice versa.
Let $T_a$ denote the generators of the group
$SU(2)$ in an arbitrary representation, corresponding to
${\half} \sigma_a$.
Then a general UIR is specified by
the spin value $s$, defined by
$T_a T_a = s (s+1)$. The matrix corresponding to
$u$ is given by
\beq
\D^{(s)}_{m, m'} (u) = \bra{s, m} {\hat u} \ket{s, m},
\hskip .2in
{\hat u} = e^{ i T_a \xi^a}
\label{par13}
\eeq
The states within the representation are labeled by
$m$, $m'$ which are the eigenvalues of
$T^3$ and take the values
$m, m' = -s , -s +1,  \cdots, s$.

The matrix-valued functions $\D^{(s)}_{m, m'}(u)$
form a complete set for $SU(2)$, so that any function 
on $SU(2)$ can be expanded as
\beq
f (u) = \sum_{s, m, m'} C^{(s)}_{m m'} \, \bra{s, m} {\hat u} \ket{s, m'}
\label{par14}
\eeq
The action of the $U(1)$ transformation $u \rightarrow u h$,
$h = e^{i \xi^3\sigma_3/2}$ is represented as
\beq
f(u h) = \sum_{s, m, m'} C^{(s)}_{m, m'} \, \bra{s, m} {\hat u}
e^{i T_3 \theta^3} \ket{s, m'}
= \sum_{s, m, m'} C^{(s)}_{m,m'} \, \bra{s, m} {\hat u} \ket{s, m'} e^{i m' \xi^3 }
\label{par15}
\eeq
Functions on the coset $SU(2)/U(1)$ must be invariant under these
transformations. Therefore they have a similar mode expansion with
the state on the right side $\ket{s, m'}$ having $m' = 0$.
Thus, a function on $\mathbb{CP}^1$ has the expansion,
\beq
f(u) =  \sum_{s, m} C^{(s)}_m \, \bra{s, m} {\hat u} \ket{s, 0}
\label{par16}
\eeq
The coefficients $C^{(s)}_m$ define the function.
(The mode functions $ \bra{s, m} {\hat u} \ket{s, 0}$ are proportional to the
usual spherical harmonics, so this expansion is the classic expansion of a function on the sphere in terms of the spherical harmonics.)

To define derivative operators, we define
the right translation operators $R_a$ by
\beq
R_a \, u  = u \, {\sigma_a \over 2}
\label{par17}
\eeq
This can be lifted to any representation by using
${\hat u}$ and $T_a$ in this equation. Further,
the left-invariant one-forms $E^a$ on $SU(2)$ are
given by
\beqar
u^{-1} d u &=& - i {\sigma_a\over 2} E^a_k d\theta^k
\nonumber\\
E^1&=&  i {dz - d\bz \over 1+\bz z}, \hskip .2in
E^2 =  -{dz + d\bz \over 1+\bz z}, \hskip .2in
E^3 =  i  {z d\bz - \bz dz\over 1+\bz z}
\label{par18}
\eeqar
$E^1$, $E^2$ are the frame fields for the coset space $\mathbb{CP}^1$.
From this equation, we see that we can realize $R_a$ as
the differential operators
\beq
R_a = i (E^{-1})^k_a {\del \over \del \theta^k}
\label{par19}
\eeq
In particular, we find
\beq
R_+ = (R_1 + i R_2 ) =  (1+ \bz z) \del , \hskip .2in
R_- = (R_1 - i R_2) = - (1+ \bz z) \bdel
\label{par20}
\eeq
From (\ref{par19}) we see that $R_3$ generates the $U(1)$ transformation
on the right of $u$. It corresponds to the isotropy group and is thus the
analog of the Lorentz group for Minkowski space.
In particular, while functions are invariant under $R_3$, vectors
should transform nontrivially, with the same transformation properties as
$R_\pm$. Since $[R_3, R_\pm ] = \pm R_\pm$, a vector corresponding to
holomorphic components will have the mode expansion
\beq
A_+ =  \sum_{s, m} a^{(s)}_m \, \bra{s, m} {\hat u} \ket{s, 1}
\label{par21}
\eeq
Since the state $\ket{s, 1}$ can be obtained from
$\ket{s,0}$ as $\ket{s,1} \sim R_+ \ket{s, 0}$, we can write
(\ref{par21}) as
\beq
A_+ = R_+ \sum_{s, m} a^{(s)}_m \, \bra{s, m} {\hat u} \ket{s, 0}
= - R_+ \Theta
\label{par22}
\eeq
where $\Theta$ is the function $ -\sum_{s, m} a^{(s)}_m \, \bra{s, m} {\hat u} \ket{s, 0}$. This $A_+$ is written using a tangent frame.
Using (\ref{par20}) and going to the coordinate frame, (\ref{par22})
becomes
\beq
A = - \del\, \Theta
\label{par23}
\eeq
This is adequate for an Abelian gauge potential, with
$\bA = - (A)^\dagger$. The generalization to the nonabelian case follows
the arguments given after (\ref{par3}) and we arrive at
\beq
A = - \del M M^{-1}, \hskip .2in \bA = M^{\dagger -1} \bdel M^\dagger
\label{par24}
\eeq
These are still on the space $\mathbb{CP}^1$ in terms of components in the coordinate frame. (For the components in the tangent frame, these will be multiplied by
$(1+ \bz z)$.) If we now scale $z \rightarrow z/r$ and take the large $r$
limit, $\mathbb{CP}^1$ approximates to the flat
space $\mathbb{R}^2$ and we recover the parametrization
(\ref{par6}) for the flat case as well.

We close this section with a comment on what we shall refer to
as the holomorphic ambiguity or holomorphic invariance.
From the definition in (\ref{par6}) it is clear that, for a given
$A$, $M$ is not unique. It is easy to see that
$M$ and $M {\bar V}$,
where ${\bar V} = {\bar V} (\bz )$ is an $SL(N, \mathbb{C})$-matrix whose matrix elements are antiholomorphic functions, lead to the same potential. Similarly, $M^\dagger$ and
$V(z) M^\dagger$ lead to the same $\bA$, where
$V$ is holomorphic in its dependence on the coordinates.
For the two-sphere or for the Riemann sphere,
the only (nonsingular and globally defined)
antiholomorphic/holomorphic function is a constant by Liouville's theorem, 
Thus ${\bar V}$ has to be constant. We can eliminate the ambiguity
by requiring a condition like $M \rightarrow 1$ at spatial
infinity.

However, in general, this global view is not adequate. The $(M, M^\dagger)$ or
$H = M^\dagger M$ corresponding to given potentials
$(A, \bA)$ can have singularities.
To avoid these and obtain a nonsingular description, one has to
resort to a patchwise definition of $(M, M^\dagger )$ with
transition functions on the intersections of coordinate patches.
Notice that $(A, \bA)$ are themselves defined 
only patchwise in general, with gauge transformations
acting as the transitions on intersections. By using $H$ which is
gauge-invariant we avoid this issue, but we may still need to modify
$(M, M^\dagger )$ or $H$ as we move from one coordinate patch to another.
The values on coordinate patches ${U}_1$ and ${U}_2$ will be related on the intersection by
$M_1= M_2 {\bar V}_{12},$ etc., or $H_1 = V_{12} H_2 {\bar V}_{12}$.
Since this is an ambiguity of choice of field  variables, all observable
results must be invariant under this. In particular, we will choose
regularizations in such a way as to preserve this invariance.
This holomorphic ambiguity in the choice of $H$
and the need for antiholomorphic/holomorphic transition 
functions also play a role in connection with the 
Gribov problem, we discuss this briefly in Appendix \ref{TopC}.

\section{The volume element for the gauge-orbit space}\label{Vol}
\setcounter{equation}{0}
\begin{quotation}
\fontfamily{bch}\fontsize{10pt}{16pt}\selectfont
\noindent We calculate the volume element for the physical configuration space
starting with the volume for the space of gauge potentials.
The change of variables from $A$, $\bA$ to $M$, $M^\dagger$ has a Jacobian determinant $\det (- D \bD )$, where $D$, $\bD$ are covariant derivatives. This is calculated exactly in terms of the Wess-Zumino-Witten action. The volume for $M$, $M^\dagger$ is the Haar measure for
the complex group and is calculated by writing the top rank differential form.
Gauge transformations can be exactly factored out to obtain the volume for
the gauge-invariant space, given in (\ref{vol31}). This volume defines the inner product for the wave functions.
\end{quotation}
\fontfamily{bch}\fontsize{12pt}{16pt}\selectfont
The next logical step for us should be to make the change of variables from
$A$, $\bA$ to $M$ and $M^\dagger$ and obtain the volume element of the 
configuration space $\C$.
Our strategy will be to start with the space of gauge potentials $\A$ and divide out the volume of gauge transformations. 
(The calculation we present is from
\cite{Gaw-K, Bos-N, KKN1}. See also \cite{KKN3} for more details
regarding regularization.)
As mentioned earlier, $\A$ is an affine space and we would expect the metric on this space to be
the standard flat Euclidean one. We can confirm that this is indeed the relevant metric for the dynamics by considering the Yang-Mills action.
With $A_0 =0$, we have
\beq
S_{\text{Y-M}} = \int dt d^2x\, \left[
{1\over 2} {\del A^a_i \over \del t}{\del A^a_i \over \del t}
- {1\over 2} B^2 \right]
\label{vol1}
\eeq
A field theory can be thought of as describing the dynamics of a
point-particle moving in an infinite dimensional ambient space
of fields. Thus comparing (\ref{vol1}) to the action for a point-particle,
namely,
\beq
S = \int dt \left[ {1\over 2} g_{\mu\nu} {dx^\mu \over dt} 
 {dx^\nu \over dt}  - V\right],
 \label{vol2}
 \eeq
we see that (\ref{vol1}) does indeed correspond to the case where the
ambient space has the Euclidean metric\footnote{Our convention is
$A = {\half} (A_1+ i A_2)^a (-i t_a)$, $\bA = {\half} (A_1- i A_2)^a (-i t_a)$,
with $\Tr (t_a t_b ) = {\half} \delta_{ab}$.}
\beq
ds^2 = \int d^2x\, ( \delta A^a_i \, \delta A^a_i )
= - 8 \int d^2x\, \Tr ( \delta A \, \delta \bA )
\label{vol3}
\eeq
This is our starting point. Now we can use the parametrization
(\ref{par6}) to write
\beqar
\delta A &=& - \left( \del (\delta M M^{-1} ) + [ - \del M M^{-1}, \delta M M^{-1}]
\right)\nonumber\\
&=& - D (\delta M M^{-1} ) 
\label{vol4}\\
\delta \bA &=& \bD (M^{\dagger -1} \delta M^\dagger )
\nonumber
\eeqar
where $D$, $\bD$ denote covariant derivatives
$D \phi = \del \phi + [A, \phi ]$, $\bD \phi = \bdel \phi
+ [\bA , \phi ]$.
Using these expressions we find
\beqar
ds^2&=& 8 \int d^2x\, \Tr \left[ D (\delta M M^{-1}) \bD (M^{\dagger -1} \delta M^\dagger )\right]\nonumber\\
&=& 8 \int d^2x\, \Tr \left[ (\delta M M^{-1}) (- D \bD ) (M^{\dagger -1} \delta M^\dagger ) \right]
\label{vol5}
\eeqar
As shown in section {\ref{Par}, $M$ and $M^\dagger$ can be thought of as elements of
$SL(N, \mathbb{C})$. The Cartan-Killing metric for
$SL(N, \mathbb{C} )$ viewed as the complexification of
$SU(N)$ is of the form
$\Tr (\delta M M^{-1}\, M^{\dagger -1} \delta M^\dagger )$.
In extending this to $SL(N, \mathbb{C})$-valued functions on $\mathbb{R}^2$, we must include an integral over all space
(which is the continuum version of summing over indices), so
the metric is given as
\beq
ds^2_{SL(N, \mathbb{C})}
= 2 \int d^2x\, \Tr (\delta M M^{-1}\, M^{\dagger -1} \delta M^\dagger )
\label{vol6}
\eeq
We now see that, given the structure of (\ref{vol5}) and the
$SL(N, \mathbb{C})$ metric, the volume element for $\A$ can be written as
\beq
d\mu (\A) = \det (- D \bD ) \, d\mu(M, M^\dagger )
\label{vol7}
\eeq
where $d\mu (M, M^\dagger )$ is the volume element
associated with the metric (\ref{vol6}) for $M$, $M^\dagger $.
(We have ignored some possible
constant multiplicative factors. These are
irrelevant for us, since we will be using this to normalize
the wave functions. Any such factor will cancel out in matrix elements.)

There are two further simplifications to be done.
We must write $d\mu (M, M^\dagger )$ in terms of
$H = M^\dagger M$ and a unitary part which corresponds to the
$SU(N)$ gauge degrees of freedom.
Secondly, we have to calculate the Jacobian
determinant $\det (- D \bD )$ arising from the change of
variables from $A$, $\bA$ to $M$, $M^\dagger$.

The volume element for $SL(N, \mathbb{C})$ is given by the top-rank differential form constructed from $dM M^{-1}$ and
$M^{\dagger -1} d M^\dagger$.
It is given by
\beqar
dV(M,M^{\dagger}) &\propto & \epsilon _{a_1...a_n}  (dM M^{-1})_{a_1}
\wedge \cdots  \wedge (dMM^{-1})_{a_n}  \nonumber\\
&&\hskip .1in \times  \epsilon _{b_1...b_n} (M^{\dagger -1} d M^{\dagger})_{b_1} \wedge \cdots
\wedge  (M^{\dagger -1} d M^{\dagger})_{b_n}  
\label{vol8}
\eeqar
where $n={\rm dim} G= N^2 -1$. (Again we use a proportionality relationship, some
constant numerical factors, which are irrelevant for us, are ignored.)
The components indicated are of the form
$(dM M^{-1})_{a} = 2 \Tr ( t_a dM M^{-1} )$,
$(M^{\dagger -1} d M^{\dagger})_{b} = 2 \Tr (t_b M^{\dagger -1} d M^\dagger )$.

We now use a polar decomposition for the matrices
$M$, $M^\dagger$, given as
$M = U \rho$, $M^\dagger = \rho U^\dagger$, where $\rho$ is hermitian and
$U$ is unitary. Since gauge transformations act on $M$ as
$M^g = g M$, we see that $U$ corresponds to the gauge degree of freedom
in $M$.
By direct substitution of $M = U\rho$, (\ref{vol8}) becomes
\beqar
dV(M, M^{\dagger}) &\propto  & \epsilon _{a_1...a_n} (d\rho \rho^{-1} +
\rho^{-1} d\rho)_{a_1} \wedge ... \wedge (d\rho \rho^{-1} + \rho^{-1} d\rho)_{a_n}  \nonumber\\
&&\hskip .1in \times  \epsilon _{b_1...b_n} (U^{ -1} d U)_{b_1} \wedge ... \wedge (U^{ -1} d
U)_{b_n}  \nonumber\\
&\propto & \epsilon _{a_1...a_n} (H^{-1}dH)_{a_1} \wedge ...  \wedge (H^{-1}dH)_{a_n}  
dV_U
\label{vol9}
\eeqar
Here $dV_U$ is the Haar measure for $SU(N)$. 
If we parametrize $H$ as $H = e^{t_k \vf^k}$
in terms of the real functions
$\vf^k$
we can also
write the $H$-dependent terms in (\ref{vol9}) as
\beq
 \epsilon _{a_1...a_n} (H^{-1}dH)_{a_1} ... (H^{-1}dH)_{a_n}   =  (\det r )\,\,d\vf^1 d\vf^2 \cdots d\vf^n
 \label{vol10}
\eeq
where $H^{-1}dH = d\vf^a r_{ak} (\vf )\, t_k$. 
This is the volume element for $SL(N, \mathbb{C})/ SU(N)$
obtained by reduction from the Cartan-Killing metric for
$SL(N, \mathbb{C})$.

An important feature of (\ref{vol9}) is that the volume of 
$SU(N)$, namely, $dV_U$ factors out from the terms
involving $H$. There is no topological
obstruction to this factorization, because
$SL(N, \mathbb{C})/SU(N)$ is a contractible space.

Upon taking the product of  $dV_U$ and the expression
in (\ref{vol10}) over all points of space to convert to 
a functional integration measure for $SL(N, \mathbb{C})$-valued fields,
we can write
\beqar
d\mu (M, M^{\dagger}) &=& \prod_{x} dV(M, M^{\dag}) = \bigl[ (\det r )\,
d\vf^1 d\vf^2 \cdots d\vf^n\bigr]\,
~ \prod_x dV_U \nonumber\\
& = & d\mu (H) \, d\mu (U)
\label{vol11}
\eeqar
$d\mu (H)  = \prod _{x}  (\det r) d\vf^1 d\vf^2 \cdots d\vf^n$ is the Haar measure for hermitian matrix-valued fields.
We also note that
 $d\mu (U) = \prod_x dV_U$ gives the volume of $\G _*$.
The volume element in (\ref{vol7}) can now be written as
\beq
d\mu(\A) = \det (- D \bD) \, d\mu(H)\, d\mu (U)
\label{vol12}
 \eeq
 It is now straightforward to factor out the volume of
 gauge transformations ($d\mu (U)$) and write the
 volume element for
 $\C = \A /\G_*$ as
 \beq
 d\mu(\C ) = \bigl[d \mu (\A ) / d\mu (U) \bigr]
 = \det(-D \bD ) \, d\mu (H)
 \label{vol13}
 \eeq
 The real advantage of our parametrization (\ref{par6}) is in this expression
 where we can factor out the volume of gauge transformations exactly.
The remaining task is to calculate the determinant of the operator
 $(- D \bD)$.
 Towards this, we start with
 \beq
 \Gamma = \log \det \bD = \Tr \log \bD
 \label{vol14}
 \eeq
 Taking a variation of $\bA$ we find
 \beq
 \delta \Gamma = \Tr (\bD^{-1} \delta \bA )
 =
 \int d^2x\, \Tr \left[ (\bD^{-1})_{x,y} \delta \bA(y) \right]_{y\rightarrow x}
 \label{vol15}
 \eeq
 (Here $\Tr$ on the right hand side denotes the trace over the Lie algebra while
 $\Tr$ on the left hand side of (\ref{vol15}) denotes the full functional trace.)
 We see from this equation that the result for $\delta \Gamma$ will depend on the coincident point
 limit of the Green's function $\bD^{-1}$. It is easy to verify that
 \beqar
{\bar G} (x, y) &=& (\bdel )^{-1}_{x,y} ={1\over \pi (x- y)},
 \hskip .2in x= x_1 -i x_2, ~y= y_1- i y_2\nonumber\\
  {G} (x, y) &=& (\del )^{-1}_{x,y} ={1\over \pi (\bx- \by)}
  \label{vol16}
  \eeqar
  For the gauge-covariant Green's functions we then find
  \beqar
  (\bD^{-1})_{x, y} &=& M^{\dagger -1} (x) \left[ {1\over \pi (x - y)}\right] M^\dagger (y)\nonumber\\
    (D^{-1})_{x, y} &=& M (x) \left[ {1\over \pi (\bx - \by)}\right] M^{-1} (y)
 \label{vol17}
 \eeqar
 The coincident point limit of $\bD^{-1}$ is singular and so we need regularized expressions in place of (\ref{vol17}).
 We will take up this issue in more detail later, but for now, notice that
 for small infinitesimal but nonzero separations
 \beq
(\bD^{-1})_{x, y}    \approx  {1\over \pi (x-y)} 
+ {1\over \pi} \del M^{\dagger -1}\, M^\dagger (y)
+ {\bx - \by \over \pi (x -y)} \bdel M^{\dagger -1} M^\dagger (y) + \cdots
\label{vol18}
\eeq
Since $\Tr\, \delta\bA = 0$, the use of this expression in (\ref{vol15}) gives
\beq
\delta \Gamma = {1\over \pi }\int d^2x\, \left\{ \Tr \left[ ( \del M^{\dagger -1} M^\dagger)
\delta \bA (y) \right] +  {\bx - \by \over (x -y)} 
\Tr\left[ \bdel M^{\dagger -1} M^\dagger \delta \bA (y) \right] +\cdots
\right\}_{y\rightarrow x}
\label{vol19}
\eeq
If we now take the limit $y \rightarrow x$ in a rotationally symmetric fashion,
(so that $(\bx - \by )/(x- y) \rightarrow 0$), we find
\beqar
\delta \Gamma &=& - {1\over \pi} \int d^2x\, \Tr \left[ M^{\dagger -1} \del M^\dagger ) \delta \bA \right]\nonumber\\
&=& {1\over \pi} \int d^2x\, \Tr \left[ \bD (M^{\dagger -1} \del M^\dagger )\,
M^{\dagger -1} \delta M^\dagger \right]
\label{vol20}
\eeqar
The Wess-Zumino-Witten action for a matrix-valued field
$M$ is defined as
\beq
S_{\rm wzw} (M) = {1\over 2 \pi} \int \Tr \left( \del M \, \bdel M^{-1}
\right) + {i \over 12 \pi} \int \Tr \left( M^{-1} d M \right)^3
\label{vol21}
\eeq
The first term on the right hand side involves
 the integral over the 2-manifold while the last term is the integral of
 the 3-form over a 3-manifold whose boundary is the 2-manifold of interest.
 By direct calculation we can verify that
 \beq
 S_{\rm wzw}(N M) =  S_{\rm wzw}(N)  +  S_{\rm wzw}(M) 
 - {1\over \pi} \int \Tr \left( N^{-1} \bdel N\, \del M M^{-1}\right)
 \label{vol22}
 \eeq
 This result is known as the Polyakov-Wiegmann identity \cite{PW}. The key point about it is the chiral splitting in the last term; $N$ has only the antiholomorphic derivative, $M$ has only the holomorphic derivative.
 By taking $NM \rightarrow  M^\dagger (1+ \theta)$,
 we find
 \beqar
 S_{\rm wzw}(M^\dagger (1 + \theta )) - S_{\rm wzw}(M^\dagger)
 &=& -{1\over \pi} \int \Tr \left( M^{\dagger -1} \bdel M^\dagger\, \del \theta\right)
 \nonumber\\
 &=&{1\over \pi} \int \Tr \left( \del ( M^{\dagger -1} \bdel M^\dagger )\, \theta\right)\nonumber\\
 &=&{1\over \pi} \int \Tr \left( \bD (M^{\dagger -1} \del M^\dagger ) \,
 M^{\dagger -1} \delta M^\dagger \right)
 \label{vol23}
 \eeqar
 where we have used the identity 
 \beq
\del ( M^{\dagger -1} \bdel M^\dagger ) - \bD (M^{\dagger -1} \del M^\dagger )  = 0
\label{vol24}
\eeq
and the fact that $\theta = M^{\dagger -1} \delta M^\dagger$.
Comparing with (\ref{vol20}), we see that we can identify
\beq
\delta \Gamma = 2\, c_A\, \delta S_{\rm wzw} (M^\dagger )
\label{vol25}
\eeq
where $c_A$ is the value of the quadratic Casimir operator
in the adjoint representation.
(The trace in (\ref{vol20}) is over the adjoint representation, while we wrote the WZW action using traces in the fundamental representation.
The identity $\Tr (T_a T_b)_A = 2\, c_A\, \Tr (t_a t_b)_F$ leads to
the factor $2 c_A$ in (\ref{vol25}).)
The integrated version of (\ref{vol25}) then gives the result
\beq
\Gamma = \Tr \log \bD = 2\, c_A \, S_{\rm wzw}(M^\dagger),
\label{vol26}
\eeq
up to an additive constant. Although we used a simple expansion of
$\bD^{-1}$, what we have is really an anomaly calculation, namely, the change of
$\det\bD$ under an $SL(N, \mathbb{C})$ transformation.
So, as with anomaly calculations,
 the answer is robust and is obtained by other regularizations as well.
In a similar way to how we arrived at (\ref{vol26}), we get
\beq
\Tr \log D = 2\, c_A\, S_{\rm wzw}(M)
\label{vol27}
\eeq
If we write $\Tr \log (-D \bD ) = \Tr \log D + \Tr \log \bD$
with (\ref{vol26}), (\ref{vol27}), the result is not gauge-invariant.
Basically, the regularization we used is not gauge-invariant.
However, as with the calculation of effective actions from quantum corrections, changing regularizations is equivalent to adding local counterterms. In the present case we can add the local counterterm
\beq
S_{\rm counter} = {2 \,c_A \over \pi} \int \Tr ( \bA\, A) =
- {2\, c_A \over \pi} \int \Tr \left( M^{\dagger -1} \bdel M^\dagger\, \del M M^{-1} 
\right)
\label{vol28}
\eeq
With this counterterm, or with the corresponding choice of regularization,
\beqar
\log \det(- D \bD ) &=&
\Tr \log D + \Tr \log \bD + {2\, c_A \over \pi}\int \Tr (\bA\,A)\nonumber\\
&=& 2\, c_A \left[ S_{\rm wzw}(M) + S_{\rm wzw}(M^\dagger )
- {1\over \pi} \Tr \left( M^{\dagger -1} \bdel M^\dagger\, \del M M^{-1} 
\right)\right]\nonumber\\
&=& 2\, c_A \, S_{\rm wzw} (M^\dagger M) \nonumber\\
&=& 2\, c_A S_{\rm wzw}(H)
\label{vol29}
\eeqar
where we have used the Polyakov-Wiegmann identity again to combine terms. Since $H$ is gauge-invariant, we have a gauge-invariant
result for the determinant. Since we used the variation of the determinant,
this calculation does not fix an overall multiplicative constant
for the determinant. The constant can be evaluated
by considering the case
where $M = M^\dagger = \mathbb{1}$, i.e., to $\det(-\del \bdel )$.
Combining all results, we can then write
\beq
\det( - D \bD ) = \left[ {\det' (-\del \bdel) \over \int d^2x}\right]^{{\rm dim}\,G}
e^{2\, c_A S_{\rm wzw} (H) }
\label{vol30}
\eeq
The prime on $\det'(-\del \bdel)$ indicates that the constant modes,
which are zero modes of the Laplacian are not to be included
in the
determinant. The division by $\int d^2x$ is to take account of the normalization of the same zero modes. Using this back in
(\ref{vol13}), we get the volume for the gauge-orbit space as
\beq
\boxed{
d\mu (\C) = {\cal N} \, d\mu (H)\, e^{2\, c_A S_{\rm wzw} (H) },
\hskip .3in 
{\cal N} = \left[ {\det' (-\del \bdel) \over \int d^2x}\right]^{{\rm dim}\,G}
}
\label{vol31}
\eeq

A worthwhile remark regarding this result is that
$S_{\rm wzw} (V H {\bar V})
= S_{\rm wzw} (H)$. This follows from the
Polyakov-Wiegmann identity (\ref{vol22}).
We also have $d\mu (V H {\bar V}) =
d\mu (H)$, since $V$, ${\bar V}$ are matrices of unit determinant.
Thus the volume element (\ref{vol31}) has the required holomorphic invariance.

 \section{The Hamiltonian for the Yang-Mills theory}\label{Ham}
 \setcounter{equation}{0}
\begin{quotation}
\fontfamily{bch}\fontsize{10pt}{16pt}\selectfont
\noindent 
The kinetic energy $T$, which involves functional derivatives with respect to
$A$, $\bA$, is first written in terms of derivatives with respect to the
parameters of $M$, $M^\dagger$. We then argue that the wave functions can be taken to be functions of a current $J \sim \del H H^{-1}$ and write
$T$ in terms of derivatives with respect to $J$ using the chain rule for differentiation. A different argument is also given, using the Gauss law
to eliminate one of the components of the electric field, and setting
$M^\dagger$ to $1$ by a complex gauge transformation.
The potential energy is also written in terms of $J$. The final result for the Hamiltonian, in a form appropriate for the Schr\"odinger equation, is in (\ref{ham26}).
\end{quotation}
\fontfamily{bch}\fontsize{12pt}{16pt}\selectfont
In section {\ref{Vol}}, we obtained the volume element of the gauge orbit space. As discussed in section {\ref{gaugeprinciple}}, the wave functions for the physical states must obey the invariance condition
$G_0 (\theta ) \Psi = 0$. The inner product is then given by
integration with $d\mu (\C)$, see (\ref{G32}). For the present case,
with the volume element from section \ref{Vol}, it can be written out as
\beq
 \braket{1|2} = \int d\mu (H)\, e^{2\, c_A S_{\rm wzw} (H) }\,
 \Psi_1^* \, \Psi_2
 \label{ham1}
 \eeq
 
The next step is to work out the
expression for the Hamiltonian $\H$.
It has the form given in (\ref{G33}) or (\ref{G41}). Since it involves
 products of operators at the same point, a regularized version has to be defined, consistent with all the symmetries which have to be maintained.
 (The construction of the Hamiltonian, including regularization issues, is discussed in detail
 in \cite{KKN3}.)
Towards this, we first define translation operators on the 
$SL(N, \mathbb{C})$ group elements
$M$ and $M^\dagger$ by
\beqar
[p_a (\vec{x}), M(\vec{y}) ] &=& M(\vec{y}) (-it_a )\, \delta^{(2)}(\vec{x}-\vec{y})\nonumber\\
{}[ \bp_a (\vec{x}) , M^\dagger (\vec{y}) ]  &=& (-it_a) M^\dagger (\vec{y}) \,\delta^{(2)}(\vec{x}-\vec{y})
\label{ham2}
\eeqar
Here $M$ and
$M^\dagger$ are taken to be 
$N\times N$ matrices, corresponding to the fundamental representation
of $SL(N, \mathbb{C})$.
Correspondingly, $t_a$ are hermitian $N \times N$ matrices which form a basis for the Lie algebra of $SU(N)$. We take them to be normalized as
$\Tr (t_a t_b ) = {\half } \delta_{ab}$.
Parametrizing $M, M^{\dag}$ in terms of $\Theta ^a (\vx),~{\bar \Theta}^a (\vx)$
respectively, we can write
\beq
M^{-1}\delta M~ = \delta \Theta^a R_{ab}(\Theta )\, t_b, \hskip .2in 
\delta M^\dag M^{\dag -1} ~= \delta {\bar \Theta}^a R^*_{ab}({\bar \Theta})t_b
\label{ham3}
\eeq
These equations define $R_{ab}(\Theta )$ and 
$R^*_{ab}({\bar \Theta})$.\footnote{They are basically the frame fields on the group $SL(N, \mathbb{C})$. }
From the parametrization of the gauge potentials, we can work out the variation of $A$, $\bA$ as
\beq
\delta A = - D (\delta M M^{-1}), \hskip .3in
\delta \bA = \bD (M^{\dagger -1} \delta M^\dagger )
\label{ham4}
\eeq
Using these relations, we can solve for $\Theta^a$ and ${\bar\Theta}^a$
in terms of $\delta A^a$ and $\delta \bA^a$ and identify
the functional derivatives (which are the electric fields up to a factor of $e^2$)
as
\beqar
 -{i \over 2} {\delta \over {\delta \bA _k (\vx)}} &=& {i \over 2}
M^{\dag} _{ak} (\vx) \int_y \bG (\vx,\vy)\, \bp _a (\vy) \nonumber\\
 -{i \over 2} {\delta \over \delta A _k (\vx)} &=& - {i \over 2} M _{ka} (\vx)
\int_y G (\vx,\vy)\, p _a (\vy) 
\label{ham5}
\eeqar
where $M_{ab}= 2\,\Tr (t^a M t^b M^{-1})$ is the adjoint representation of $M$. The kinetic energy operator in (\ref{G41}) can now be written down as
\beqar
 T&=& - {e^2 \over 2}  \Delta_\C =  -{e^2 \over 2} \int_x {\d ^2 \over {\d A_k (\vx) \d \bA _k (\vx)}}\nonumber\\
 &=& {e^2
\over 2} \int_x K_{ab} (\vx) (\bG \bp _a) (\vx) (G p _b) (\vx) 
\label{ham6}
\eeqar
where $K_{ab} = M^{\dag}_{ak} M_{kb} = 2\, \Tr (t^a H t^b H^{-1})$ and
$Gp_b(\vx)\equiv \int_y G(\vx, \vy)p_b(\vy)$, etc. 

Another way to write $T$, which shows explicitly that it is a symmetric
operator, is to write its matrix element as
\begin{align}
\la 1|T |2\ra =& {e^2 \over 4} \int d\mu (H) e^{2 c_A S_{\rm wzw} (H)} 
\left[{\delta \Psi_1^*\over \delta \bA_k} {\delta \Psi_2\over \delta A_k} 
+ {\delta \Psi_1^*\over \delta A_k} {\delta \Psi_2\over \delta \bA_k} 
\right]\nonumber\\
=& {e^2 \over 4} \int d\mu (H) e^{2 c_A S_{\rm wzw} (H)} \left[ {\overline {G p_a
\Psi _1}}\, K_{ab}\, (G p_b \Psi _2) + {\overline { \bG \bp _a \Psi _1}}\, K_{ba} \, (\bG \bp _b \Psi _2) \right]\nonumber\\
=& \la T1|2\ra \nonumber\\
=& {e^2\over 4} \int d\mu (H) e^{2 c_A S_{\rm wzw}  (H)}\Psi_1^* \biggl[ \int_x e^{-2c_A S_{\rm wzw}  (H)} \Bigl[ {\bar
G}\bp_a(\vx) K_{ab}(\vx) \,e^{2c_A S_{\rm wzw}  (H)} Gp_b(\vx) \nonumber\\
& \hskip 1.8in + Gp_a(\vx) K_{ba}(\vx)\,
e^{2c_AS_{\rm wzw}  (H)} {\bar G}\bp_b(\vx)\Bigr]\biggr] \Psi_2
 \label{ham7}
\end{align}
 In this expression, if we try to move $\bG \bp _a$ through $K_{ab} \,e^{2 c_A \S}$ to act on $Gp_b (\vx) \Psi_2$, we will
encounter the singular commutator $[\bG \bp _a(\vx), K_{ab}(\vx)]$. 
The regularized version of (\ref{ham7}) should be such that
it agrees with (\ref{ham6}).

The regularization of a field theory in the Schr\"odinger formulation
in terms of the Hamiltonian and wave functions is more involved 
(and less well-known) than the case of covariant perturbation theory.
We have discussed this and related  issues in some detail separately
in Appendix \ref{AppC}. But for now, we make an observation about observables 
and the wave function. Since we are considering the gauge theory
without matter fields, the Wilson loop operators $W(C)$, over all closed curves
$C$, constitute a complete (in fact, overcomplete) set of observables.
These are given by
\beq
W(C) = \Tr\, \P e^{- \oint_C A dz + \bA d\bz }
= \Tr\, \P e^{ \oint_C \del H H^{-1} dz } 
= \Tr\, \P e^{ (\pi /c_A) \oint_C J dz }
\label{ham8}
\eeq
Here $\P$ signifies path-ordering of the matrices in the exponent and $J$ is the current given by
\beq
J = {c_A \over \pi} \del H \, H^{-1}
\label{ham9}
\eeq
This is also the current associated with the WZW action 
$S_{\rm wzw} (H)$ which is part of the volume of the gauge orbit space.
The result (\ref{ham8}) implies a simplification of the nature of
the wave function. {\it A priori}, we
are starting with wave functions which are functions of 
$A$, $\bA$, or equivalently, $M$ and $M^\dagger$. Since they must be gauge-invariant by the Gauss law condition 
$G_0 (\theta ) \Psi = 0$, we can take them to be functions of
$H$. But since all observables can be given in terms of $J$,
we can further assume $\Psi$'s to be functions of $J$.
Thus it is advantageous to 
express the Hamiltonian entirely in terms of $J$.
The kinetic energy operator then takes the form
\begin{empheq}[box=\fbox]{align}
 T\, \Psi(J) &=m  \left[ \int J_a (\vz) {\d \over {\d J_a (\vz)}} + \int _{z,w}
 \Omega_{ab} (\vz,  \vw) 
 {\d \over {\d J_a (\vw)}} {\d \over {\d
J_b (\vz)}}\right] \, \Psi (J)\nonumber\\
 \Omega_{ab} (\vz,  \vx) &= \left( { c_A\over \pi^2}  
 {1\over (z-w)^2} + i f_{abc} {J^c(w) \over \pi (z-w)}\right) 
 + {\cal O} (\e ) 
 \label{ham10}
\end{empheq}
where $m = (e^2 c_A /2 \pi )$. We have done the regularization using
$\e$ as a short-distance cutoff. Although a detailed discussion of the regularization is given in Appendix \ref{AppC},
we will just state here that our regularization amounts to
a point-splitting where the Dirac $\delta$-function is replaced by
\beq
\sigma (\vx,\vy;\e) ={{e^{-|\vx-\vy|^2/\e}} \over {\pi \e}}
\label{ham10a}
\eeq
This shows that the regularization parameter $\e$  is essentially 
a short-distance cutoff.
We recover the
$\delta$-function as $\e \rightarrow 0$. This has to be augmented by
certain factors involving $K_{ab}$ to preserve various invariances,
as discussed  later.
The terms displayed in (\ref{ham10}) are the 
finite regularized terms, with ${\cal O}(\e )$ indicating terms which are
negligible as the cutoff $\e \rightarrow 0$.
(In Appendix \ref{AppC}, we show the equivalence between the regularized forms of (\ref{ham6}) and (\ref{ham7}) and how the expression reduces to
what is given (\ref{ham10}) when acting on functions of $J$.)

The two terms appearing in the expression for $T$ are of some interest in
their own right. The first  term is essentially due to the anomaly
in the two-dimensional case. We can see this by calculating
\beqar
 T~J_a(\vx) &=& -{e^2\over 2}\int d^2y {\delta^2 J_a(\vx)\over 
\delta \bA^b(\vy) \delta A^b(\vy)} ~={e^2c_A\over 2\pi} M^{\dag} _{am} \Tr \left[ T^m
\bD ^{-1}(\vy,\vx) 
\right]_{\vy \rightarrow \vx}\nonumber\\
 &=& m ~J_a(\vx)
 \label{ham11}
 \eeqar
The coincident point limit of $\bD ^{-1}(\vy,\vx) $ which appears here is exactly the same as in the calculation of the gauge-invariant measure of integration. Therefore, calculating it exactly as in that case, i.e., using 
(\ref{vol19}), leads to the second line in (\ref{ham11}).
The result in (\ref{ham10}) then follows by the chain rule for functional differentiation.

The second term involving $\Omega_{ab} (\vz, \vx)$ gives the
singular pole terms in the operator product expansion for the current of the WZW model $S_{\rm wzw}(H)$, from a conformal field theory point of view.
Its appearance is again very natural.

There is another way to obtain the result (\ref{ham10}) for the 
operator $T$, which is also illuminating in some ways.
For this we first write the Gauss law operator,
defined in (\ref{G23}) as
\beqar
G_0 (\theta) &=& \int d^2x \,\theta^a {(D_iE_i)^a \over e^2}
= \int d^2x\, \theta^a {I^a}\nonumber\\
I^a &=&  {(D_i E_i)^a\over e^2} = {2 \over e^2} ( D{\bar E} + \bD E )^a
\label{ham12}
\eeqar
The idea then is to regard ${\bar E}^a$ and $I^a$ as independent invariables
and eliminate $E^a$. We can solve (\ref{ham12}) for $E$ in terms of $({\bar E}^a, I^a)$
as
\beq
E(\vx ) = \int_y ( \bD^{-1})_{x,y}  \left( {e^2\over 2} I - D{\bar E} \right)
\label{ham13}
\eeq
The fundamental commutation rules are
\beqar
[E^a (\vx), \bA^b (\vy) ] &=& [\bE^a (\vx) ,A^b(\vy)] = -{{i e^2\over 2}}
\d^{ab}\d (\vx-\vy)\nonumber\\
{}[I^a(\vx), A^b(\vy)] &=& -i D^{ab}_x \d (\vx-\vy)
\label{ham14}
\eeqar
It is easy to check that this is consistent with the solution for
$E$, so that we may take (\ref{ham13}) as an operator identity.
We can thus write the kinetic energy operator as
\beq
T= {2\over e^2} \int_x E^a (\vx) \bE^a (\vx)~= {2\over e^2} \int _{x,y} \left[
(\bD^{-1})^{ab}(\vx,\vy) \left( {e^2\over 2} I -D{\bar E}
\right)^b(\vy)\right]~\bE^a(\vx)
\label{ham15}
\eeq
We then notice that we can move the Gauss law operator to the right end
of this expression; this gives
\beqar
{1\over 2} \int_y (\bD^{-1})^{ab}(\vx,\vy) I^b(\vy) \bE^a(\vx) &=& {1\over 2} \int_y 
(\bD^{-1})^{ab}(\vx,\vy) \bE^a(\vx)
I^b(\vy)\nonumber\\
&&\hskip .2in -{{i\over 2}}\int_{y} (\bD^{-1})^{ab}(\vx,\vy) f^{abc} \bE^c(\vy) 
\d (\vx-\vy)\nonumber\\
&=& {1\over 2}  \int_y (\bD^{-1})^{ab}(\vx,\vy) 
\bE^a(\vx)
I^b(\vy)\nonumber\\
&&\hskip .2in  -{1\over 2}  \Tr \left[ T^c(\bD^{-1}) (\vx,\vy) 
\right]_{\vy\rightarrow \vx}
\bE^c(\vx)
\label{ham16}
\eeqar
Notice that, once again, the coincident point involved is exactly what we had for
the calculation of the volume element and in (\ref{ham11}) as well.
We can evaluate it as done previously to write
\beq
 -{1\over 2}  \Tr \left[ T^c(\bD^{-1}) (\vx,\vy) 
\right]_{\vy\rightarrow \vx}
\bE^c(\vx) = {i c_A \over 2 \pi} \left( A - M^{\dagger -1} \del M^\dagger \right)^c
\label{ham17}
\eeq
We can now write $T$ from (\ref{ham15}) as
\beqar
T&=& 2im  \int_{x} (A-M^{\dag -1} \del M^\dag )^a (\vx) \bE^a (\vx)~-~ 2e^2 
\int_{x,y} \left[(\bD^{-1}) (\vx,\vy)  D\bE (\vy)\right]^a \bE^a (\vx)\nonumber\\
 &&+~ e^2  \int_{x,y} 
(\bD^{-1})^{ab}(\vx,\vy) \bE^a(\vx)
I^b(\vy)
\label{ham18}
\eeqar
where $m = e^2 c_A /(2 \pi)$. On physical states annihilated by the Gauss law, the last term gives zero.
This simplification for $T$ did not require any wave function, and is valid for both the $E$-diagonal and $A$-diagonal representation.\footnote{While we do not discuss the $E$-diagonal representation 
in detail here, see \cite{KKN3} for some useful comments on this.}
We can reduce this expression for $T$ further if we choose wave functions in the
$A$-representation. Towards this, write the parametrization of the fields as
\beq
A = M^{\dagger -1} (-\del H H^{-1}) M^\dagger + M^{\dagger -1} \del M^\dagger, \hskip .2in \bA = M^{\dagger -1} \bdel M^\dagger
\label{ham19}
\eeq
This displays our parametrization for $(A, \bA)$ as a complex gauge transformation of $(-\del H H^{-1}, 0)$, by the $SL(N, \mathbb{C})$ group element $M^\dagger$.
We may therefore take the wave function $\Psi (A, \bA)$ as
a function of $J = (c_A/\pi) \del H H^{-1}$ and $M^\dagger$.
Since a change of $M^\dagger$ is equivalent to an $SL(N, \mathbb{C})$
gauge transformation, 
we may write, for infinitesimal $\theta$,
\beq
\Psi (M^\dag e^\theta ,J) \approx \Psi (M^\dag ,J) + \int \theta^a I^a 
~\Psi (M^\dag ,J)
\label{ham20}
\eeq
This shows that, even though $\theta$ is complex, the Gauss law condition
is enough for us to conclude that
\beq
\Psi (M^\dag e^\theta ,J) = \Psi (M^\dag ,J)
\label{ham21}
\eeq
We see that, by a sequence of such transformations, we can set
$M^\dagger $ to the identity. 
(In two spatial dimensions, all configurations $M^\dagger$ are homotopic to the identity, since $\Pi_2 ( SL(N, \mathbb{C})) = 0$. So there is no obstruction to this procedure of compounding infinitesimal
transformations.)
In other words, we can take
the physical wave functions to be functions of $J$.
In this case, we can take $A= - \del H H^{-1}$,
$\bE\sim (\delta/\delta J )$ and $T$ simplifies to
the expression given in (\ref{ham10}).

It is a simpler task to write the potential energy term in terms of
the current $J$. From the structure of the parametrization of fields as in
(\ref{ham19}), we see that
\beq
B^a t_a = M^{\dagger -1} \left[  - {2 \pi \over c_A} \bdel J^a t_a \right] M^\dagger
\label{ham23}
\eeq
so that we have
\beq
 \int {B^a B^a \over 2 e^2}  = {\pi \over {m c_A}} \int_x : \bdel J^a (\vx) \bdel J^a  (\vx) : 
 \label{ham24}
 \eeq
 The normal-ordering indicates the subtraction of the short-distance
 singularity. We can write this more explicitly as
 \beq
 \int {B^a B^a \over 2 e^2}  = {\pi \over {m c_A}} \left[ \int_{x,y} \sigma (\vx,\vy;\e) \bdel J_a (\vx) \left[K(x,\by)
K^{-1} (y,\by)\right]_{ab} \bdel J_b (\vy) - {{c_A {\rm dim} G} \over {\pi^2
\e^{2}}} \right] 
\label{ham25}
\eeq

Finally, we can combine the expression for $T$ from
(\ref{ham10}) and the potential energy from (\ref{ham24})
to write the full Hamiltonian as
\begin{empheq}[box=\fbox]{align}
\H &=m  \left[ \int J_a (\vz) {\d \over {\d J_a (\vz)}} + \int _{z,w}
 \Omega_{ab} (\vz,  \vw) 
 {\d \over {\d J_a (\vw)}} {\d \over {\d
J_b (\vz)}}\right] \nonumber\\
&\hskip .4in +{\pi \over {m c_A}} \int_x : \bdel J^a (\vx) \bdel J^a  (\vx) : 
+ \, {\cal O} (\e )
\label{ham26}
\end{empheq}
We will use this Hamiltonian to set up the Schr\"odinger equation
and solve it for the vacuum state
in a systematic expansion scheme in section \ref{SchE}.
 \section{A propagator mass for the gluon}\label{promass}
 \setcounter{equation}{0}
\begin{quotation}
\fontfamily{bch}\fontsize{10pt}{16pt}\selectfont
\noindent 
We consider the small field version of the Hamiltonian and the measure of integration and show that it is equivalent to a massive scalar field theory.
This gives a gauge-invariant description of the gluon in a partially resummed perturbation theory. The motivation for this analysis is two-fold: It sets the stage for an alternate argument for the lowest order solution of the Schr\"odinger equation discussed in the next section. It also serves as a theory which can be used to calculate the Casimir energy in section \ref{Res}.
\end{quotation}
\fontfamily{bch}\fontsize{12pt}{16pt}\selectfont
We have obtained the Hamiltonian in terms of the current $J$. We also have the volume element for the gauge orbit space, which is what defines the inner product for wave functions. Thus we are now in a position to write down the Schr\"odinger equation and solve it, in some suitable approximation. 
However, before we do that, we will discuss the theory from the perturbative limit as it can provide some useful insights.
From standard perturbation theory in terms of Feynman diagrams, the effective action $\Gamma$ (which is the generating function for
one-particle irreducible vertices), calculated to one-loop order, 
has the form\footnote{This is a standard one-loop
calculation in Yang-Mills theory. It can also be read off from the
calculations in \cite{AN} by keeping just the contributions from the
Yang-Mills vertices.}
\beq
\Gamma = - {1\over 4 e^2} \int F^a_{\mu\nu} (x) \left[
1 - {7 e^2 c_A \over 32} {1\over \sqrt{-\nabla^2}} \right]_{x,y}
F^{a\mu\nu}(y) + \cdots
\label{promass1}
\eeq
There is no renormalization of the coupling constant, so predictions for 
string tension, masses, etc. can be made without worrying about
the scale at which $e^2$ is to be defined. Secondly, the correction shows clearly that the expansion parameter is $e^2/ \sqrt{-\nabla^2} \sim  (e^2/k)$
where $k$ is the momentum of the field $F_{\mu\nu}$. 
Thus in a Fourier decomposition, the modes of the field for 
momenta high compared to $e^2$ can be treated perturbatively, while the
low momentum modes with $k \ll e^2$ have to be treated nonperturbatively.
There is no real expansion parameter for the theory as a whole, 
$e^2$ is only a marker to signify which modes can be,
and which modes cannot be, treated perturbatively.

Based on our Hamiltonian, we can take this a step further and 
consider an improved perturbation theory where a partial resummation has been carried out.
(This has been discussed in \cite{KKN1, KKN3}.) Towards this, we write
$H = e^{t^a\vf^a}$ in terms of a set of fields $\vf^a$. Then we have
\beqar
J &=& - {c_A \over \pi} \del H \, H^{-1}
= - {c_A \over \pi} \int_0^1 d\alpha \, e^{\alpha t\cdot\vf}  (t\cdot \del\vf)
e^{-\alpha  t\cdot \vf } \nonumber\\
&\approx& - {c_A \over \pi} t_a \left[ \del \vf^a
+ {i \over 2} f^{abc} \vf^b \del\vf^c + \cdots \right]
\label{promass2}
\eeqar 
In perturbation theory, interaction vertices arise from commutators
and carry factors of $f^{abc}$. With this in mind, we can consider a simplification
of the Hamiltonian where we keep only the leading term in (\ref{promass2}). 
With $J^a\simeq {c_A\over\pi}\del\vf ^a$, the
Hamiltonian has the expansion
\beqar
\H &\simeq& m \left[ \int \vf _a {\d \over {\d \vf _a}} + {\pi \over
c_A} \int
\Omega (\vx,\vy) {\d \over {\d \vf _a (\vx)}} {\d \over {\d \vf _a
(\vy)}} \right]\nonumber\\
&&\hskip .2in + {c_A \over m\pi} \int \del 
\vf _a (-\del
\bdel) \bdel \vf _a~+{\cal O}(\vf^3 )
\label{promass3}\\
\Omega (\vx, \vy) &=& - \int
{d^2 k \over (2\pi)^2} ~e^{i k \cdot (x-y)}{1\over k{\bar k}}
\nonumber
\eeqar
The first term in the Hamiltonian, namely,
$\int\vf _a {\d / {\d \vf _a}}$ shows that every $\vf$ in a
wave function will get a contribution of $m$
to the energy. This is basically the origin of the mass gap.
To the same order, with
$H= e^{t_a\vf^a} \approx 1+t_a \vf^a$, 
the volume element becomes
\beq
 d\mu (\C ) = d\mu (H) \, e^{ 2 c_A S_{\rm wzw}(H)} \simeq [d \vf ]\, e^{- {c_A \over 2\pi} \int \del
\vf ^a \bdel \vf ^a}~\left(1~+{\cal O}(\vf^3 )\right)\label{promass4}
\eeq
The exponential factor with the WZW action,
can be absorbed into the wave function by defining
\beq
\Phi = e^{c_A S_{\rm wzw}(H) }\Psi \simeq e^{-{c_A \over 4\pi} \int
\del \vf \bdel
\vf} \Psi
\label{promass5}
\eeq
In terms of the wave functions
$\Phi$, the inner product is given by
\beq
\la 1 \vert 2\ra \approx \int  [d \vf ] ~\Phi_1^* (H) \,\Phi_2
(H)\label{promass6}
\eeq 
We defined $\H$ to act on the $\Psi$'s. 
As an operator acting on the
wave functions $\Phi$, the Hamiltonian should be
\begin{empheq}[box=\fbox]{align}
\H'&=  e^{c_A S_{\rm wzw}(H)} \, \H \,e^{c_A S_{\rm wzw}(H)} 
\simeq
e^{-{c_A \over 4\pi} \int
\del \vf \bdel \vf} \, \H \,  e^{{c_A \over 4\pi} \int
\del \vf \bdel \vf} \nonumber\\
&\simeq {1\over 2} \int_x \left[- {\d ^2 \over {\d \phi _a
^2 (\vx)}} + \phi _a (\vx)  
\bigl( m^2 -
\nabla ^2 \bigr)  \phi _a (\vx)\right] + \cdots
\label{promass7}
\end{empheq}
 where $\phi _a (\vk) = \sqrt {{c_A k \bar{k} }/ (2 \pi m)}~
\vf _a (\vk)$.  This is exactly the free part of a Hamiltonian for a
field of mass $m= e^2 c_A / 2\pi$. 
Thus, to this order, the gauge-invariant version of the gluons
are represented by $\phi_a$ and behave as a field of mass
$m$. It is then straightforward to realize that
the propagator corresponding to $\phi_a$ is\footnote{Here $\T$ denotes the usual time-ordering.}
\beq
\la \T \phi_a (x)\, \phi_b (y) \ra = \delta_{ab} \int {d^3 k \over (2\pi)^3}
e^{-i k \cdot (x-y) } {i\over k^2 - m^2 + i \e}
\label{promass8}
\eeq
Since $m = (e^2 c_A /2 \pi)$, this is not the result at the lowest order in the usual perturbation theory. We must expand this in powers of
$m$ to make the comparison.
The terms of order $(m^2)^n$ in such an expansion may be viewed as
arising from the diagrams of order $(e^2)^n$ in perturbation theory, so that (\ref{promass8}) can be taken to be the result of a selective resummation of the perturbation expansion, where a set of specific terms (and, in fact, a particular kinematic limit of such terms) are summed up.

Thus in setting up perturbation theory using our Hamiltonian and expanding $H$ in powers of $\vf$ to any order, what we get is an
``improved" perturbation theory, where a selective resummation has been done even at the lowest order. The theory at this lowest order is 
a free scalar field theory of mass $m$. This does give a useful starting point for some calculations. In fact, we will use this version later to calculate the Casimir energy for a parallel plate geometry in the nonabelian  theory and compare with lattice simulations.
 \section{The Schr\"odinger equation: An expansion scheme}\label{SchE}
 \setcounter{equation}{0}
\begin{quotation}
\fontfamily{bch}\fontsize{10pt}{16pt}\selectfont
\noindent 
In this section, we set up a systematic expansion scheme for the solutions of the Schr\"odinger equation in powers of the current $J$
where (the nonpertubative) mass generation is included exactly,
but some interaction terms are treated in a power series.
After setting up the general scheme, the solution is obtained for the
vacuum wave function to the lowest and next-to-lowest order in the expansion. An alternate argument for the lowest order result, using
the results of section \ref{promass}, is also given.
\end{quotation}
\fontfamily{bch}\fontsize{12pt}{16pt}\selectfont
In this section, we shall return
 to the full version of the Hamiltonian in terms of the currents, write down the Schr\"odinger equation and develop a recursive scheme for solving it 
for the vacuum wave function. 
{\it A priori} this is a difficult task since there is no natural expansion parameter in the theory. 
As explained earlier, the modes of, say, $J$
with momenta $k \ll m$ can be considered low momentum modes
and those with $k \gg m$ can be considered as high momentum modes,
with the coupling constant $e^2$ only serving to separate the modes into these two domains.
Our aim will be to focus on the nonperturbative part due to the low
$k$ modes.
Towards this, 
we will adopt the following strategy to set up the expansion
scheme.
We will consider an extension of the theory defined by the Hamiltonian
as in (\ref{ham26}) with $m$ and $e$ considered as independent parameters.
This will require a rescaling of the current as explained below.
We can then develop a series expansion for the vacuum wave function,
writing $\Psi_0 = e^{{\half}\F}$ where $\F$ is a power series in $e$.
Mathematically, this framing of the problem, with $m$ and $e$ treated as independent parameters, gives us a way to systematize the solution for the
vacuum wave function.
At the end, we will set
$m = (e^2 c_A/2\pi)$ to regain the gauge theory of interest.
(The solution for the vacuum wave function to the lowest order was given in
\cite{KKN4} and used to calculate the string tension.
The systematic expansion scheme and the solution with the first set
of corrections were given in \cite{KNY}.)

It is worth emphasizing again that this is
very different from perturbation theory since $m$ is included exactly in the lowest order result for $\F$. Further in the present case, we are not expanding $J$ in terms of $\vf$ either.
The resulting recursive procedure will still be
some sort of resummed theory. The
resummation involves collecting $A, \bA$ in an appropriate series to define $J$ and then including $m$ at the lowest order which is another series.
Getting to details, we first do a scale transformation on $J$ as
$J \rightarrow (e c_A /2\pi) J$. In terms of the new $J$, we can write
 \beqar
 {\cal H}&=& {\cal H}_0 ~+~{\cal H}_1\nonumber\\
{\cal H}_0 &=& m  \int J_a (\vz) {\d \over {\d J_a (\vz)}} + {2\over \pi} \int _{w,z} 
 {1\over (z-w)^2} {\d \over {\d J_a (\vw)}} {\d \over {\d
J_a (\vz)}} \nonumber\\
&&\hskip .3in+{1\over 2} \int_z : \bdel J^a(z) \bdel J^a(z): \label{SchE1}\\
{\cal H}_1&=&
+ i e \int_{w,z} f_{abc} {J^c(w) \over \pi (z-w)} {\d \over {\d J_a (\vw)}} {\d \over {\d
J_b (\vz)}} \nonumber
\eeqar
As stated before, in the expression for $\H$, we take $m$ and $e$ to be independent parameters for now. The interaction term
$\H_1$ is to be treated as a perturbation. In the
vacuum wave function $\Psi_0 = e^{{\half}\F}$, $\F$ is an arbitrary functional of $J$. Therefore it can, in general, be taken to be of the form
\beqar
\F &=& \int f^{(2)}_{a_1 a_2}(x_1, x_2)\ J^{a_1}(x_1) J^{a_2}(x_2) ~+~
\frac{e}{2}\int  f^{(3)}_{a_1 a_2 a_3}(x_1, x_2, x_3)\ J^{a_1}(x_1) J^{a_2}(x_2) J^{a_3}(x_3)
\nonumber\\
&&\hskip .2in~+~
\frac{e^2}{4}\int  f^{(4)}_{a_1 a_2 a_3 a_4}(x_1, x_2, x_3, x_4)\ J^{a_1}(x_1) J^{a_2}(x_2) J^{a_3}(x_3)
J^{a_4}(x_4)~+~\cdots
\label{SchE2}
\eeqar
In accordance with the idea of treating $\H_1$ perturbatively, each of the coefficient functions will also be taken to have an expansion in powers of
$e^2$, so that we can write
\beqar
f^{(2)}_{a_1 a_2}(x_1, x_2) &=& f^{(2)}_{0~a_1 a_2}(x_1, x_2) +
e^2 f^{(2)}_{2~a_1 a_2}(x_1, x_2) +\cdots\nonumber\\
f^{(3)}_{a_1 a_2 a_3}(x_1, x_2, x_3)&=& f^{(3)}_{0~a_1 a_2 a_3}(x_1, x_2, x_3) + e^2 f^{(3)}_{2~a_1 a_2 a_3}(x_1, x_2, x_3)
+\cdots\label{SchE3}\\
f^{(4)}_{a_1 a_2 a_3 a_4}(x_1, x_2, x_3, x_4) &=& f^{(4)}_{0~a_1 a_2 a_3 a_4}(x_1, x_2, x_3, x_4) +\cdots\nonumber
\eeqar
The Schr\"odinger equation for the vacuum wave function takes the expected form
\beq
(\H_0 + \H_1) \, \Psi_0 = 0
\label{SchE4}
\eeq
We can now substitute for $\Psi_0$ with $\F$ as in
(\ref{SchE2}) into the Schr\"odinger equation (\ref{SchE4}) and equate
the coefficients of terms with similar powers of $J$ to obtain
a set of recursion relations.
The term with zero powers of $J$ is a constant which can be removed by
a suitable normal-ordering of the Hamiltonian. In fact, we have already taken account of this as indicated by the normal-ordering of the potential
energy term. Terms with only one power of $J$ will vanish by color contractions. The lowest nontrivial relation pertains to 
$f^{(2)}_{a_1 a_2}(x_1, x_2)$; it is given by
\beqar
&& 2m~ f^{(2)}_{a_1 a_2}(x_1, x_2) + 4 \int_{x,y}  f^{(2)}_{a_1 a}(x_1, x) (\bar{\Omega}^0)_{ab}(x,y) f^{(2)}_{b a_2}(y, x_2) +V_{ab} \nonumber \\
&&+e^2 \left[ 6 \int_{x,y} \!\! f^{(4)}_{a_1 a_2 a b }(x_1, x_2, x,y) (\bar{\Omega}^0)_{ab}(x,y) + 3 \int_{x,y} \!\! f^{(3)}_{a_1 a b  }(x_1, x,y) (\bar{\Omega}^1)_{ab a_2}(x,y, x_2)\right] 
= 0\nonumber\\
 \label{SchE5}
\eeqar
where, for brevity, we have used the definitions
\beqar
(\bar{\Omega}^0)_{ab}(x,y) &=& \delta_{ab} \partial_y \bar{G}(x,y) 
= \delta_{ab} \,{1\over \pi (x-y)^2}\nonumber \\
(\bar{\Omega}^1)_{abc}(x,y,z) &=& -\frac{i}{2}\ f^{abc} \left[ \delta(z-y) + \delta(z-x)\right] \bar{G}(x,y) \nonumber \\
V_{ab}(x,y) &=& \delta_{ab} \int_z \bar{\partial}_z \delta(z-x) ~\bar{\partial}_z \delta(z-y) \label{SchE6}
\eeqar
In this equation, we have also used (\ref{vol16}) for $\bG (x, y)$.
For the higher point functions, the recursion relation is given by
\beqar
&&m p f^{(p)}_{a_1\cdots a_p} + \sum_{n=2}^{p} n(p+2-n) f^{(n)}_{a_1\cdots a_{n-1} a}(\bar{\Omega}^0)_{ab} f^{(p-n+2)}_{b a_n\cdots a_p} \nonumber \\
&&+ \sum_{n=2}^{p-1} n(p+1-n) f^{(n)}_{a_1\cdots a_{n-1}a} (\bar{\Omega}^1)_{ab a_p} f^{(p-n+1)}_{b a_n\cdots a_{p-1}} \nonumber\\
&&+ e^2 \left[ \frac{(p+1)(p+2)}{2}\ f^{(p+2)}_{a_1\cdots a_p a b}(\bar{\Omega}^0)_{ab} +\frac{p(p+1)}{2}\ f^{(p+1)}_{a_1\cdots a_{p-1} a b} (\bar{\Omega}^1)_{ab a_p}\right] =0 \label{SchE7}
\eeqar
This applies for $p \ge 3$. We must solve (\ref{SchE5}) and
this set of equations (\ref{SchE7}) to calculate the vacuum wave function in our scheme.
\subsection{The lowest order solution}
With each $f$ having a series expansion in powers of $e^2$, the lowest order solution to (\ref{SchE5})  is
\beqar
f^{(2)}_{a_1 a_2}(x_1, x_2) \approx f^{(2)}_{0\ a_1 a_2}(x_1, x_2)  &=& \delta_{a_1 a_2}
\left[- {{\bar q}^2 \over m +E_q}\right]_{x_1, x_2}\nonumber\\
&=& -\delta_{a_1 a_2} \int {d^2 q\over (2\pi)^2} \left[ {{\bar q}^2 \over m +E_q}\right]\, e^{i \vq \cdot (\vx_1-\vx_2)}
\label{SchE8}
\eeqar
where $E_q = \sqrt{m^2 + q^2}$ and ${\bar q} = {\half} (q_1 - i q_2)$.
Using this expression, we get the vacuum wave function to this order as
\cite{KKN4}
\beq
\boxed{
\Psi_0 \approx {\cal N} \, \exp\left[ - {1\over 2}
\int_{x,y} \bdel J^a(x) \left( {1\over m + \sqrt{m^2 - \nabla^2}}\right)_{x, y} \bdel J^a(y)\right] }
\label{SchE9}
\eeq
where ${\cal N}$ is a normalization factor. Even with this lowest order result, we can extract some predictions regarding physical quantities.
This will be taken up in the next section, but for now, we will give the first set of corrections to this expression.
\subsection{The first order corrections to the vacuum wave function}
For the first order corrections to
$\Psi_0$, we will need the lowest order results for $f^{(3)}$ and $f^{(4)}$.
Then using them, we can get $f^{(2)}_{2\ a_1 a_2}(x_1, x_2)$,
which is the term in $f^{(2)}_{a_1 a_2}(x_1, x_2)$ at order
$e^2$.

The expressions for the kernels $f^{(3)}$ and $f^{(4)}$ obtained by solving the recursion rules (\ref{SchE7}) to the lowest order are
\beqar
f^{(3)}_{0\ a_1 a_2 a_3}(k_1, k_2, k_3) &=& -\frac{f^{a_1 a_2 a_3}}{24}\ (2\pi)^2 \delta (k_1+k_2+k_3)\  g^{(3)}(k_1,k_2,k_3)\label{SchE10}\\
f^{(4)}_{0\ a_1 a_2; b_1 b_2}(k_1, k_2; q_1, q_2) &=& \frac{f^{a_1 a_2 c} f^{b_1 b_2 c}}{64}\ (2\pi)^2 \delta (k_1+k_2+q_1+q_2)\ g^{(4)}(k_1, k_2; q_1, q_2) \nonumber\\
\label{SchE11}
\eeqar
where
\beq
g^{(3)}(k_1,k_2,k_3) = \frac{16}{E_{k_1}\! + E_{k_2}\! + E_{k_3}}\left \{ \frac{\bar k_1 \bar k_2 (\bar k_1 - \bar k_2)}{(m+E_{k_1})(m+E_{k_2})} + {\rm cycl.\ perm.} \right \}
\label{SchE12}
\eeq
\begin{align}
g^{(4)}(k_1, k_2; q_1, q_2) =& \vspace{.2in}  \frac{1}{E_{k_1}\! + E_{k_2}\! + E_{q_1}\! + E_{q_2}}\times \nonumber\\
\vspace{.2in}
& \left \{ g^{(3)}(k_1, k_2, -k_1-k_2)\ \frac{k_1 + k_2}{\bar k_1 +\bar k_2}\ g^{(3)}(q_1, q_2, -q_1-q_2) \right. \nonumber\\
&-  \left [ \frac{(2\bar k_1 + \bar k_2)\,\bar k_1}{m + E_{k_1}} - \frac{(2\bar k_2 + \bar k_1)\,\bar k_2}{m + E_{k_2}}\right ]\frac{4}{\bar k_1+\bar k_2}\  g^{(3)}(q_1, q_2, -q_1-q_2) \nonumber\\
&-  \left .   g^{(3)}(k_1, k_2, -k_1-k_2)\ \frac{4}{\bar q_1+\bar q_2}\left [ \frac{(2\bar q_1 + \bar q_2)\,\bar q_1}{m + E_{q_1}} - \frac{(2\bar q_2 + \bar q_1)\,\bar q_2}{m + E_{q_2}}\right ] \right\} 
\label{SchE13}
\end{align}
We have displayed the kernels in terms of their Fourier transforms
\beqar
f^{(3)}_{a_1 a_2 a_3}(x_1, x_2, x_3)&=& \int d\mu (k)_3 \exp\left( i \sum_i^3 k_i x_i\right) \ f^{(3)}_{a_1 a_2 a_3}(k_1, k_2, k_3)\nonumber\\
f^{(4)}_{a_1 a_2 a_3 a_4}(x_1, x_2, x_3, x_4) &=& \int d\mu (k)_4 \exp\left( i \sum_i^4 k_i x_i\right)\ f^{(4)}_{a_1 a_2 a_3 a_4}(k_1, k_2, k_3, k_4)
\label{SchE14}\\
d\mu (k)_n &=& {d^2k_1\over (2\pi )^2}\cdots {d^2k_n\over (2\pi )^2}
\label{SchE15}
\eeqar
Note also that $f^{(4)}_{a_1 a_2; b_1 b_2}(k_1, k_2; q_1, q_2)$ as defined in (\ref{SchE11},\ref{SchE13}) is symmetric under independent exchange of the first and second pairs of indices as well
as under the simultaneous exchange $(\{a_1,k_1\}, \{a_2, k_2\}) \leftrightarrow (\{b_1,q_1\}, \{b_2, q_2\})$. It could have been made completely symmetric but it is notationally simpler to leave it as it is 
for now.

Finally, using the expressions (\ref{SchE10})-(\ref{SchE13})
for $f^{(3)}_0$, $f^{(4)}_0$ in the recursion rule (\ref{SchE5}), 
the term of order $e^2$ in $f^{(2)}$ is
 given by \cite{KNY}
\beq
f_2^{(2)}(q) = \frac{m}{E_q}\left(\int \frac{d^2 k}{32\pi}\ \frac{1}{\bar k}\ g^{(3)}(q,k,-k-q)\ +\ \int \frac{d^2 k}{64\pi}\ \frac{k}{\bar k}\ g^{(4)}(q,k;-q,-k)\right)\label{SchE16}
\eeq
This completes the calculation of $\F$ to order $e^2$.
The kernels $f^{(n)}$, $n\geq 5$, are zero to this order,
becoming nonzero starting
only at the next order in $e^2$.

To summarize, to the lowest order in our expansion scheme, the vacuum wave function is given in (\ref{SchE9}). 
Equations(\ref{SchE10}-\ref{SchE13}) and (\ref{SchE16})
then give the first set of corrections to the wave function, i.e., 
 to order $e^2$.
\subsection{Another route to the vacuum wave function}
We have already seen in section \ref{promass} how we can define
an improved perturbation theory where the lowest order result gives the
Hamiltonian for a free massive scalar field.
The Hamiltonian given in (\ref{promass3}) has the term
$m \vf^a (\delta/\delta \vf^a)$, which assigns a mass $m$ to
each power of $\vf$. The existence of this term is directly related to the integration measure 
\beq
 d\mu (\C ) = d\mu (H) \, e^{ 2 c_A S_{\rm wzw}(H)} \simeq [d \vf ]\, e^{- {c_A \over 2\pi} \int \del
\vf ^a \bdel \vf ^a}~\left(1~+{\cal O}(\vf^3 )\right)\label{SchE17}
\eeq
Given this integration measure, the term $m \vf^a (\delta/\delta \vf^a)$
is necessary for self-adjointness of the Hamiltonian.
We can now use this to give an alternative approach to the wave function
(\ref{SchE9}).

Since it corresponds to a free massive scalar field, the
Hamiltonian $\H'$ from (\ref{promass7}) has the vacuum wave function
\beq
\Phi_0 \sim \exp\left( - {1\over 2} \int \phi^a \sqrt{m^2 - 
\nabla^2}\, \phi^a\right), \hskip .3in
\phi _a (\vk) = \sqrt {{c_A k \bar{k} }/ (2 \pi m)}~
\vf _a (\vk)
\label{SchE18}
\eeq
Converting this back to $\Psi = e^{{c_A \over 4\pi} \int\del \vf \bdel\vf} \Phi$, we find
\beq
\Psi_0 \sim \exp \left[ - {c_A \over \pi m} \int (\del\bdel \vf^a ) \left(
{1\over m + \sqrt{m^2 - \nabla^2}} \right) (\del\bdel \vf^a )\right]
\label{SchE19}
\eeq
The key argument is then the following. We know, from other considerations, 
that the wave functions can be taken to be functionals of the current
$J$. For small $\vf$, $H \approx 1 + \vf$ and
$J \approx {c_A \over \pi} \del \vf$.
So we ask: Is there a functional of $J$ which reduces to
the form (\ref{SchE19}) for $J \approx {c_A \over \pi} \del \vf$?
There is a unique answer, it is given by
(\ref{SchE9}),
\beq
\Psi_0 \approx {\cal N} \, \exp\left[ - {1\over 2}
\int_{x,y} \bdel J^a(x) \left( {1\over m + \sqrt{m^2 - \nabla^2}}\right)_{x, y} \bdel J^a(y)\right] 
\label{SchE20}
\eeq

It is useful to restate the simple logic leading to this result.
The measure of integration for
the inner product is exact, being determined by an anomaly
calculation. This in turn fixes the form of $\H$ for the small $\vf$
version of the Hamiltonian, and gives the wave function 
(\ref{SchE19}). The requirement that $\Psi$ be a function of the current
 then ties it down to the form (\ref{SchE20}).
This argument shows that there is a certain robustness to the form of $\Psi_0$ in (\ref{SchE20}).

 \section{Analytic results, comparison with numerics}\label{Res}
 \setcounter{equation}{0}
\begin{quotation}
\fontfamily{bch}\fontsize{10pt}{16pt}\selectfont
\noindent 
We obtain results for some physical quantities and compare with numerical
estimates. An analytic formula for the string tension is calculated, the result is given in (\ref{Res10}). This is compared, for various groups and representations, with numerical estimates; the agreement is within about
$2\%$. There has been some discussion in the literature on Casimir scaling of the string tension versus what is called the sine-law and 
possible incompatibility with expectations from the diagrammatic
 ${1\over N}$-expansion. We comment on these problems and 
 the resolution of compatibility with the ${1\over N}$-expansion.
The Casimir energy for a parallel plate 
arrangement is then calculated, based on the massive scalar field version given in section \ref{promass}. The result is given in (\ref{Res21});
it is then compared with numerical estimates based on lattice simulations.
There is also a significant body of literature on the propagator mass.
We discuss various calculational methods briefly and
compare with the result of our analysis from 
section \ref{promass}.
\end{quotation}
\fontfamily{bch}\fontsize{12pt}{16pt}\selectfont
In the previous section we have obtained the solution of the Schr\"odinger equation for the vacuum wave function up to the lowest two orders in our expansion scheme. In this section we will use this result to
calculate the string tension and compare it with
numerical studies.
In section \ref{promass}, we
have also identified an approximate description of the gluons by a massive scalar field. The generation of mass 
is a nonperturbative effect, even though we expect the approximate description given in section \ref{promass} to be valid in a kinematic 
regime where the momenta are in some intermediate range,
high enough to neglect some of the interactions
from the vertices but not so high as to neglect the mass terms.
In this section we will also use this
approximate description in terms of the massive scalar field for
the calculation of the Casimir energy and compare it with numerical
estimates.
\subsection{String tension}
Since confinement has been the aspirational goal of many attempts at 
the nonperturbative analysis of gauge theories,
first, we will consider the calculation of the string tension $\sigma_R$
for the representation $R$.
As explained in section \ref{Conf}, this is related to the vacuum expectation value of the Wilson loop operator as
\beq
\la W_R (C) \ra \approx {\cal N} e^{ - \sigma_R A_C }
\label{Res1}
\eeq
Since we are interested in loops of large area, we will consider the
vacuum wave function for the low momentum
modes of the fields. From
(\ref{SchE9}), the lowest order result for this is
\beqar
\Psi_0 &\approx& {\cal N} \, \exp\left[ - {1\over 2}
\int_{x,y} \bdel J^a(x) \left( {1\over m + \sqrt{m^2 - \nabla^2}}\right)_{x, y} \bdel J^a(y)\right]\nonumber\\
&\approx&{\cal N} \, \exp\left[ - {1\over 4 m}
\int_{x} \bdel J^a \,\bdel J^a\right] 
= {\cal N} \, \exp\left[ - {1\over 8 m e^2}
\int_{x} F^2\right] 
\label{Res2}
\eeqar
where, in going to the second line, we have simplified the kernel
as it applies to the low momentum modes, with
momenta $k\ll m$.
The expectation value of the Wilson loop operator $W_R (C)$, where $C$ is purely spatial, can be written as
\beqar
\la W_R (C) \ra &=& \int d\mu (\C) \, \Psi_0^* \Psi_0 \, W_R (C)
\nonumber\\
&=& {\cal N}' \, \int d\mu (\C) \, W_R (C) \, \exp\left[
- {1\over 4 g^2} \int F^2 \right]
\label{Res3}
\eeqar
where $g^2 = m e^2$. This is exactly the Euclidean path-integral version of the expectation value in a two-dimensional Yang-Mills theory, 
with a coupling constant $g^2$.
By the arguments presented in section {\ref{Conf}},
we can also calculate this as the interaction energy
for a heavy particle-antiparticle pair in the 1+1 dimensional Yang-Mills theory.
Using $\phi$ and $\chi$ to represent the heavy particles as in
section {\ref{Conf}}, the action we need is
\beq
S =  \int d^2x\left[ - {1\over 4 g^2}  F^2 + i \phi^\dagger D_0\phi
+ i \chi^\dagger D_0\chi \right]
\label{Res4}
\eeq
This is a fairly trivial theory to investigate. Since the canonical momentum
$\Pi_0^a$ is zero, we can choose $A^a_0 = 0$ as the conjugate constraint and eliminate the pair. There is no magnetic field in 1+1 dimensions,
so the Hamiltonian in the $A^a_0 = 0$ gauge is
\beq
\H = {1\over 2 g^2} \int dx\, E^a E^a
\label{Res5}
\eeq
We also have the Gauss law constraint
\beq
(D E)^a + g^2( \phi^\dagger T^a \phi - \chi^\dagger {\tilde T}^a \chi ) = 0
\label{Res6}
\eeq
(This is the same as (\ref{Conf6}), but now for the 1+1 dimensional
theory which defines the equal-time matrix elements for the 2+1 dimensional
YM theory.) 
As the conjugate constraint, 
we can take $\del_x A = 0$. If we take $A$ to vanish at spatial
infinity, the only solution is $A = 0$.
The Gauss law (\ref{Res6}) then constrains $E^a$ in terms of the charge densities. Thus there are no propagating degrees of freedom 
associated to the Yang-Mills field.
There will be just the Coulomb interaction. To identify this term, notice that
the solution of the Gauss law condition is $E^a = \del f^a$,
with
\beq
f^a (x) = - {g^2\over 2}  \int_y \vert x-  y \vert \, 
( \phi^\dagger T^a \phi - \chi^\dagger {\tilde T}^a \chi )(y)
\label{Res7}
\eeq
The Hamiltonian now becomes
\beq
\H = - {g^2\over 4} \int_{x, y} ( \phi^\dagger T^a \phi - \chi^\dagger {\tilde T}^a \chi )(x) \, \vert x- y\vert\, ( \phi^\dagger T^a \phi - \chi^\dagger 
{\tilde T}^a \chi )(y)
\label{Res8}
\eeq
Acting on the state $\ket{0, L} = \phi_i^\dagger (0) \chi_i^\dagger (L) \ket{0}$
we find
\beqar
\H \ket{0,L} &=& {g^2 \over 2} L \,(\phi^\dagger T^a)_i ( \chi^\dagger {\tilde T}^a)_i \ket{0} = {g^2 c_R \over 2} L \ket{0,L} \nonumber\\
&=& {e^4 c_A c_R \over 4 \pi} \, L\, \ket{0,L}
\label{Res9}
\eeqar
where $c_R$ is the value of the quadratic Casimir operator for the representation
$R$. The string tension can now be read off from this result as
\beq
\boxed{
\sigma_R = {e^4 c_A c_R \over 4 \pi} }
\label{Res10}
\eeq
This is an analytic prediction for the string tension for the Wilson loop operators in any representation \cite{KKN4}.

An interesting observation regarding this result is that we have not used any simplification
of the gauge theory that might arise from the large $N$ approximation.
However, the final result (\ref{Res10}) is consistent with large
$N$ expectations. For example, for the fundamental representation of
$SU(N)$, we find 
\beq
\sigma_F = {(e^2 N)^2 \over 4\pi} {N^2 -1 \over 2 N^2}
\rightarrow {\lambda^2 \over 8\pi}, \hskip .2in {\rm as}~ N \rightarrow \infty
\label{Res10a}
\eeq
where $\lambda = e^2 N$ is the 't Hooft coupling constant.
\subsection{Comparison of string tension with numerical estimates}
Even though we obtained the result (\ref{Res10}) for the string tension
 using the wave function to the lowest order in our expansion scheme, it is useful at this stage to pause and compare the values given by (\ref{Res10}) 
\begin{table}[!t]
\begin{center}
\renewcommand{\arraystretch}{.9}
\begin{tabular}{ p{1.3cm}| p{1.7cm} p{1.5cm} p{1.5cm} p{1.5cm} p{1.5cm} p{1.5cm} }
Group&\multicolumn{6}{c}{Representations}\\
\hlineB{3}
&k=1& k=2 & k=3 & k=2 &k=3 &k=3\\
&Fund.&antisym&antisym&sym&sym&mixed\\
\hlineB{3}
\multirow{3}{*}{$SU(2)$}&0.345&&&&&\\
&{\color{red}0.335}&&&&&\\
\hline
\multirow{3}{*}{$SU(3)$}&0.564&&&&&\\
&{\color{red}0.553}&&&&&\\
\hline
\multirow{3}{*}{$SU(4)$}&0.772&0.891&&1.196&&\\
&{\color{red}0.759}&{\color{red}0.883}&&{\color{red}1.110}&&\\
\hline
\multirow{3}{*}{$SU(5)$}&0.977&&&&&\\
&{\color{red}0.966}&&&&&\\
\hline
\multirow{3}{*}{$SU(6)$}&1.180&1.493&1.583&1.784&2.318&1.985\\
&{\color{red}1.167}&{\color{red}1.484}&{\color{red}1.569}&{\color{red}1.727}&{\color{red}2.251}&{\color{red}1.921}\\
\hline
\multirow{2}{*}{$SU(N)$}&0.1995 $N$&&&&&\\
\raisebox{-7pt}{$N\!\!\rightarrow \!\!\infty$}&\color{red}0.1976 $N$&&&&&\\
\hlineB{3}
\end{tabular}\\
\vskip .1in
\caption{Comparison of $\sqrt{\sigma}/e^2$ as predicted by (\ref{Res10}) (upper entry) and lattice estimates
(lower entry, in red) from \cite{teper, meyer, bringoltz}. $k$ is the rank of the representation.}
\label{sigmas}
\end{center}
\end{table}
with numerical simulations. In the Table \ref{sigmas},
 we show the results for
 a number of different gauge groups and representations carried out by
Teper and collaborators \cite{teper, meyer, bringoltz}. It is clear that the values are very close to the
 predictions from (\ref{Res10}), the difference being less than $3\%$.
 In addition to these, there has been a high precision calculation for 
 the fundamental representation ($k = 1$) of
 $SU(2)$ which gives value of $0.33576(24)$ for $\sqrt{\sigma}/e^2$
 \cite{haridass}.
Again this compares favorably with our value of $0.3455$.
An independent numerical estimate of the large $N$ result has also been
carried out in \cite{kiskis}, giving a value of $0.1964\, N$.

An especially fascinating group
is $G_2$, since all representations of this group are screenable. 
Lattice-based calculations of the string tension for
the representations $\underline{7}$, $\underline{14}$,
$\underline{27}$, $\underline{64}$, $\underline{77}$,
$\underline{77'}$, $\underline{182}$ and $\underline{189}$
have been carried out in \cite{wellegehausen}.
They have verified the relation $\sigma_R = \sigma_7 (c_R/c_7)$
(which follows from (\ref{Res10}) ) to within $1\%$.
The value of $\sqrt{\sigma_7}$ itself agrees with
(\ref{Res10}) to within $1.8\%$.

The fact that the predictions from (\ref{Res10}) and the results of the numerical calculations do match rather well
is very nice, but one could ask whether there are corrections and, if so, 
whether they do remain small so as not to vitiate the agreement
we find here.
We will consider the corrections due to the terms of the next
order (i.e., to order $e^2$)
in the wave function and show that the corrections are indeed small.
Since these calculations are rather long, and all too technical, we will
defer this to Appendix \ref{AppD}. For now, we will make some comments regarding the string tension and then move on to the Casimir effect and to
the propagator masses.
\subsection{Comments regarding string tension}
There are a couple of interesting and important comments to be made about the string tension.

As mentioned earlier, $G_2$ is a group for which all representations are screenable. The fundamental representation of $G_2$ is $7$-dimensional 
while the adjoint representation is
$14$-dimensional. The product (${\rm 7} \times {\rm Adjoint} \times {\rm Adjoint} 
\times {\rm Adjoint} $) contains a singlet or the trivial representation,
ensuring that all representations are screenable.
Generally the form of the potential for static sources
in screenable representations
will show a linear increase with distance up to a certain critical value
$R_b$ and will become flat for $r > R_b$.
The distance $R_b$ is referred to as the string-breaking distance.
The lattice estimate of the string tension for $G_2$ (and for
screenable representations for other groups) is the slope of the linearly rising part, before the flattening, i.e., for $r < R_b$.
These are the values for which we make the comparison for the screenable cases. 

In a larger context, we can ask whether it makes sense to
consider the string picture of confinement in a situation where the string can eventually break. 
The lattice simulation in \cite{bonati-CM} considered 3d Yang-Mills theory coupled to a number of scalar fields in the fundamental representation, so that all representations are screenable (by suitable binding with
the scalar fields). The results
show that an effective string description is still valid for the confining part of the potential; even boundary terms and higher order corrections from the Nambu-Goto string action can be correctly reproduced by the simulation.

The second comment is about Casimir scaling versus the sine-law for the string tension, an issue which took some time to be clarified.
Here one considers the $k$-string corresponding to the antisymmetric rank $k$ representation of $SU(N)$. The value of the quadratic Casimir operator for this representation is easily calculated as
\beq
c_k = {N+1 \over 2 N} \, k (N-k)
\label{Res10b}
\eeq
If the string tensions are proportional to $c_k$ (as we found in (\ref{Res10})),
and as argued by others as well, then
\beq
{\sigma_k \over \sigma_F} = {k (N-k)\over N-1} 
\label{Res10c}
\eeq
This is the Casimir scaling law. An important feature of this ratio is that, in a large $N$ expansion, we have
\beq
{\sigma_k \over \sigma_F} \approx k \left( 1 - {k-1\over N} - {k+1\over N^2} + \cdots
\right)
\label{Res10d}
\eeq
Thus one can get odd powers of $1/N$ in this case.

The sine-law for the $k$-string is the statement that
\beq
{\sigma_k \over \sigma_F} = {\sin (\pi k /N) \over \sin(\pi/N)}
\approx k \left( 1- {(k^2-1)\pi^2 \over 6 N^2} + \cdots \right)
\label{Res10e}
\eeq
In this case, we have only even powers of $1/N$, evident
from the symmetry of the ratio of the sines under
$N \rightarrow - N$.

The sine-law was recognized as a possibility that one needs to consider
following the work of Douglas and Shenker who derived it in
${\cal N} = 1$ supersymmetric Yang-Mills (SYM) theory in 4 dimensions
\cite{Doug-S}.
This theory can be obtained by adding a supersymmetry breaking term to the
${\cal N} = 2$ SYM theory whose nonperturbative
analysis was carried out by Seiberg and Witten, and who obtained the exact
low energy effective action \cite{Sei-W}.
A similar result was obtained in \cite{Han-SZ} using a 5-brane construction in $M$-theory, the so-called MQCD.
Within the context of holography, one can obtain the $k$-string tension as
the value of the Hamiltonian for a classical supergravity configuration in the holographic dual description.
The sine-law is then obtained for the 4d SYM for the Maldacena-Nunez dual and an approximate sine-law for the Klebanov-Strassler background
\cite{Herz-K}.

While these results were obtained for the supersymmetric theory
using the gravity dual, a general argument for the sine-law was suggested
in \cite{Arm-S}, see \cite{Shif} for a review.
The basic argument is the following.
Since representations with zero $N$-ality can all be screened,
the asymptotic formula for the string tension should
depend only on the $N$-ality of the representation.
The rank $k$ antisymmetric representation and
the rank $(N-k)$ antisymmetric representation are conjugates of each other.
Therefore we should further expect the tensions to be invariant
under $k \rightarrow N-k$. This means that the ratio $\sigma_k/\sigma_F$, which is a dimensionless function depending only on
$k$ and $N$, should
be a function of $\vert \sin (\pi k /N)\vert$; we can represent it as
a power series of the form
\beq
{\sigma_k \over \sigma_F } = c_1 \vert \sin (\pi k /N)\vert
+ c_2 \vert \sin (\pi k /N)\vert^2 + \cdots
\label{Res10f}
\eeq
Further, we know that counting powers of $N$ in terms of diagrams in
perturbation theory show that at fixed $k$, one should only have even powers of $1/N$. 
The limit $N \rightarrow \infty$ with fixed $k$ should also exist,
as we expect confinement at large $N$ with a finite and nonzero
string tension. (This is after everything is expressed in terms of the 
't Hooft coupling $e^2N$ as in (\ref{Res10a}).) These properties require that $c_{2n+1} \sim N^1$
and $c_{2n} \sim N^0$.
The terms with odd powers have the property that,
in the limit $k\rightarrow \infty$, $N \rightarrow \infty$ with
$k/N$ fixed, $\sigma_k/ k$ is a function of $x = \pi k /N$.
The authors of \cite{Arm-S} refer to this as the saturation property.
Keeping only such terms, one ends up with an odd series
in $\vert \sin (\pi k /N)\vert$, with $c_{2n +1} \sim N$.
By comparison with the gravity dual arguments and fitting to some numerical data, one can then argue that a single power of $\vert \sin (\pi k /N)\vert$
suffices. The emerging suggestion from this line of reasoning
was that Casimir scaling should be ruled out as not being compatible with the
$(1/N)$-expansion of Yang-Mills theory.

However, the data from lattice simulations were fairly decisively in favor of Casimir scaling. 
This follows from the results of \cite{teper}-\cite{kiskis} and also from the specific check of Casimir scaling done for $G_2$ in \cite{wellegehausen}.
Simulations done for the high temperature (${\mathsf{T}}$) limit of 4d Yang-Mills
theory, which should reduce to the zero-temperature
3d theory with a redefined coupling $e^2= g^2 {\mathsf{T}}$, also shows Casimir scaling
\cite{Meis-Og}. A calculation using the gravity dual for the 3d SYM also supports Casimir scaling \cite{Herz}.
Detailed analyses with the gravity dual, for the string tension and for the Luscher term, were carried out in
\cite{Rodg1}, \cite{Rodg2}; the results seem to lie in between the
Casimir law and the sine-law, and close to both cases.

It would seem from the previous two paragraphs that there is
a possible conflict between the standard $(1/N)$-expansion and
Casimir scaling (which seems to hold for a number of  cases and which can include odd powers of $(1/N)$).
However, this is not the case, there is a loophole in the arguments
presented in \cite{Arm-S}, as shown by \cite{green-LP}.
The essence of this resolution is that, for the string tension, one is
calculating a matrix element of the form
$\la 0\vert F ~e^{-HT} F^\dagger \vert 0\ra $, as shown in section \ref{Conf}.
Using a complete set of energy eigenstates, we can write this in the form
\beq
\la 0\vert F ~e^{-HT} F^\dagger \vert 0\ra 
= \sum_\alpha \C_\alpha \, e^{ - E_\alpha T}
\label{Res10g}
\eeq
where $\C_\alpha$ and $E_\alpha$ are functions of the coupling constant, $N$, etc. Generally there is also a representation dependence arising from the choice of $F$. Consider now the $(1/N)$-expansion of various terms
in the sum.
It is possible for the
individual $\C_\alpha$ and $E_\alpha$ to have
odd powers of $(1/N)$. When we expand in $(1/N)$ at finite $T$,
there can be cancellation of the odd powers between different terms
in the sum, thus rendering the $(1/N)$-expansion of the correlator
consistent with expectations from the diagrammatic side. 
However, if we take large $T$ first, then
the term which dominates is the term with the
lowest energy, say, $e^{-E_0 T}$.
This is what is done both in our analytic calculation and in the lattice simulations, with the string tension extracted from
$E_0$. 
As mentioned earlier, $E_0$ can have odd powers of $(1/N)$,
but in taking the large $T$ limit first, the possible cancellants of the
odd powers of $(1/N)$ from higher $\C_\alpha$, $E_\alpha$ are
discarded, so the odd powers in $E_0$ are retained.
This argument shows that there does not have to be any contradiction between Casimir scaling and the $(1/N)$-expansion. 

The point is that the two limits, namely large $T$ and large $N$, do not necessarily commute.
(In \cite{green-LP} the authors give a specific example of how such a scenario
can be realized, with the cancellation  of the odd powers in the correlator, while retaining odd powers in $E_0$, in a lattice model in the strong coupling expansion.)
The conclusion is that Casimir scaling is compatible with
the expectations from the $(1/N)$-expansion in terms of diagrams.

\subsection{Casimir effect: Calculation}
In  section \ref{promass} we argued that our analysis leads to an
``improved" perturbation theory where, at the lowest order
the gauge-invariant version of the gluon is described by a scalar field $\phi^a$ with mass $m=(e^2 c_A /2\pi)$. The Hamiltonian for this was given in (\ref{promass7})
and it corresponds to the action
\beq
S = \int d^3x \, {1\over 2}\left[  {\dot \phi}^a {\dot \phi}^a - {(\nabla \phi^a ) (\nabla \phi^a ) }
- {m^2 \phi^a \phi^a }  \right] + \cdots
\label{Res11}
\eeq 
We can now use this to calculate the Casimir energy for the nonabelian gauge theory, in the usual classic set-up of two parallel conducting
plates or, rather, wires since we are in two spatial dimensions.
This calculation is from \cite{KN5}, the numerical estimate of the
Casimir energy is from
\cite{chernodub}.
We consider the fields in a square box of side $L$, with two
parallel wires separated by a distance $R$.
Eventually, we can take $L, \, b_1, \, b_2 \rightarrow \infty$
keeping $R$ fixed.
The relevant geometry is shown in Fig.\,\ref{figure1}.
 \begin{figure}[!b]
 \centering{
\scalebox{1}{\includegraphics{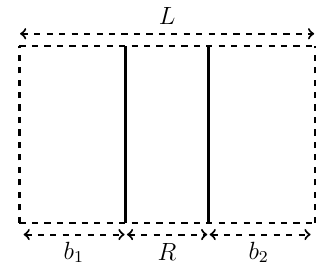}}
\caption{The set-up for Casimir effect}
\label{figure1}
}
\end{figure}
In the small $\vf^a$ expansion, the gauge potentials have the form
\beq
A^a_i \approx {1\over 2} \left[ - \del_i \theta^a + \epsilon_{ij} \, \del_j \vf^a + \cdots\right], \hskip .2in
M = \exp\left( - {i\over 2} t_a (\theta^a + i \vf^a )\right)
\label{Res12}
\eeq
(The field $\phi^a$ is related to $\vf^a$ as
$\phi _a (\vk) = \sqrt {{c_A k \bar{k} }/ (2 \pi m)}~~ \vf _a (\vk)$.)
The boundary condition appropriate to
perfectly conducting wires
is that the tangential component of the electric field should vanish; i.e.,
\beq
\epsilon_{ij} \, n_i F_{0j}^a = 0 ,
\label{Res13}
\eeq
where $n_i$ is the unit vector normal to the wire.
For small $\vf^a$, we see that this is equivalent to
the condition
\beq
n _i \epsilon_{ij} \epsilon_{jk} \del_k {\dot \vf^a} = 
- n_i \del_i {\dot \vf^a} = 0
\label{Res14}
\eeq 
Since the time-derivative does not affect the 
spatial boundary conditions, this can be satisfied by
imposing the
Neumann boundary condition $n\cdot \del \vf^a = 0$ on the
scalar field $\vf^a$ or, equivalently, on $\phi^a$.
This gives us a simple strategy for calculating the Casimir energy
within our improved perturbation theory: We just calculate
the Casimir energy of a free
massive scalar field, of mass $m$, with Neumann boundary conditions on the wires. (It may be worth re-emphasizing that, even though we use a free field theory, interactions and some nonperturbative effects are folded in
since there is a nonzero mass $m$.) Accordingly, the field in the region between the wires has the expansion
\beq
\phi^a = \int {d k\over 2\pi}  \sum_{n=0}^\infty C^a_{n, k} \, \sqrt{2 \over R}
\cos\left( {n \pi x_1 \over R}\right) \, e^{ik x_2} ,
\label{Res15}
\eeq
consistent with the Neumann boundary conditions.
The action is then obtained as
\beq
S = \int dt\,{d k \over 2 \pi} \sum_n {1\over 2} \left[
{\dot C^a_{n,k} } {\dot C^a_{n,k} } - \Omega_{n,k}^2 \, {C^a_{n,k} }{C^a_{n,k} }
\right] + \cdots
\label{Res16}
\eeq
where $\Omega^2_{n,k} = k^2 + (n \pi /R)^2 + m^2$.
Here $n$ is an integer $\geq 1$. (Notice that $n =0$ will not give an $R$-dependent term.)
 The unrenormalized zero-point energy can be easily read off as
\beqar
\E &=& {L \over 2} {\rm dim}G \, \int {d k \over 2 \pi} \sum_n 
\Omega_{n,k} \nonumber\\
&=&  {L \over 2} {\rm dim}G \, \int {d k \over 2 \pi} \sum_n 
 \left( {\del^2 \over \del x_0^2 }\right)
\int {dk_0 \over \pi} { e^{ik_0 x_0} \over k_0^2 + \Omega_{n,k}^2}  \biggr]_{x_0 =0}
\label{Res17}
\eeqar
The summation can be done using the formula
\beq
\sum_{n=1}^\infty {1\over k_0^2 + \Omega^2_{n,k}}
= - {1\over 2 \omega^2} + {R \over 2 \omega}
+ {R \over \omega} {1\over e^{2 \omega R} - 1} 
\label{Res18}
\eeq
where $\omega^2 = k_0^2 + k^2 + m^2$.
Thus $\E$ splits into three terms. The contribution from the first term on the right hand side of (\ref{Res18}) is independent of $R$ and will disappear
when we take $\E (R) - \E (R \rightarrow \infty )$ to obtain the
renormalized energy. As for the second term, there will be similar contributions from the regions of extent $b_1$ and $b_2$, so that together
we get $(R + b_1 + b_2)/(2\omega ) = L/(2\omega)$. So its contribution is also independent of $R$.
The expression for the energy now becomes
\beqar
\E &=& - {L R \over 4\pi} {\rm dim}G\,\int_0^\infty dp \, {p^3 \over \sqrt{p^2 + m^2} }\,
{1\over e^{ 2 R \sqrt{p^2 + m^2} } - 1}\nonumber\\
&=&- {\rm dim}G\, {L \over 4\pi R^2} (m R)^3 \int_1^\infty d z  { (z^2 -1) \over e^{2 m R  z} - 1}\nonumber\\
&=&- {\rm dim}G\, {L \over 16\pi R^2}  
\left[  2m R ~ {\rm Li}_2 (e^{- 2 m R}) + {\rm Li}_3(e^{- 2 m R} )\right]
\label{Res19}
\eeqar
In going from (\ref{Res17}), (\ref{Res18}) to the first line of this equation, we have carried out the angular integration over the angle
$\alpha$, taking $k_0 = p \cos \alpha$,
$ = p \sin \alpha$, where $p = \sqrt{k_0^2 + k^2}$.
By using $p = m \sinh q$, and $z = \cosh q$ we get to the second line.
The expansion of this in powers of $e^{- 2m R z}$ leads to the last line
of (\ref{Res19}) in terms of the polylogarithms,
\beq
{\rm Li}_s (w) = \sum_{1}^\infty  {w^n \over n^s}
\label{Res20}
\eeq
Using the expression (\ref{Res10}) for the string tension, we can re-express (\ref{Res19})  in terms of $x = \sqrt{\sigma_F}\, R$ as
\beq
\boxed{
{\E \over L \sigma_F } = - {{\rm dim}G \over 16\pi} \left[ {2\sqrt{c_A/\pi c_F} \over x}\,
{\rm Li}_2 \left( e^{- 2\sqrt{c_A/\pi c_F} \,x} \right) + { 1 \over x^2} {\rm Li}_3 \left( e^{- 2\sqrt{c_A/\pi c_F}\, x} \right)
\right] }
\label{Res21}
\eeq
This is the analytic result we get for the Casimir energy
as a function of the separation of the wires.
\subsection{Casimir effect: Comparison with lattice data}
The formula for the Casimir energy given in (\ref{Res21}) is
in a form that can be compared to the lattice simulations.
 In fact, for the case of $G = SU(2)$, such a simulation and extraction of the Casimir energy for the parallel wire geometry
  have been carried out in \cite{chernodub}.
Using the appropriate values of $c_A$ and $c_F$, the specialization
of formula (\ref{Res21}) to $SU(2)$ is
\beq
{\E \over L \sigma_F } = - A\,{{\rm dim}G \over 16\pi} \left[ { 1.84 \over x}\,
{\rm Li}_2 \left( e^{- 1.84\,x} \right) + {1 \over x^2} {\rm Li}_3 \left( e^{- 1.84\, x} \right)
\right]
\label{Res22}
\eeq
We have also included a prefactor $A$.
The motivation for a possible change of the prefactor
\begin{figure}[!b]
\begin{center}
\scalebox{1}{\includegraphics{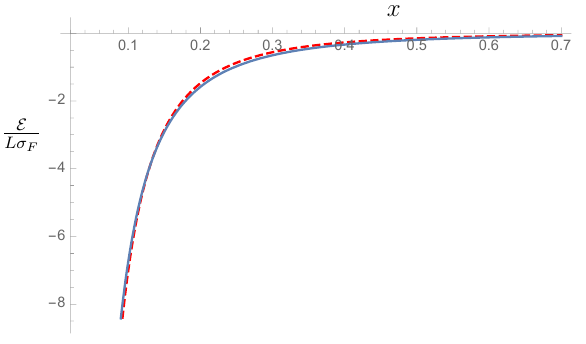}}
\caption{Comparison of (\ref{Res22}) (solid blue line) and (\ref{Res23}) (dashed red line). }
\label{figure2}
\end{center}
\end{figure}
(from the value of $A = 1$ as in
(\ref{Res21}))
is the following.
The prefactor is really a measure
of the number of degrees of freedom. This is clear from the
${\rm dim}G$ factor. However, lattice simulations of Yang-Mills
theories have shown that
the number of degrees of freedom do not quite reach a value corresponding to a gas of free gluons even at very high temperatures, where we expect a
deconfined gluon plasma. This has known for a fairly long time.
(A recent review which 
gives updated results is \cite{pasechnik}; in particular,
see figure 4 of this reference.) 
There could also be higher order effects (interactions among the
$\vf^a$ fields, corrections to the wave function, etc.) which could
 contribute to
$A$. The exponential fall-off is however controlled by the mass
$m$. So we do not tamper with that; the value from our analysis, namely $m = (e^2 c_A /2\pi)$,
has been used in (\ref{Res22}).

As for the comparison with lattice data, the authors of
\cite{chernodub} fitted the data points to a phenomenologically
motivated formula
\beq
{\E \over L \sigma_F} = - {\rm dim}G\, { \zeta (3)\over 16 \pi}\, x^{-\nu} \,
e^{- M\, x / \sqrt{\sigma_F}},
\label{Res23}
\eeq
the best fit values being 
$\nu = 2.05$ and $M = 1.38\, \sqrt{\sigma_F}$.
in Fig.\,\ref{figure2}, we show the curve corresponding to 
(\ref{Res23}) as the dashed red line, using the best fit values
quoted above.
It is very clear that our formula (\ref{Res22}) is at least as good a fit 
to the lattice data as the
phenomenological formula (\ref{Res23}).
We have used only one fitting parameter, namely $A$. Its best fit value is
$A= 0.9715$. If we used (\ref{Res21}) without allowing for a change of
the prefactor (which means $A = 1$), the agreement would still be rather good, since this would only give a small change in the
overall coefficient.
Notice that the exponential factors are just as predicted
from (\ref{Res21}).
Even a small error in the mass $m$ could give a significant deviation 
in the profile of the function since it
is in the exponent.

Why is our result for the Casimir energy so accurate considering that it is obtained using the ``free theory", albeit including the mass 
which is nonperturbatively generated? 
Obtaining a lattice estimate of the Casimir energy
at large separations is problematic because of the exponential damping.
The numerical values are lost in the noise. 
At the other end, for very short distances, lattice artifacts get in the way.
So the lattice estimates are by necessity
confined to a certain range (roughly between $x = 0.1$ and $x = 0.7$
in the graph). 
This is the kinematic regime which we might expect to be more or less accessible by
perturbation theory, but improved to incorporate a mass
which is
necessary to include the exponential fall-off.
\subsection{Propagator mass: Alternate approaches}
The Casimir effect, as discussed above,
may be the most accurate way to test the prediction about the
propagator mass for gluons. But there are a few other ways
to attempt the calculation or the numerical estimate of this quantity.

First of all, since we argued that our analysis, in the high momentum
regime, could be viewed as a resummation of a select series of Feynman diagrams, one could attempt a direct resummation within standard covariant
perturbation expansion.
In such an approach, the difficulty we might face is that the selection of the terms to be resummed has to respect gauge invariance or BRST invariance.
This means that the chosen set of terms should form a closed set
with respect to the relevant Ward-Takahashi identities.
Ensuring this feature can be cumbersome in practice.
However, since we are primarily interested in the mass, not full-blown
off-shell amplitudes, there is a simpler method we can use.
The idea is to first construct a possible gauge-invariant mass term
for the gauge fields. This will be of the form
\beq
S_{\rm mass}= {1\over 2} \int {d^3k \over (2\pi )^3} 
 A_i^a(-k) A_j^a(k)\biggl(\delta_{ij}-{k_ik_j\over
{\vec k}^2}\biggr)  +{\cal O}(A^3)
\label{Res25}
\eeq
The quadratic term shows that this is truly a mass term for the transverse gauge fields, but is not gauge-invariant. However, one can add to it a suitable series involving $A$'s to get a gauge-invariant completion of this term.
Since the completion is not uniquely defined, we can have different
possible choices for $S_{\rm mass}$.
All such mass terms are necessarily sums of nonlocal monomials
of the fields. Once we have chosen a mass term
$S_{\rm mass}$, we consider the action\footnote{For the rest of this section, we use the Euclidean theory in keeping with the fact that
this signature is used in diagrammatic perturbation theory.}
\beq
S=S_{\rm YM} + M^2 S_{\rm mass} - \Delta ~S_{\rm mass}
\label{Res26}
\eeq
where $S_{\rm YM}$ is the usual Yang-Mills action.
We take $\Delta$ to have a loop expansion starting at the 1-loop order,
writing
\beq
\Delta= \sum_1 \hbar^n \, \Delta^{(n)} 
\label{Res27}
\eeq
After adding gauge fixing and ghost terms, we can calculate the effective action $\Gamma$. This will have the form
\beq
\Gamma = S_{\rm YM} + M^2 S_{\rm mass}
- \sum_1 \hbar^n \Delta^{(n)} \, S_{\rm mass}
+ \hbar \Gamma^{(1)} (\hbar) + 
\hbar^2 \Gamma^{(2)} (\hbar) + \cdots
\label{Res28}
\eeq
Notice that while $\Gamma^{(1)}$ is obtained from terms which are graphically of the 1-loop topology, it contains terms of arbitrary order
in $\hbar$ since the propagators involve
$\Delta$ and are of the form
$(k^2 + M^2 - \Delta )^{-1}$. 
Thus $\Gamma^{(1)}$ has the expansion
\beq
\Gamma^{(1)} = \Gamma^{(1)}_0 + \sum_1 \hbar^n \, \Gamma^{(1)}_n
\label{Res29}
\eeq
In a similar way, additional
$\hbar$ dependence arises for the 2-loop vertices. The terms of
order $\hbar^2$ in
$\Gamma^{(2)}$ are due to\\
a) 2-loop graphs with
the propagators $\sim (k^2 + M^2)^{-1}$\\
b) terms from $\Gamma^{(1)}_1$.\\
Collectively we will denote these terms by $\Gamma^{(2)}_0$.
Just to clarify, we may note that the second set of terms can arise from 1-loop diagrams with vertices
from $S_{\rm YM} + M^2 S_{\rm mass}$ (plus similar ghost vertices)
and propagators expanded to $\Delta^{(1)}$ order or from
1-loop diagrams with propagators $\sim (k^2 + M^2)^{-1}$,
but with vertices from $\Delta^{(1)} S_{\rm mass}$.
Similar reasoning will hold for higher order terms.
The expansion (\ref{Res28}) thus takes the form
\beqar
\Gamma &=&
S_{\rm YM} + M^2 S_{\rm mass} + \hbar \left( \Gamma^{(1)}_0 - \Delta^{(1)} S_{\rm mass}\right)  + 
\hbar^2 \left( \Gamma^{(2)}_0 - \Delta^{(2)} S_{\rm mass}\right) +\cdots
\nonumber\\
&& + \textrm{gauge-fixing terms} + \textrm{ghost terms}
\label{Res30}
\eeqar
We take $M^2$ to be the propagator mass; this is so at the tree level
with the pole appearing at $k^2 + M^2 = 0$.
It can be kept at the same point to order $\hbar$ by choosing
$\Delta^{(1)}$ to cancel any shift of the pole induced by
$\Gamma^{(1)}_0$. This will determine
$\Delta^{(1)}$ as a function of $M$.
Likewise, we choose $\Delta^{(2)}$ to cancel any shift of the pole
at order $\hbar^2$, etc.
Finally, we are not seeking to change the theory, it should still be
the usual Yang-Mills theory. So we should at the end set
\beq
M^2 = \Delta = \Delta^{(1)} + \Delta^{(2)} +  \cdots ,
\label{Res31}
\eeq
so that the starting action (\ref{Res26}) is just the Yang-Mills action.
This equation becomes a gap equation for the theory determining $M$ in terms of the coupling constant.

The procedure we have outlined gives a systematic and gauge-invariant way to carry out a resummation of the select set of terms and identify the
propagator mass. It can also be implemented
order by order; for example, $M^2 = \Delta^{(1)} $ will be the 1-loop
gap equation, $M^2 = \Delta^{(1)} +\Delta^{(2)} $ is the 2-loop
gap equation, etc.
The series of terms which are selected to be resummed is determined by
the choice of $S_{\rm mass}$, with different choices
corresponding to different series being resummed.
This method of obtaining the gap equation has been explained in some
detail in \cite{AN, J-Pi}.

Calculations along these lines have been carried out for several different choices of $S_{\rm mass}$. In \cite{BP}, a Higgs field was used to generate a gauge-invariant mass term, in a way similar to how the
Higgs mechanism gives a mass to vector bosons.
In \cite{AN}, we
have used a different mass term inspired by the 2d-WZW action and
also by the Debye screening mass term in 4d-Yang-Mills theory at high temperatures. This mass term and its properties are discussed in
\cite{AN2}.) There is also has an interesting 
geometrical side to it, in terms of Sasakian structures on three-dimensional
spaces \cite{nair-Jfest}.
Jackiw and Pi used a more conventional mass term of the form
$F(1/D^2)F$ in their analysis in \cite{J-Pi}.
A Chern-Simons term, although parity-violating, has also been used
\cite{hipp-horgan}.
In the references cited so far, the calculations were done to 1-loop
order and the resulting gap equation was then solved to obtain the value of the propagator mass. The calculations of \cite{BP} have
also been extended to obtain the 2-loop gap equation
in \cite{eberlein, Phil2}.
The values of the propagator mass obtained using these 
mass terms are given in Table \ref{magmass}.

\begin{table}[!b]
\begin{center}
\vskip .05in
\begin{tabular}{c | c l }
Group&$m/e^2$ & Method\\
\hlineB{3}
\multirow{14}{*}{$SU(2)$}&0.38& Resummation, 1-loop \cite{AN}\\
&0.28& Resummation, 1-loop \cite{BP, J-Pi}\\
&0.35& Lattice, common factor for glueball masses \cite{BP}\\
&0.34&Two-loop gap equation \cite{eberlein}\\
&0.33&Two-loop gap equation \cite{Phil2}\\
&~~~0.25~~~~~& Resummation of perturbation theory \cite{JMC}\\
&0.51&Lattice, maximal abelian gauge \cite{karsch1}\\
&0.52&Lattice, Landau gauge \cite{karsch1}\\
&0.44&Lattice, $\lambda_3 =2$ gauge \cite{karsch1}\\
&~0.456&Lattice, Landau gauge \cite{heller}\\
&0.37&Gauge-invariant lattice definition \cite{owe2}\\
&0.36&Gauge-invariant correlation length \cite{owe3}\\
&&\\
&0.32&Calculation via our Hamiltonian method\\
\hline
\multirow{6}{*}{$SU(3)$}&~0.568& Resummation, 1-loop \cite{AN}\\
&~\,0.42~~~& Resummation, 1-loop \cite{BP, J-Pi}\\
&~0.515&Lattice, Landau gauge \cite{nakamura1}\\
&~0.482& Lattice, Landau gauge \cite{nakamura2}\\
&&\\
&0.48&Calculation via our Hamiltonian method\\
\hlineB{3}
\end{tabular}
\vskip.05in
\caption{Comparison of propagator mass calculations}
\label{magmass}
\end{center}
\end{table}
A second method
is to use the Schwinger-Dyson formulation of the theory
\cite{ABP1}.
This is effectively a reorganization and resummation of the 
perturbation expansion and so it is ideologically related to 
the resummation method discussed above. By combining the 
Schwinger-Dyson equation with the pinch technique \cite{pinch, ABP2},
it is possible to get gluon propagators and identify the mass.
(The pinch technique is a way of adding a certain kinematic limit
(the pinching limit)
of some Feynman diagrams to other $n$-point functions to obtain
gauge-invariant $n$-point functions. See the references quoted for the details of how this can be implemented.)
The Schwinger-Dyson equations are nonlinear and in the end, in this approach, some numerical work is needed to solve them.
The result seems to give a value close to what is obtained by
the other methods, as seen from Table \ref{magmass}.

Yet another approach is to use lattice simulations again.
Arguably, the best feature of lattice gauge theory is the ease of preserving manifest gauge invariance. However, to define the gluon propagators and the propagator mass one needs to do gauge-fixing. 
This will also bring in issues like the Gribov problem.
Nevertheless, calculations have been done using
the Landau gauge, the maximal Abelian gauge and for what we shall
refer to
as the $\lambda_3 = 2$ gauge. (See the quoted reference for details.)

Finally, we point out that Philipsen \cite{owe3} has also calculated the propagator
mass by considering the correlation length and fall-off of gauge-invariant
partonic correlators in the 4d-Yang-Mills theory at high temperatures
(which is equivalent to the 3d-theory with a redefinition of the coupling constant).

The values obtained by all these different methods vary
by some amount, but they are not very far from what we obtain using 
our Hamiltonian method.
\section{Screenable representations and string breaking}\label{s-break}
 \setcounter{equation}{0}
\begin{quotation}
\fontfamily{bch}\fontsize{10pt}{16pt}\selectfont
\noindent 
For screenable representations, the interaction energy between a particle and antiparticle should become independent of the separation beyond a certain point, as each can combine with a number of gluons and form composites (glue lumps) of zero charge, a process referred to as string-breaking. We obtain an approximate Schr\"odinger equation for the glue lump, calculate the string-breaking energy and compare with lattice estimates, commenting on some of the subtleties
 in interpreting the lattice data.
We also give a qualitative argument about
 how string-breaking can appear in the calculation of the expectation value of the Wilson loop operator.
\end{quotation}
\fontfamily{bch}\fontsize{12pt}{16pt}\selectfont
Screenable representations are representations $R$ such that
$R \otimes {\rm Adjoint}  \otimes {\rm Adjoint} \cdots$ contains
the trivial (or singlet) representation. For $G = SU(N)$, these are representations of zero $N$-ality. As discussed in section \ref{Conf},
if we consider a particle-antiparticle pair in a screenable representation, the potential between them will flatten out at some point as the separation between the two is increased. Thus, for screenable representations $R$, the formula for the string tension should apply only up to this critical separation.

The process of the flattening out of the potential can be visualized as follows.
Since $R \otimes {\rm Adjoint}  \otimes {\rm Adjoint} \cdots$ contains a singlet,
we could have a bound state of zero charge made of the particle with a certain number of gluons.
This bound state is usually referred to as a glue lump.
Similarly one can have a glue lump for the antiparticle with some gluons.
As we increase the separation between the particle and the antiparticle,
at some point, it becomes energetically favorable to convert the
interaction energy into creating a glue lump pair.
Once this is achieved, there is no further energy cost to separating the
lumps, since each of them has zero charge. The energy
of the pair remains what it was at the point of the glue lump formation.
This is the flattening, which can also be thought of as the breaking up of the string connecting the particle and the antiparticle.

The ideal scenario theoretically would be to see this directly in the calculation of the expectation value of the Wilson loop.
This has not been possible so far using the vacuum wave function we have obtained, namely, the result to the lowest two orders within our expansion scheme.
The reason for this will be clear and commented upon later in this section.
But as a first attempt, what we will do is an approximate calculation of the ground state energy of a glue lump and then argue that the energy needed to create a glue lump pair in their ground state is the critical value $V_*$ of the
potential energy at the string-breaking point.

\subsection{A Schr\"odinger equation for the glue lump}

We will first outline the calculation of the glue lump ground state energy.
Recall that we can consider the Wilson loop as a process involving the propagation of a heavy particle-antiparticle pair, each, say, of mass $M$.
The simplest case to consider is when the representation $R$ for the particle
is the adjoint one, so that the
glue lump is the bound state of this particle (or antiparticle) with
the gluon \cite{AKN1}. If $\phi^a$ is the field representing the heavy particle,
then its gauge-invariant version, denoted by $\Phi^a$, is given by
\beq
\Phi^a t_a = M^\dagger \, \phi^a t_a \, M^{-1}
\label{s-break1}
\eeq
The wave function of a glue lump state will then have the form
\beqar
\Psi_G &=& \int_{x, y} f(\vx, \vy) \, \bdel J^a(x) {\widetilde W}^{ab} (x,y) \Phi^{b}(y) \, \Psi_0\nonumber\\
{\widetilde W}^{ab} (x,y) &=&\left[K(x, \by) K^{-1} (y, \by )\right]^{ab}
\label{s-break2}
\eeqar
Here $\Phi^a$ represents the particle and $J^a$ the gluon.
$f(\vx, \vy)$ is the two-body wave function for the gluon and the heavy particle. $\Psi_0$ is the wave function for the ground state, which is given by
\beq
\Psi_0 = {\cal N} \exp\left( - {1\over 2} \F \right)\, \exp\left({ - {1\over 2} \mathsf{M} \int \Phi^a \Phi^a}\right)
\label{s-break3}
\eeq
with $\F$ as given in section \ref{SchE}.
The second exponential is the ground state wave function for a heavy 
particle of mass $\mathsf{M}$.

Our aim now is to act on this wave function with the Hamiltonian.
In some approximation, as explained below, this will lead to an
ordinary two-body Schr\"odinger equation for $f(\vx, \vy )$.
We can then estimate the ground state energy as we do in
quantum mechanics.
The Hamiltonian is given by $\H$ from section \ref{Ham}, or
the expression (\ref{SchE1}), with the Hamiltonian for the scalar field
added to it. The result is $\H = \H_{\rm YM} + \H_\Phi$,
with
 \beqar
{\cal H}_{\rm YM}  &=& m  \int J_a (\vz) {\d \over {\d J_a (\vz)}} + {2\over \pi} \int _{w,z} 
 {1\over (z-w)^2} {\d \over {\d J_a (\vw)}} {\d \over {\d
J_a (\vz)}} \nonumber\\
&&+{1\over 2} \int_z : \bdel J^a(z) \bdel J^a(z): 
+ i e \int_{w,z} f_{abc} {J^c(w) \over \pi (z-w)} {\d \over {\d J_a (\vw)}} {\d \over {\d J_b (\vz)}} \nonumber\\
\H_\Phi&=&\int  :\left[-{1 \over 2}  \frac{\delta ^2}{\delta \Phi^a\delta \Phi ^a} +  \left( {2 \pi \over c_A} \bdel \Phi^a (\D \Phi)^a + {\mathsf{M}^2 \Phi^a \Phi^a
\over 2} \right)\right]\nonumber\\
 &&+i m \int_{z,w} \Lambda_{cd}(\vw, \vz) f^{abc}\Phi^a(\vw)
{\delta \over \delta \Phi^b(\vw)} {\delta \over \delta J^d(\vz)}\, :
\label{s-break4}
\eeqar
where
\beqar
({\cal{D}}_{w})_{ab}  &=& {c_A\over \pi}\partial_w \delta_{ab} +if_{abc}J_c (\vw)
\nonumber\\
\Lambda_{cd}(w,z)&=&- \del_z\left[ \int_x \bar{\G} _{ac} (\vx,\vw) K_{ab}(\vx) \G _{bs} (\vx,\vz)\right] \, K^{-1}_{sd}(\vz)\nonumber\\
\bar{\G} _{ma} (\vx,\vy)  &=& {1\over \pi (x-y)}   \Bigl[ \d _{ma} - e^{-|\vx-\vy|^2/\e} \bigl(
K(x,\by) K^{-1} (y, \by) \bigr) _{ma}\Bigr] \nonumber\\
\G _{ma} (\vx,\vy)  &=&  {1\over \pi (\bx - \by )} \Bigl[ \d _{ma} - e^{-|\vx-\vy|^2/\e} \bigl(
K^{-1}(y,\bx) K (y, \by) \bigr) _{ma}\Bigr]
\label{s-break5}
\eeqar
The propagators given in the last two lines of (\ref{s-break5}) are the
regularized form of the corresponding propagators.
$\H_\Phi$ is normally ordered so that there is no zero-point energy, as
indicated in (\ref{s-break4}).

We can now consider the action of $\H$ on $\Psi_G$.
Before we give details, we will make some observations which are useful
in understanding the genesis of various terms in the resulting Schr\"odinger equation. We have already seen that
$T J^a = m J^a$ in (\ref{ham11}).
When we include $\Psi_0$ as well, we find
\beq
T\, J^a(x) \Psi_0 = \left( m - {\nabla^2 \over 2 m} \right) J^a(x) \Psi_0
+ {2 i \pi \over m c_A} f_{abc} J^b \bdel J^c \Psi_0 + \cdots
\label{s-break6}
\eeq
If we ignore the terms of order $J^2$, this is like an eigenvalue equation,
the eigenvalue itself being $(m - {\nabla^2 /2 m})$, which is the nonrelativistic version of $\sqrt{m^2 + k^2}$.\footnote{Actually one can recover the fully relativistic expression $\sqrt{m^2 + k^2}$ by summing up a series in $1/m$.
The situation is very similar to what happens with quantum fluctuations around a static soliton. A series of terms produced by the zero modes of the Hessian of the action at the soliton solution has to be summed up to get the relativistic formula.}

Notice that we also have
\beq
\int  :\left[- {1 \over 2}  \frac{\delta ^2}{\delta \Phi^a\delta \Phi ^a} +  {1\over 2}{\mathsf{M}^2 \Phi^a \Phi^a} \right]: \Phi^b \Psi_0
= \mathsf{M} \,\Phi^b \Psi_0
\label{s-break7}
\eeq
The action of the operator on the left also produces a term ${\half} \mathsf{M}\,
\Phi^b \Psi_0$, but a similar result is obtained for just $\Psi_0$ as well; it is the vacuum energy which is removed by the normal ordering.
From these two statements and the fact that $\Psi_G$ contains
$\bdel J^a$ and $\Phi^b$, we expect that the action of the
Hamiltonian will produce a contribution of 
$(\mathsf{M}+ m - {\nabla^2 /2 m})$ to the eigenvalue. (One could also have
$-\nabla^2/ 2\mathsf{M}$, but this is negligible as we take $\mathsf{M}$ very large.)
There will also be the energy of the interaction between the
charged factors $\bdel J^a$ and $\Phi^b$ in $\Psi_G$.
These expectations are born out by the explicit calculations, with the
result
\beq
\H ~\Psi_G  =
\int \left[ M+ m  - {\nabla_x^2 \over 2m} + \sigma_A \vert \vx -\vy\vert
\right] f(\vx,\vy)~ \bdel J^a (\vx) {\widetilde W}^{ab}(\vx, \vy)  \Phi^b (\vy)
\Psi_0
~+ \cdots
\label{s-break8}
\eeq
Here $\sigma_A$ is the string  tension for the adjoint representation
and the ellipsis stands for a number of terms we have neglected.
The approximations involved in arriving at this equation are
the following.
\begin{enumerate}
\item First of all, there is the obvious approximation of 
using the leading solution (\ref{SchE9}) for the vacuum wave function;
i.e., $\F$ in (\ref{s-break3}) is taken to be the leading kernel for the
quadratic term in the $J$'s.
\item There are terms which correspond to new operator
structures, i.e., they are not of the form
of $\Psi_G$ and have more powers of $J$. 
These are possible new states in the theory.
The glue lump state $\Psi_G = \bdel J \widetilde{W} \Phi \, \Psi_0$
given in (\ref{s-break2})
can have overlap with such states since they have
the same quantum numbers.

 The fact that the Hamiltonian acting on $\Psi_G$ can produce
these other structures shows that there can be mixing. 
But, as is well known in quantum mechanics, the effect of nondiagonal terms in ${\cal H}$ comes with energy denominators and can be taken to be small if the differences between the energies of the lowest glue lump state (which is what we are interested in) and higher states are large enough. We expect this to be the case at large enough coupling, since the differences must go like $m$. But ultimately it is to be justified {\it a posteriori}.\footnote{The transition amplitude from the Wilson loop without string-breaking to the glue lumps is seen to be very small from lattice data, even for $SU(2)$ \cite{Phil-wittig, krat}. This is another indication for the expectation that the off-diagonal elements mentioned above should be small.}
For more details on these issues, see \cite{AKN1}.
\end{enumerate}

Accepting the caveats mentioned above, we can now see that
the glue lump state (\ref{s-break2}) will be an eigenstate of the Hamiltonian
if $f(\vx, \vy)$ obeys the ordinary Schr\"odinger equation
\beq
\left[ \mathsf{M} + m  - {\nabla_x^2 \over 2m} + \sigma_A \vert \vx -\vy\vert
\right] f(\vx,\vy) = E ~f(\vx,\vy)
\label{s-break9}
\eeq
Removing the center of mass motion (which is zero as $M \rightarrow \infty$), we see that this equation is the one obtained by minimizing the energy functional
\beqar
E &=&  \mathsf{M} + m + {1\over {\cal N}} \int d^2x~ \left[ {\vert \nabla f \vert^2 \over 2m}  ~+ \sigma_A
\vert \vx\vert ~\vert f\vert^2 \right] \nonumber\\
{\cal N} &=& \int d^2x~ \vert f \vert^2
 \label{s-break10} 
\eeqar 
The simplest way to proceed further is to use a variational procedure.
We consider an ansatz of the form
\beq
f = \exp ( -\beta \vert x\vert^\mu )
\label{s-break11}
\eeq
where $\beta$ and $\mu$ are to be treated as variational parameters.
Calculating $E(\beta, \mu)$ and extremizing it we find that the minimum occurs at $\beta = \beta_*$, with
\beqar
E(\beta_*, \mu) &=& M + m +\Biggl[  \frac{2^{-(2\mu +1)/\mu}}{2m\Gamma (2/\mu)}\left(\frac{2^{(\mu
-3)/\mu}2m\sigma_A \Gamma(3/\mu)}{\mu ^2}\right)^{-1/3}\nonumber\\
&&\hskip .2in\times \left(8m\sigma_A
\Gamma(3/\mu ) + 8^{1/\mu }\mu^2 \left(\frac{2^{(\mu
-3)/\mu}2m\sigma_A \Gamma(3/\mu)}{\mu^2}\right)\right)\Biggr]
\label{s-break12}\\
\beta_* &=& \left( \frac{2^{(\mu -3)/\mu}(2 m
\sigma_A )\Gamma(3/\mu )}{\mu ^2}\right)^{\mu/3}\nonumber
\eeqar
The minimization with respect to $\mu$ has to be done numerically.

A seemingly more general ansatz for $f$ is
\beq
f = \frac{e^{-\beta \vert x
\vert^\mu}}{(1+ \vert x\vert )^\nu }
\label{s-break13}
\eeq
with $\beta$, $\mu$, $\nu$ as variational parameters.
This may seem better motivated since the
solutions of the Schr\"odinger equation
with a linear potential involve the Airy functions ${\rm Ai}$
and we have \cite{AKN1}
\beq
\frac{{\rm Ai}\left((2m \sigma_A)^{1/3}\vert x \vert\right)}{\sqrt{\vert x \vert
}}\approx \frac{\exp \left[ {-\frac{2}{3} \left((2m\sigma_A)^{1/3}\vert x
\vert\right)^{3/2}}\right]}{\vert x \vert ^{3/4}} \label{s-break14}
\eeq
for large $\vert x\vert$.
However the minimization of $E(\beta, \mu, \nu )$ shows that the minimum
occurs at $\nu =0$, so, in the end, this is equivalent to
(\ref{s-break11}).
\subsection{Comparison with lattice simulations}
The calculation of string-breaking given above is admittedly rather crude.
As stated, there are several terms which have been neglected.
Even after this, the Schr\"odinger equation (\ref{s-break9})
has only the linear potential $\sigma_A \vert \vx - \vy\vert$. At short distances
one should expect a Coulomb potential (logarithmic in 2+1 dimensions)
as in perturbation theory. In the glue lump, since the wave function
has some nonzero probability at short separation between the constituents, the Coulomb potential can have an impact on the energy. There is also the
L\"uscher term $(\pi /24 \vert \vx - \vy\vert)$ for the potential energy
\cite{luscherterm}.
Finally we are only carrying out a variational estimate of the ground state energy. So, all things considered, the calculation presented above
should be viewed as primarily being of qualitative value, demonstrating
the possibility of string-breaking. Nevertheless, it is interesting to compare with lattice estimates, keeping in mind all the caveats mentioned above.

Minimizing the energy in the formula (\ref{s-break12})
with respect to $\mu$ and 
using the value of the adjoint string tension given by
(\ref{Res10}), namely, $\sigma_A = (e^4 c_A^2/4\pi) = \pi m^2$,
we find $E(\beta_*, \mu_*) = M + 3.958 m$. Since we had two masses 
$M$ initially, the extra energy needed to create a glue lump pair is
$2\times 3.958 m = 7.916 m$. Thus we should expect that the string breaking
should occur when the interaction energy is $V_{* \rm cal} = 
7.916 m$. 

A point of internal consistency of the calculation is the following.
Taking the value $7.916 m$ for $V_{* \rm cal}$,
the separation between the constituents of the glue lump at
the point of breaking is $r_* = (V_*/\sigma_A) \gtrsim 2.52/m$ and so, the
typical momentum at this separation $\lesssim (m/2.52)$.
While this does not make an unassailable case for the nonrelativistic treatment, it is not inconsistent.

Turning to the lattice numbers, an early estimate of string-breaking in $SU(2)$ by Philipsen and Wittig 
\cite{Phil-wittig} gave a value of the breaking separation $R_b$ as
$13.6/M_g$ where $M_g$ is the mass of the lightest glueball.
Taking this to be the $0^{++}$ with a mass of $5.17 \, m$
as given by \cite{teper}, this translates to
$V_{*{\rm lat}} = 8.26 m$. 
The calculated value agrees with the lattice estimate
up to $\vert V_{*{\rm lat}} - V_{*{\rm cal}} \vert /V_{*{\rm lat}}
= 4.16\%$.\footnote{String-breaking was also clearly demonstrated in
\cite{kne-som}, but the authors considered the theory with
quarks in the fundamental representation. This is different from our case of adjoint static charges, so a comparison is not obtained.}

In the lattice estimate by Kratochvila and de Forcrand \cite{krat},
the value of $V_*$ for $G = SU(2)$ was obtained as
$V_{*{\rm lat}} = 2.063 a^{-1}$, while the fundamental string
tension is $\sigma_F = 0.0625 a^{-2}$, where 
$a$ is a lattice spacing used in the simulation. This implies
$V_{*{\rm lat}} = 2.06 m \sqrt{(\pi c_F /0.0625 c_A)}
= 8.68 m$, fairly close to the value obtained by
\cite{Phil-wittig}. The deviation of this
from the calculated value of $7.916 m$ is
approximately
$8.76\%$.
\begin{figure}[!t]
\begin{minipage}{7.6cm}
\begin{center}
\scalebox{.5}{\includegraphics{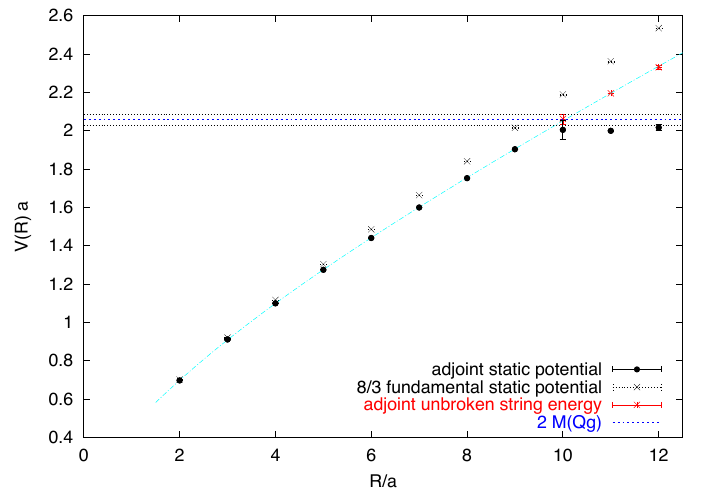}}
\caption{The adjoint and fundamental static potentials $V (R)$ (the latter multiplied by the Casimir factor $8/3$) versus $R$ using Wilson loops only. The adjoint static potential remains approximately constant for 
$R \geq R_b = 10 a $ proving string breaking. The unbroken-string state energy is also drawn. The horizontal line at $2.06(3)a^{-1}$ represents twice the mass of a glue lump. This is graph is from \cite{krat}.}
\label{breaking1}
\end{center}
\end{minipage}
\hskip .1in
\begin{minipage}{7.5cm}
\begin{center}
\null\vspace{-42pt}
\scalebox{.8}{\includegraphics{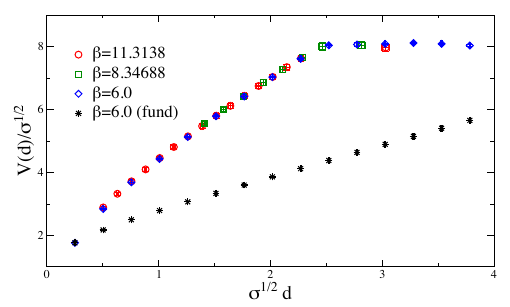}}
\caption{Continuum scaling of the adjoint static potential: data have been obtained on lattices with $L = 32 (\beta = 6.0)$, $L=48(\beta=8.34688)$ and $L=64(\beta=11.3138)$. The potential between two fundamental charges for 
$\beta = 6.0$ is also reported for comparison.  This is graph is from
\cite{bonati}.}
\label{breaking2}
\end{center}
\end{minipage}
\end{figure}
However, the form of the potential which emerges from the simulations 
in \cite{krat} do display the Coulomb term and the L\"uscher term in addition to the linear
potential, so a direct comparison with (\ref{s-break12}) is not
really appropriate. A more meaningful quantity
might be the distance at which breaking occurs, signaled by the
flattening of the potential. This happens at about $10 a$ 
for $\beta = 6$, where
$\beta$ is the parameter which appears in the lattice action (Wilson action)
and $a$ is the lattice spacing. These are related to
the coupling $e^2$ as $a = (2 N /\beta e^2)$.
We then find $\sigma_A \times (10 a)  = 6.67 m$. 
This is about $19\%$ below our estimate.
(The flattening of the potential is very clearly seen in many lattice simulations. We display two examples, just to illustrate this point,
in Fig.\,\ref{breaking1} and Fig.\,\ref{breaking2}.)

Another recent estimate is the one reported in \cite{pepe-W}.
These authors show that the breaking occurs at a value of
$1.3$ (which is twice the mass of a ``constituent gluon" in their terminology).
The units used are such that the $0^{++}$ glueball has a mass of
$1.198$. Taking $M_{0^{++}} =  5.17 m$ as in \cite{teper},
we find that $V_{*{\rm lat}}= (1.3/1.198)\times 5.17 m=
5.61 m$.

We also mention \cite{trottier}, where the authors again demonstrate 
string-breaking for a bound state of a quark and an antiquark.
As in the case of \cite{kne-som}, the quarks being in the fundamental
representation, we are not able to obtain a comparison with the
value calculated here.

An interesting recent simulation for $SU(2)$ lattice gauge theory
considered the longitudinal and transverse
chromoelectric fields as the distance between the particle and antiparticle static sources is increased \cite{bonati}.\footnote{I thank Claudio Bonati for a useful comment and for sharing some of their data.}
 The longitudinal fields suddenly drop to almost zero
as the separation approaches the string-breaking value.
This happens at about $10 a$ for $\beta = 6$, 
in agreement with \cite{krat}. So, as in that case,
$\sigma_A \times (10 a)  = 6.67 m$. 
It is however worth mentioning that the
focus of \cite{bonati} was not so much on the value of the energy.
The key result is about how the breaking occurs, signaled by
the rapid decay of
of the longitudinal contribution to the energy (for which they 
find clear evidence).
\subsection{A comment on $\la W_R(C)\ra$ and string breaking}
We will close this section with a comment on string-breaking
as it should appear in the calculation of the expectation value of the Wilson loop operator.
\begin{figure}[!b]
\begin{center}
\scalebox{1}{\includegraphics{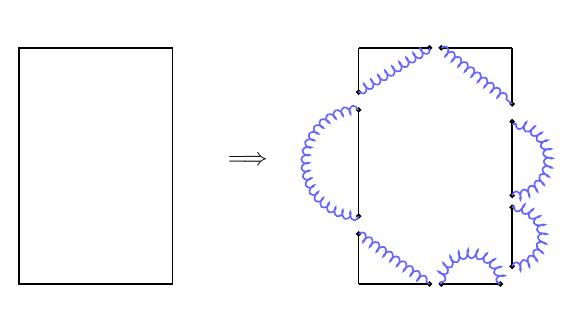}}
\caption{Showing an example of how $\la W_R (C)\ra$ can break up
into the disconnected product of expectation values of separate segments.
The helical lines represent possible contractions of the currents $J$ from integration over $\Psi_0^* \Psi_0$.}
\label{W-break}
\end{center}
\end{figure}
The formation of the glue lump shows how a particle-antiparticle pair
can break up as it propagates forward in time. This will appear naturally
in the computation of
$\la W_R(C)\ra$ where $C$ traces out a spacetime area.
The question that arises naturally is how a purely spatial Wilson loop will show the effect of string-breaking.
The glue lump tells us that, for screenable representations,
 a segment of a Wilson loop can form a 
state of zero charge. Therefore we should expect that in calculating the expectation value of a purely spatial Wilson loop, we will find disconnected terms where segments of the loop combine with the currents $J$ to
give a nonzero expectation value.
In other words, as shown in Fig.\,\ref{W-break}, the expectation value
of $W_R(C)$ breaks up into a number of disconnected diagrams.

This argument also shows why we did not see string-breaking when we calculated the string tension in section \ref{Res}.
We did not consider possible disconnected expectation values.
And that is the correct procedure for representations which are not screenable,
for which we cannot have disconnected expectation values.
However, for screenable representations, we must allow for 
such a possibility. But since there is a minimum energy needed for a glue lump, we expect a similar lower cutoff on the size of the individual segments which can have a nonzero expectation value. We do thus obtain a string tension even for screenable representations which should be valid
for separations smaller than the string-breaking distance.
\section{Alternate candidates for the vacuum wave function}\label{Alter}
 \setcounter{equation}{0}
\begin{quotation}
\fontfamily{bch}\fontsize{10pt}{16pt}\selectfont
\noindent 
There have been some other proposals for
candidate vacuum wave functions, partially motivated by our solution.
 We review two of them here, commenting on similarities and differences
with our solution (\ref{SchE9}). A cautionary comment regarding the use of variational calculations in a field theory is also given.
\end{quotation}
\fontfamily{bch}\fontsize{12pt}{16pt}\selectfont
The procedure we have described in section \ref{SchE} gives a systematic expansion scheme for solving the Schr\"odinger equation.
The solution was then used in arriving at the formula for the string tension.
But there have been a few other suggestions regarding the vacuum wave function for the 3d Yang-Mills theory. We will briefly review some of them here.

In a couple of very interesting papers, Leigh, Minic and Yelnikov (LMY) considered an alternate method of solving the Schr\"odinger equation \cite{Leigh-MY}.
The starting observation was to note that the kernel
given in the solution (\ref{SchE9}) can be expanded in powers of
$-\nabla^2$ as
\beq
\left( {1\over m + \sqrt{m^2 - \nabla^2}}\right)
= {1\over 2m } \left[ 1- {1\over 4} \left({-\nabla^2 \over m^2}\right)
+ {1\over 8}  \left({-\nabla^2 \over m^2}\right)^2 + \cdots \right]
\label{Alter1}
\eeq
As a result, the exponent of the wave function (\ref{SchE9}) can be viewed as a sum of terms involving monomials of the form
\beq
{\cal O}'_n = \int \bdel J^a (\del \bdel )^n \bdel J^a
\label{Alter2}
\eeq
Based on this observation, LMY introduced a set of operators \cite{Leigh-MY}
\beq
{\cal O}_n = \int \bdel J^a \bigl[(\D \bdel )^n\bigr]^{ab} \bdel J^b,
\hskip .2in
{\cal{D}}^{ab}  = {c_A\over \pi}\partial\, \delta^{ab} +i f^{abc}J^c
\label{Alter3}
\eeq
(The operator $\D^{ab}$ was previously introduced in
(\ref{s-break5}).) 
The motivation for this has to do with invariance under
holomorphic transformations of $H$, namely,
under $H \rightarrow V(z) H {\bar V} (\bz )$. The operator 
$\D$ is a covariant derivative for this and leads to manifest
invariance for ${\cal O}_n$ under the holomorphic transformations.
The next step in \cite{Leigh-MY} was to
postulate an ansatz for the wave function of the form $\Psi_0 = e^{{\half }\F}$, with
\beq
\F = \sum_n c_n {\cal O}_n
\label{Alter4}
\eeq
To evaluate the action of the Hamiltonian, they assumed that
${\cal O}_n$ are eigenvectors of the kinetic operator,
\beq
T \, {\cal O}_n = (2 +n ) m \, {\cal O}_n
\label{Alter5}
\eeq
This relation can actually be proved for $n = 0, 1$, but there are additional terms in general for higher $n$. Nevertheless, if one neglects any correction to
(\ref{Alter5}), one can solve the Schr\"odinger equation and arrive at
a wave function
\beq
\Psi_0 = \exp \left[ - {1\over 4m} \int \bdel J ~K[L]~ \bdel J \right],
\hskip .3in K[L] = {J_2 (4 \sqrt{L})\over \sqrt{L}~ J_1(4\sqrt{L}) }
\label{Alter6}
\eeq
where $L = -{\cal D}\bdel/m^2$ and $J_1, ~J_2$ are Bessel functions of orders
$1$ and $2$ respectively. Given the possibility of extra terms in
(\ref{Alter5}), we may take this as an approximate solution, maybe even a good approximate solution, of the
Schr\"odinger equation.

\begin{figure}[!t]
\begin{center}

\scalebox{1}{\includegraphics{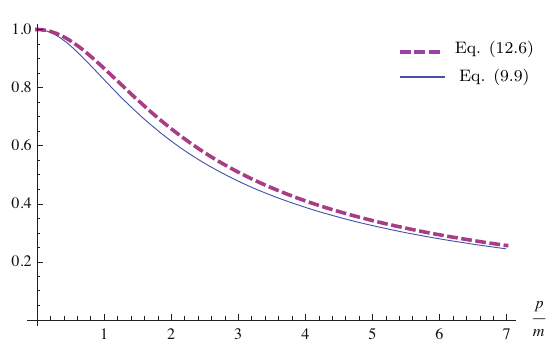}}

\caption{Comparison of the kernel from (\ref{SchE9}) and
(\ref{Alter6}) for the Gaussian term of the vacuum wave function as a function of the momentum $p$}
\label{LMY-graph}
\end{center}
\end{figure}
It turns out that the kernel $K$ which appears in (\ref{Alter6}) is, despite appearances, very close to the kernel in our solution
(\ref{SchE9}). In Fig.\,\ref{LMY-graph}, we show a comparision of the two kernels.\footnote{I thank A. Yelnikov for this comparison graph.}. 

\begin{table}[!b]
\begin{center}
\begin{tabular}{ p{2cm} c p{3.2cm}  }
State&~~~~~LMY Calculation~~~~~~~& Lattice\\
\hlineB{3}
$0^{++}$&$4.098$& $ 4.065\pm 0.055$\\
$0^{++*}$&$5.407$&$ 6.18\pm 0.13$\\
$0^{++**}$&$6.716$&$ 7.99\pm 0.22$\\
$0^{++***}$&$7.994$&$9.44 \pm 0.38$\\
\hline
$0^{--}$&$6.15$&$ 5.91\pm 0.25$\\
$0^{--*}$&$7.46$&$7.63\pm 0.37$\\
$0^{--**}$&$8.77$&$ 8.96\pm 0.65$\\
\hline
$2^{++}$&$6.72$&$ 6.88\pm 0.16$\\
$2^{++*}$&$7.99$&$ 8.62\pm 0.38$\\
$2^{++**}$&$9.26$&$ 9.22\pm 0.32$\\
\hline
$2^{+-}$&$8.76$&$ 8.04\pm 0.50$\\
$2^{--}$&$8.76$&$ 7.89\pm 0.35$\\
$2^{+-*}$&$10.04$&$ 9.97\pm 0.91$\\
$2^{--*}$&$10.04$&$9.46\pm0.46$\\
\hlineB{3}
\end{tabular}
\caption{Comparison of glueball mass estimates from \cite{Leigh-MY} and lattice calculations from \cite{teper, meyer}.}
\label{glueball}
\end{center}
\end{table}
The value of the string tension does not change compared to 
(\ref{SchE9}), since the low momentum limit of the kernel is the same.
However, LMY were able to use this wave function (\ref{Alter6}) to calculate a number of glueball masses as well.
These were obtained from the two-point function for different color-singlet composite operators,
characterized by spin, parity and charge conjugation properties (${\rm J}^{\rm PC}$-notation). The results, in units of $\sqrt{\sigma_F}$, are shown in 
Table\,\ref{glueball}. Again, there is reasonable agreement with the lattice data
of references \cite{teper, meyer}.

Since two-dimensional Yang-Mills theory leads to an area law for the Wilson loop, it has long been suspected that a similar form for the wave function might be applicable in higher dimensions \cite{Green1}. 
The form of the wave function we find, namely 
(\ref{SchE9}), is in accordance with this. We find
$\F \sim F^2$ for modes of the field with low momenta, while 
$\F \sim F(1/\sqrt{-\nabla^2}) F$ at high momenta, in agreement with
the expected perturbative behavior. An interpolation between these two limiting behaviors, different from our result (\ref{SchE9}), was suggested
by Samuel \cite{Samuel} and used to estimate the $0^{++}$ glueball mass.
Essentially the same form (with a small variation) was suggested more recently as a candidate variational ansatz for the wave function \cite{Green-O}.
Specifically it reads $\Psi_0 = e^{{\half}\F}$, with
\beq
\F = - {1\over 2} \int_{x,y} F_{ij}^a(x) 
\left( {1\over \sqrt{-D^2 - \lambda_0 + M^2}}\right)^{ab}_{x,y} F^b_{ij}(y)
\label{Alter7}
\eeq
Here $D^2$ is the square of the covariant derivative in the adjoint representation, $\lambda_0$ is the lowest eigenvalue of
$-D^2$ and $M$ is a parameter with the dimension of mass, treated as
a variational parameter. A number of quantities can be calculated after fixing $M$ by minimizing the ground state energy.
In \cite{Green2}, Monte Carlo simulations of the kernel in a Gaussian ansatz for the wave function were carried out and then compared against
kernels in our wave function (\ref{SchE9}) and the variational ansatz (\ref{Alter7}), see Fig.\,\ref{wave-comp}.
\begin{figure}[!t]
\begin{center}
\scalebox{1.1}{\includegraphics{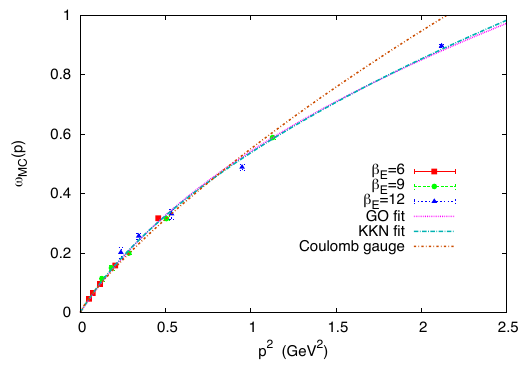}}
\caption{Comparison of cumulative data for $\omega_{MC}$ versus $p^2$
on lattices of extensions $L= 16, 24, 32, 40, 48$ and Euclidean lattice couplings $\beta_E = 6, 9, 12$. The curve labeled
``GO fit" refers to the ansatz in \cite{Green-O} while the curve
labeled ``KKN-fit" refers to our result from (\ref{SchE9}). This graph is from
\cite{Green2}.}
\label{wave-comp}
\end{center}
\end{figure}
For the two cases, the kernel for the quadratic Gaussian part may be written in terms of momentum variable $p$ as
\beqar
\omega_{\rm KKN} &=& {1\over e^2} { p^2 \over \sqrt{p^2 + m^2} + m}, \hskip .2in
(\textrm{from (\ref{SchE9})}\nonumber\\
\omega_{\rm GO} &=& {1\over e^2} {p^2 \over \sqrt{p^2 + M^2}}\hskip .6in
(\textrm{from \cite{Green-O}})
\label{Alter8}
\eeqar
The fit to the Monte Carlo data is designated as $\omega_{\rm MC}$.
From the figure, it seems clear that both agree rather well with the simulations.

The close match between the wave function we calculated, namely
(\ref{SchE9}), and the other candidate functionals is rather nice, but
the variational approach comes with a word of caution.
The exact vacuum wave function has zero energy, for reasons of Lorentz invariance, as already explained in the Introduction.
For the variational approaches, one calculates the expectation value of the energy and minimizes it with respect to the variational parameters.
This is almost always nonzero, unless one is so lucky as to hit on the exact 
vacuum functional as the guess for the variational ansatz.
Recall that the variational estimate is an upper bound on the true ground state energy, so we generally have
$E_{\rm var, min} > 0$.
One could try to subtract this out by using
$\H - E_{\rm var, min}$ as our notion of a ``normally ordered" Hamiltonian. 
While this will give $(\H - E_{\rm var, min}) \Psi_{\rm var} = 0$,
for the true ground state we will get
\beq
(\H - E_{\rm var, min} ) \Psi_0 = - E_{\rm var, min}\, \Psi_0
\label{Alter9}
\eeq
This is problematic regarding Lorentz invariance.
So $\H - E_{\rm var, min}$ cannot be taken to be the true normally-ordered Hamiltonian to be used in the Schr\"odinger equation.
So, basically, this means that we have to live with a nonzero vacuum energy
in the variational approach.
Then the question is: Is this acceptable?
When we set up a relativistic field theory, the aim is to solve for the vacuum state preserving all the isometries of the spacetime on which it is defined, namely, full Poincar\'e invariance in flat space.
So, with variational ans\"atze, there is a contradiction between the premise and the end result.
(One cannot view $E_{\rm var, min}$ as a cosmological constant or anything of that sort; that would require a spacetime with a different group of isometries, such as the de Sitter or anti-de Sitter space.
This would again imply a discord between the starting spacetime and the one
consistent with the final results.)
This is not to say the variational approach is not useful, but it should be used with caution.

\section{A short aside on the Chern-Simons theory}\label{CStheory}
 \setcounter{equation}{0}
\begin{quotation}
\fontfamily{bch}\fontsize{10pt}{16pt}\selectfont
\noindent 
In this section, we consider the pure Chern-Simons theory. The main motivation is to derive the finite renormalization
of the level number $k$ as $k \rightarrow k+ c_A$, entirely within the Hamiltonian analysis. A second useful result is that the
factor $e^{2 c_A S_{\rm wzw}(H)}$ in the measure for the inner product of wave functions is changed to 
$e^{(k + 2 c_A) S_{\rm wzw}(H)}$. Both these results will be useful in considerations related to supersymmetric theories in section \ref{Ext}.
\end{quotation}
\fontfamily{bch}\fontsize{12pt}{16pt}\selectfont
We will now consider some features in the Schr\"odinger 
quantization of the pure Chern-Simons theory. This is 
primarily meant as a prelude to the Yang-Mills-Chern-Simons theory as well as to the discussion of some supersymmetric extensions.

The Chern-Simons (CS) action is given by
\beq
S_{\rm CS} = -{k\over 4\pi} \int d^3x~\Tr \left( A_\mu\partial_\nu A_\alpha 
+{2\over 3} A_\mu A_\nu A_\alpha \right)\e^{\mu\nu\alpha}
\label{CS1}
\eeq
The action $S_{\rm CS}$ is the integral of the Chern-Simons (CS) 3-form.
Although it had been known in the mathematics literature
in the context of secondary characteristic classes, the CS 3-form was initially
 introduced in physics literature as a possible mass term for gauge fields in
 three dimensions
 \cite{DJT,Sch}, see the section on the YMCS theory.
 The parameter $k$ is known as the level number of the Chern-Simons term.

 Under a gauge transformation,
 the CS action changes as
 \beq
 {\rm CS}(A^g ) = {\rm CS} (A) - {k \over 4 \pi} \int d \left[ \Tr (g^{-1} d g\, A )
 \right]
 - {k \over 12\pi} \int \Tr (g^{-1} dg)^3
 \label{CS2}
 \eeq
 For transformations $g$ such that $g \rightarrow 1$ at the spacetime boundary, we see that the total derivative will integrate to zero.
The boundary condition means that $g(x)$ is equivalent to a map
from $S^3 $ to $G$.
Such maps can be homotopically nontrivial since $\Pi_3(G)$
is $\mathbb{Z}$ for a simple Lie
group, see (\ref{gorb3}).
The last term is then $2\pi$ times the winding number of this
map.
For the invariance of the theory under all such gauge transformations, 
we need invariance of $e^{i S}$. This requires that the level
number $k$ should be an integer.
(There are other related ways to understand the quantization of 
the level number, see for example \cite{Asor-M}.
An argument which is entirely in the Hamiltonian framework
is also given below.)

The action ({\ref{CS1}) is linear in the time-derivatives of $A$ because of the
$\e$-tensor, and hence the Hamiltonian defined by the usual Legendre transformation is identically zero.
Another way to see this result is to note that since $S_{\rm CS}$ is
the integral of
a differential form, it is independent of the metric and hence
the energy-momentum tensor defined by the variation of
the action with respect to the metric is zero.
For the quantization of the theory, the key ingredients are thus
the canonical
structure and the constraint of Gauss law.
The term in the action with the time-derivative of the gauge field is
given by
\beqar
S_{\rm CS} &=&   - {k \over 8\pi}\int dt\,d^2x\,  \e^{ij} A^a_i {\dot A}^a_j 
+ \cdots\nonumber\\
&=& -i {k \over 4 \pi} \int dt d^2x\, 
\left( A^a {\dot {\bar A}^a} - {\bar A}^a {\dot A}^a \right) 
\label{CS3}
\eeqar
The simplest way to quantize this theory is to use coherent states
or geometric quantization in the holomorphic polarization.
The first step is thus to read off the canonical or symplectic
structure of the
theory from (\ref{CS3}) as
\beqar
\alpha_{\rm symp} &=& -{k \over 8\pi} \int d^2x\, \e^{ij} A^a_i \delta A^a_j
= -i {k \over 4 \pi} \int d^2x\, 
\left( A^a { {\delta \bA}^a} - {\bar A}^a {\delta A}^a \right) 
\nonumber\\
\omega_{\rm symp}&=&\delta \alpha = {k \over 4\pi} \int \Tr (\delta A_i\,
\delta A_j ) dx^i\wedge dx^j
=   -i {k \over 2 \pi} \int d^2x\,  \delta A^a \, \delta {\bar A}^a
\label{CS4}
\eeqar
Here $\alpha_{\rm symp}$ is the canonical one-form and $\omega_{\rm symp}$
is the canonical two-form. As in Appendix \ref{TopC}, we
use $\delta$ to denote the exterior derivative on the space of fields.
Notice that the expression for the canonical two-form $\omega_{\rm symp}$
coincides
with $ k$ times the two-form $ \Omega$ defined in Appendix \ref{TopC}, equation
(\ref{gorb6}).
One of the requirements of consistent quantization
in the Hamiltonian framework is that
the integral of $\omega_{\rm symp}$
over closed two-surfaces in the phase space should be $2\pi$ times
an integer.
(This is the generalization of the usual Dirac quantization condition
for magnetic monopoles applied in the context of geometric quantization.)
 We have already seen in
Appendix \ref{TopC} that the integral of $\Omega$ is $2\pi$ times
 the integer $\nu$. Thus, in the present case, we see that consistent
 quantization requires that the level number $k$ should be
 an integer.
A set of configurations, which form a somewhat simpler case than the ones
discussed in Appendix \ref{TopC}, is the following.
We write
\beq
A = \tau \, g^{-1} d g = \tau \, g^{-1}\del_i g dx^i, \hskip .2in g = g(x^1, x^2, \lambda )
\label{CS4a}
\eeq
The parameters $\tau$ and $\lambda$ define a two-surface
in the space of potentials. We take
$0\leq \tau, \lambda \leq 1$. Further we take
$g(x^1, x^2, \lambda )$ to obey the boundary conditions
\beq
g(x^1, x^2, \lambda ) = 1 \hskip .2in {\rm at~} \lambda = 0, 1
\label{CS4b}
\eeq
We then see that $A = 0$ on the three sides of the square
at $\tau = 0$, $\lambda = 0, 1$. At
$\tau = 1$, we have a pure gauge $A= g^{-1} d g$. Since
$A$ is gauge-equivalent to the vacuum on the boundary of the
square, topologically, we can view 
(\ref{CS4a}) as defining a closed two-surface in $\A / \G_*$.
From (\ref{CS4a}),
\beq
\delta A = d\tau \, dx^i g^{-1} \del_i g  + \tau \, d\lambda dx^i
\del_\lambda (g^{-1} \del_i g) 
\label{CS4c}
\eeq
Using this\footnote{Wedge products are left implicit in this equation to avoid clutter of notation.}
\beqar
\int \omega_{\rm symp}&=& {k \over 4\pi} \int \Tr (\delta A \wedge
\delta A )\nonumber\\
&=&{k \over 4\pi} \int d\tau \, \tau\,
dx^i d\lambda dx^j \, \Tr  \left[(g^{-1} \del_i g) 
\del_\lambda (g^{-1} \del_j g) 
+ \del_\lambda (g^{-1} \del_i g) (g^{-1} \del_j g) \right]\nonumber\\
&=& {k \over 4\pi} (- 2) \int d\tau \, \tau\,
d\lambda  dx^i dx^j \, \Tr \left[ \del_\lambda (g^{-1} \del_i g) (g^{-1} \del_j g) 
\right]\nonumber\\
&=&- {k \over 4\pi} {2\over 3} 
\int d\tau \, \tau\,
dx^\mu dx^\nu dx^\a \, \Tr \left[ \del_\mu  (g^{-1} \del_\nu g) (g^{-1} \del_\a g) 
\right]\nonumber\\
&=& {k \over 12 \pi} \int \Tr \left[ (g^{-1} dg )^3\right]
= 2\pi k\, Q[g]
\label{CS4d}\\
Q[g] &=&{1 \over 24 \pi^2} \int \Tr \left[ (g^{-1} dg )^3\right]
\nonumber
\eeqar
In going from the third line of this equation to the fourth line, we have treated
$\lambda$ on an equal footing with the coordinates, so we can
write it as a differential form.
Also $Q[g]$ in this equation is the winding number of the map
$S^3 \rightarrow G$, and is an integer. 
We thus see that $k$ must be an integer to satisfy the condition of the
integral of $\omega_{\rm symp}$ being $2\pi$ times an integer.
From these arguments, we see how the quantization of the
 level number emerges in a purely Hamiltonian framework.

We now turn to setting up the quantization of the theory.
There is a large body of literature on this topic;
some of the early papers are given in \cite{{witten-jones},{CS-early}}.
Geometric quantization
in general is discussed in many papers and books
\cite{geom-quant}.
Specifically for the Chern-Simons theory,
see \cite{axelrod-PW}. Our discussion follows closely \cite{Bos-N},
see also \cite{{Jackiw},{DuJT}} for the use of the Schr\"odinger representation.

We start by noting that
by inversion of $\omega_{\rm symp}$ or by constructing 
Hamiltonian vector fields
corresponding to $A$ and $\bA$, the commutation rules can be identified as
\beq
[A^a (x), A^b (y) ] = [\bA^a (x), \bA^b (y) ] = 0,
\hskip .2in
[A^a (x) , \bA^b (y) ] = {2 \pi \over k} \delta^{ab} \delta^{(2)}(x-y)
\label{CS5}
\eeq

Under a canonical transformation, $\alpha_{\rm symp}$ changes as
$\alpha_{\rm symp} \rightarrow \alpha_{\rm symp} +  \delta f$, for some function $f$
on the phase space. Equivalently, the action changes
as $S \rightarrow S + \int {\dot f}$.
The pre-quantum wave functions must correspondingly transform as
\beq
\Psi \rightarrow e^{i f } \, \Psi
\label{CS6}
\eeq
We can define covariant derivatives of $\Psi$ on the phase space
as
\beqar
\left[ {\delta \over \delta A^a} -i \alpha_A \right] \Psi
&=& \left[ {\delta \over \delta A^a} + {k \over 4\pi} \bA^a \right] \Psi
\nonumber\\
\left[ {\delta \over \delta \bA^a} -i \alpha_{\bA} \right] \Psi
&=& \left[ {\delta \over \delta \bA^a} - {k \over 4\pi} A^a \right] \Psi
\label{CS7}
\eeqar
The wave functions should only depend on half of the phase space degrees of freedom. They are obtained by restricting the 
pre-quantum wave functions  to those satisfying 
a polarization condition. The holomorphic polarization is defined
by the condition
\beq
\left[ {\delta \over \delta A^a} + {k \over 4\pi} \bA^a \right] \Psi
= 0
\label{CS8}
\eeq
The solution to this equation is obviously of the form
\beq
\Psi = {\cal C} \, e^{- {k\over 4 \pi}  \int A^a \bA^a }\, \Phi (\bA)
= {\cal C} \, e^{ {k\over 2 \pi} \int  \Tr (A \bA ) }\, \Phi (\bA)
\label{CS9}
\eeq
where $\C$ is a normalization constant.

The commutation rules (\ref{CS5}) can be implemented on 
$\Phi (\bA )$ as
\beq
A^a (x)  \, \Phi (\bA ) = {2\pi \over k} {\delta \Phi \over \delta \bA^a(x)}
\label{CS10}
\eeq
The operator corresponding to $\bA^a$ is diagonal in this basis, with
eigenvalue $\bA^a (x)$.\footnote{These results can be obtained 
in terms of prequantum operators acting on wave functions of the form
(\ref{CS9}), and reducing to action on $\Phi (\bA )$;
see \cite{geom-quant}. Alternatively, one can check that
(\ref{CS10}) gives the correct commutation rule and is the adjoint
of $\bA^a$ with the given inner product for $\Phi(\bA)$.}

The remaining step is to impose the Gauss law on the wave functions
to select out the physical states.
In the absence of any charges, this will eliminate all wave functions, except for the one for the vacuum state. The theory is thus trivial
with a one-dimensional Hilbert space corresponding to the vacuum.
We will therefore consider the theory with a number of point 
charges added.
This situation corresponds to the action
\beq
S = S_{\rm CS} + \sum_r (T^a)_r \left[ A_0^a (x_r) 
+ A^a(x_r) {\dot z}_r + \bA^a(x_r)  {\dot {\bz}}_r \right]
\label{CS11}
\eeq
Here $x_r = (z_r, \bz_r)$ correspond to the positions of the
particles, $T^a$ are matrices for the basis of the Lie algebra
in the representations corresponding to the charged particles.
The equation of motion for $A_0^a$ is then
\beq
{k \over 8 \pi} F^a_{ij} \e^{ij} + \sum_r (T^a)_r
\delta^{(2)}(x- x_r) = i {k \over 2 \pi} 
\left[ - \bD A + \del \bA \right]^a
+ \sum_r (T^a)_r
\delta^{(2)}(x- x_r) = 0
\label{CS12}
\eeq
The Gauss law condition on the wave functions become
\beq
\left[ \left( \bD {\delta \over \delta \bA}\right)^a  - {k \over2 \pi}
\del \bA^a + \sum_r (i T^a)_r \delta^{(2)}(x- x_r) \right] \Phi = 0
\label{CS13}
\eeq
The wave function $\Phi$, being a function of $\bA$, can be equivalently
taken to be a function of $M^\dagger$. 
Notice that under $M^\dagger \rightarrow M^\dagger (1+ \theta)$,
$\bA \rightarrow \bA + \bD \theta$, so that
\beq
{\delta \Phi \over \delta \theta^a}
= - \bD \left( {\delta \over \delta \bA}\right) \Phi
\label{CS14}
\eeq
The Gauss law condition is thus
\beq
{\delta \Phi \over \delta \theta^a} =
\left[- {k \over 2 \pi}\del \bA^a  + \sum_r (i T^a)_r \delta^{(2)}(x- x_r) \right] \Phi
\label{CS15}
\eeq
To solve this equation, we note that we have already seen in section \ref{Vol}
that the WZW action defined in (\ref{vol21})
gives
\beqar
 S_{\rm wzw}(M^\dagger (1 + \theta )) - S_{\rm wzw}(M^\dagger)
&=& -{1\over \pi} \int \Tr \left( M^{\dagger -1} \bdel M^\dagger\, \del \theta\right)\nonumber\\
&=&{1\over \pi} \int \Tr \left( \del \bA \, \theta \right)
= - {1\over 2 \pi} \int \theta^a \,\del \bA^a 
\label{CS16}
 \eeqar
This is basically
(\ref{vol23}) and follows from the Polyakov-Wiegmann identity given in (\ref{vol22}). Further notice that
\beq
{\delta M^{\dagger -1}\over \delta \theta^a } =
(i T^a) M^{\dagger -1},
\hskip .2in
{\delta M^{\dagger}\over \delta \theta^a } =
M^{\dagger} (- i T^a) 
\label{CS17}
\eeq
Using these results, the solution to the Gauss law (\ref{CS15}) can be worked out as
\beq
\Phi = N (\{ x_r \}) \,
e^{k S_{\rm wzw} (M^\dagger )} \, \prod_{\otimes r} 
\bigl( M^{\dagger -1} (x_r) \bigr)
\label{CS18}
\eeq
The prefactor $N$ can depend on the coordinates of the charged particles but it is independent of $\bA$.

Since we used the holomorphic polarization, the normalization should be done by integrating $\Psi^* \Psi$ with the volume for the phase space.
For the wave functions obeying the Gauss law, the volume 
has to be reduced to the gauge-invariant subspace.
Notice that for this theory the phase space is the space of the gauge potentials, the space $\A$ mentioned in Appendix \ref{TopC};
this is evident from the canonical two-form
in (\ref{CS4}).
Therefore the phase volume modulo gauge transformations is just
the volume element for $\A / \G_*$ which was obtained in 
section \ref{Vol} as
\beq
d\mu (\C) = {\cal N} \, d\mu (H)\, e^{2\, c_A S_{\rm wzw} (H) },
\hskip .3in 
{\cal N} = \left[ {\det' (-\del \bdel) \over \int d^2x}\right]^{{\rm dim}\,G}
\label{CS19}
\eeq
The normalization condition for the wave function $\Phi$
using (\ref{CS9}) and this volume element is
\beq
\vert N (\{x_r\})\vert^2 \,{\cal N} \int d\mu (H)\, e^{(k + 2\, c_A) S_{\rm wzw} (H) }
\big\vert \prod_{\otimes r}\bigl( M^{\dagger -1} (x_r) \bigr)\big\vert^2
= 1
\label{CS20}
\eeq
where we combined the terms involving the level number $k$
as
\beq
k S_{\rm wzw}(M) + k S_{\rm wzw}(M^\dagger ) + 
{k \over \pi} \int \Tr (\bA A ) 
= S_{\rm wzw}(H)
\label{CS21}
\eeq
This is again done by use of the PW identity.
The relevant integration measure now involves
$e^{(k + 2\, c_A) S_{\rm wzw} (H) }$.

Once we introduce charges, there is one more ingredient that comes into play, namely, the Schr\"odinger equation, because the Hamiltonian is no longer zero. From the action (\ref{CS11}), the Hamiltonian, 
as an opertaor on $\Phi$, is seen to be
\beq
\H = - \sum_r (T^a)_r \left[ {\dot z}_r {2\pi \over k} {\delta \over \delta \bA^a}
+ {\dot\bz}_r \bA^a \right]
\label{CS22}
\eeq
The wave function $\Phi$ must obey the Schr\"odinger equation
with this Hamiltonian. This will determine the $x_r$-dependence of
the coefficient $N (\{ x_r \} )$ in (\ref{CS18}).
However, there is a subtlety that arises in solving the Schr\"odinger equation
which leads to a finite renormalization of the level number $k$.
(This finite renormalization of $k$ has been obtained in terms of perturbation theory for the CS theory
using Feynman diagrams \cite{Pis-R}, it also appears in the context of relating
Chern-Simons theory and conformal field theory \cite{witten-jones}.)
It is interesting to see how this finite renormalization arises in
the Schr\"odinger representation, so
we will go over this in some detail.
For this it is sufficient to consider two charged particles
with opposite charges, i.e.,
in conjugate representations.
Since $(-T^a)^T$ are the matrices corresponding to
$T^a$ in the conjugate representation,
the solution for $\Phi$ takes the form
\beq
\Phi_{ij} = N (x_1, x_2) \, e^{k S_{\rm wzw}(M^\dagger )} 
\bigl( M^{\dagger -1} (x_1) M^\dagger (x_2)\bigr)_{ij}
\label{CS23}
\eeq
The relations (\ref{CS17}) can be used to verify that this is indeed a
solution
to the Gauss law (\ref{CS15}).
The Schr\"odinger equation then takes the form
\beqar
i {\del \Phi \over \del t} &=& \H \,\Phi\nonumber\\
&=& (-T^a) \left[ {\dot z}_1 {2\pi\over k} {\delta \over \delta \bA^a(x_1)}
+ {\dot\bz}_1 \bA^a (x_1) \right] \Phi
+ \left[ {\dot z}_2 {2\pi\over k} {\delta \over \delta \bA^a(x_2)}
+ {\dot\bz}_2 \bA^a (x_2) \right] \Phi \, T^a\nonumber\\
\label{CS24}
\eeqar
Using $\Phi$ from (\ref{CS23}) and equating terms
proportional to
${\dot z}_1$, we get
\begin{align}
{\del \log N \over \del z_1} &
M^{\dagger -1} (1) M^\dagger (2)
+ {\del M^{\dagger -1} (1) \over \del z_1} M^\dagger (2)\nonumber\\
&= {2\pi \over k} (i T^a) e^{-k S_{\rm wzw}(M^\dagger)} 
{\delta \over \delta \bA^a(1)} \left( M^{\dagger -1} (1) M^\dagger (2)
e^{k S_{\rm wzw}(M^\dagger)}\right)
\label{CS25}
\end{align}
Towards further simplification, notice that since $\bA = M^{\dagger -1} \bdel M^\dagger$,
we have 
\beqar
\delta \bA &=&
\bdel (M^{\dagger -1} \delta M^\dagger ) + \bA \, (M^{\dagger -1} \delta M^\dagger )  - (M^{\dagger -1} \delta M^\dagger ) \,\bA\nonumber\\
- M^\dagger \delta \bA M^{\dagger -1}
&=& \bdel ( M^\dagger \delta M^{\dagger -1} )
\label{CS26}
\eeqar
From this we can identify
\beqar
{\delta M^{\dagger -1} (x) \over \delta \bA^a (y)}
&=& - M^{\dagger -1} (x) {1\over \pi (x-y)}  M^\dagger (y)
 (-iT^a) M^{\dagger -1} (y)
\nonumber\\
{\delta M^{\dagger} (x) \over \delta \bA^a (y)}
&=&{1\over \pi (x-y)}  M^\dagger (y)
 (-iT^a) M^{\dagger -1} (y) \, M^\dagger (x)
 \label{CS27}
\eeqar
We see from the first of these equations that when we
take the functional derivative of $M^{\dagger -1} (1)$  with respect to
$\bA^a (x_1)$, we will get a singular term which is proportional to
$c_R$, the quadratic Casimir value for the representation
$R$ corresponding to the particles. Therefore we need a regularization
for the right hand side of (\ref{CS25}).
We will use a point-splitting regularization defined
by the replacement \cite{Bos-N}
\beq
{1\over k} {\delta \over \delta \bA^a(x_1)}
(i T^a \Phi) \rightarrow
{1\over \kappa} \left[ {\delta \over \delta \bA^a(x_1)}
(i T^a \Phi) + {c_R\, \Phi\over \pi (x_1 - y) }\right]_{y \rightarrow x_1}
\label{CS28}
\eeq
where $\kappa$ is a parameter to be determined.
The regularized version of (\ref{CS25}) is thus
\begin{align}
{\del \log N \over \del z_1} &
M^{\dagger -1} (1) M^\dagger (2)
+ {\del M^{\dagger -1} (1) \over \del z_1} M^\dagger (2)\nonumber\\
&= {2\pi \over \kappa} \left[(i T^a) e^{-k S_{\rm wzw}(M^\dagger)} 
{\delta \over \delta \bA^a(y)} \left( M^{\dagger -1} (1) M^\dagger (2)
e^{k S_{\rm wzw}(M^\dagger)}\right)\right.\nonumber\\
&\hskip .5in \left.+ {c_R\over \pi (x_1 - y)} M^{\dagger -1} (1) M^\dagger (2) \right]_{y\rightarrow x_1}
\label{CS29}
\end{align}
For the functional derivative of the WZW action we have the result
\beq
{\delta S_{\rm wzw} (M^\dagger) \over \delta \bA^a}
= {1\over 2\pi} \, (M^{\dagger -1} \del M^\dagger )^a ,
\label{CS30}
\eeq
Thus, using (\ref{CS27}) for the functional derivatives of $M^\dagger$
 and $M^{\dagger -1}$ and the result (\ref{CS30}),
we can reduce (\ref{CS29}) to the form
\begin{align}
{\kappa \over 2\pi} (\del_1 \log N)&
M^{\dagger -1} (1) M^\dagger (2) + {\kappa - k\over 2 \pi}
\del M^{\dagger -1} (1) M^\dagger (2) \nonumber\\
&= {c_A \over 2 \pi} \del M^{\dagger -1} (1) M^\dagger (2) 
- {c_R \over \pi (x_1 - x_2) } M^{\dagger -1} (1) M^\dagger (2) 
\label{CS31}
\end{align}
This equation is satisfied if
\beq
\boxed{
\kappa = k + c_A, \hskip .3in
N = C \, {1\over (x_1 - x_2)^{2 c_R /(k+ c_A)}} }
\label{CS32}
\eeq
where $C$ is a constant.
This is the solution of the Schr\"odinger equation for the two-particle case.
We see that the regularization of the Schr\"odinger
operator requires that $k$ should be renormalized to $\kappa = k+ c_A$.
The solution also identifies the
$x$-dependence of $N$.
We emphasize that $x_1$ and $x_2$ in the expression for
$N$ are the holomorphic coordinates. The solution $N$ is the chiral block for
two-point function of the WZW theory with level number $k$.

While the CS theory by itself
is not germane to the discussion of the YMCS theory,
the renormalization of $k$ will
be important for the supersymmetric cases.
Our main purpose here was to show how this happens
in the Hamiltonian framework, in
the spirit of staying within the Schr\"odinger representation.

We will close this section with a couple of useful remarks. 
First of all, we note that the symplectic form
$\omega_{\rm symp}$ is of the K\"ahler type, with a
K\"ahler potential
$K = {k \over 2\pi}A^a \bA^a$.
Thus the space of gauge potentials admits a K\"ahler structure;
effectively the complex structure of using complex coordinates
in the two-dimensional space $\mathbb{R}^2$ is being lifted to
the space of potentials.
The K\"ahler potential also enters in the wave function
(\ref{CS9}) as expected for quantization in the holomorphic polarization.

We have considered the Schr\"odinger equation for the case of two
charged particles. Along similar lines, one can consider
 the Schr\"odinger equation for multiple charges. This will again reduce to
 an equation for the prefactor $N(\{ x_r\})$ in 
( \ref{CS18}). This is the Knizhnik-Zamolodchikov (KZ) equation familiar
from conformal field theory.
The solutions then show that $N$ is the chiral block of the appropriate correlator of the corresponding
conformal field theory, which is the WZW theory of level number $k$.

\section{Extensions of the Yang-Mills theory and some comments}
\label{Ext}
 \setcounter{equation}{0}
\begin{quotation}
\fontfamily{bch}\fontsize{10pt}{16pt}\selectfont
\noindent 
We first stress the role of the integration measure in the inner product 
in generating a mass gap by giving a general intuitive argument
for it. We then consider the quantization of
the Yang-Mills-Chern-Simons theory obtaining the leading term of the vacuum wave function. The concordance between the integration measure and the mass gap is again verified explicitly.
An argument for an indirect calculation of the integration measure is
given. It is used to get some features of supersymmetric gauge
theories. Explicit quantization of the supersymmetric theories to verify these expectations is also carried out.
\end{quotation}
\fontfamily{bch}\fontsize{12pt}{16pt}\selectfont
\subsection{An intuitive argument for the mass gap}\label{Int-mass}
The emergence of the mass gap $m = (e^2 c_A/2\pi)$
is an important feature which allowed for the
expansion scheme for solving the Schr\"odinger equation.
It is therefore useful to understand this, not just in technical
terms, but in some intuitive way.
The key ingredient for this is the measure of integration
(\ref{vol31}).
In some ways, its role is already evident in how the propagator mass
emerged in section \ref{promass}. To bring out the connection of the
volume element (\ref{vol31}) and the mass gap, consider the
Hamiltonian of the theory written simply as
\beq
\H = {1\over 2 e^2} \int \left[ (E^a)^2 + (B^a)^2 \right]
\label{Comm1}
\eeq
There is a simple argument based on the uncertainty principle which
helps us to get a sense of the low lying excitations of this Hamiltonian.
The basic commutation rule we need is
$[E_i^a, A_j^b]= -i\delta_{ij}\delta^{ab}$. 
Let us first consider the Abelian case of electrodynamics,
with $B = \nabla \times A$. In terms of
components $E_k$ and $B_k$ of wave vector (or momentum) $k$,
the basic commutation rule becomes $[E_i, B ] = - e^2 \e_{ij} k_j$, so that the uncertainty principle reads
$\Delta E_k~\Delta B_k\sim  e^2 k$, where
$\Delta E_k, ~\Delta B_k$ stand for the root mean square of the fluctuations of the  
electric field $E_k$ 
and the magnetic field $B_k$.
The expectation value of the Hamiltonian for a state with wave function
$\Psi$ with momentum $k$ is then
\beqar
{\E}&=& {1\over 2 e^2}  \bra{\Psi_k} \left[ E^2 + B^2\right] \ket{\Psi_k}
= {1\over 2 e^2} \left[ (\Delta E_k^2) + (\Delta B_k^2) \right]
\nonumber\\
&\sim&{1\over 2} \left( {e^2 k^2\over\Delta B_k^2 } +{\Delta B_k^2 \over e^2} 
 \right) 
 \label{Comm2}
\eeqar
For low lying states, we must minimize this
${\E}$ with respect to $\Delta B^2$,  
which gives 
$\Delta B^2_{k, {\rm min}}$
$ \sim 
e^2 k$, giving ${\E} \sim k$. This corresponds to the familiar photon
of the Abelian theory.

For the nonabelian theory, this is inadequate since the expectation value
$\la \H \ra = \int \Psi^* \, \H \, \Psi$ 
involves 
the factor $e^{2\, c_A S_{\rm wzw} (H) }$. In fact, 
\beq
\la \H \ra ~\sim {1\over 2 e^2} \int d\mu (H) e^{2\, c_A\,S_{\rm wzw} (H) }~ (E_k^2 +B_k^2 ) 
\label{Comm3}
\eeq 
In terms of $B$, the behavior of the WZW action is
\beq
2\, c_A\,S_{\rm wzw} (H) \approx  -{c_A \over 2\, \pi } \int_k B_{-k}  \left({1\over k^2}\right) B_k + \cdots
\label{Comm4}
\eeq
We  see that, in the integration measure in
(\ref{Comm3}), we have a Gaussian
distribution for $B$ with a width of $\Delta B_k^2 \approx \pi k^2 /c_A$, for  
small values of $k$. Evidently, this Gaussian dominates near small $k$,
since it becomes narrower and narrower as $k \rightarrow 0$,
giving $\Delta B_k^2 \sim k^2 (\pi /c_A )$.  
Another way to see this is to notice that
$B^2 \sim \bdel J \, \bdel J$ and the currents in the WZW theory
obey
\beq
\la J^a (x) J^b (y) \ra \sim \delta^{ab} \del {1\over (x-y)} 
\label{Comm5}
\eeq
This translates to $\la J^a (k) J^b (-k) \ra \sim (k /{\bar k})$, leading to
$\Delta B_k^2 \sim k{\bar k}  (\pi /c_A ) \sim k^2 (\pi/c_A)$ again.
What this means is that, even though
${\E}$ in (\ref{Comm2}) is minimized around $\Delta B_k^2 \sim k$, 
probability is  
concentrated around $\Delta B_k^2 \sim k^2 (\pi /c_A)$. For the expectation 
value of the energy, 
we then find 
$\la {\H}\ra \sim e^2c_A/2\pi  +{\O}(k^2)$. Thus the kinetic term, in  
combination with  
the measure factor $e^{2c_A S_{\rm wzw} (H)}$, leads to a mass gap of order 
$e^2 c_A/2 \pi$.  
The argument is admittedly not rigorous, but does capture the essential physics.
The key point is that the volume element
 (\ref{vol31}) cuts off the low momentum modes. This suggests
 that the calculation of the volume element in extensions of the theory
 with matter content can, by itself, shed light on the issue of  the mass 
 gap. 
 In fact, we shall briefly analyze some extensions of the Yang-Mills theory
 from this point of view. In the case of the
 Yang-Mills theory modified by the addition of a Chern-Simons term,
 which will be considered next, in subsection \ref{YMCS}, we can carry out the simplification of the Hamiltonian and see how this is indeed realized.
We will also consider some supersymmetric extensions
 in subsection \ref{SUSY}.

\subsection{Yang-Mills-Chern-Simons theory}\label{YMCS}
We consider the Yang-Mills theory modified by the addition of a Chern-Simons (CS) term, so that the action we start with is
\beqar
S = - {1\over 4e^2}\int d^3x~ F_{\mu\nu}^a F^{a\mu\nu}
-{k\over 4\pi} \int d^3x~\Tr \left( A_\mu\partial_\nu A_\alpha 
+{2\over 3} A_\mu A_\nu A_\alpha \right)\e^{\mu\nu\alpha}
\label{Ext1}
\eeqar
The second term in $S$ is the integral of the Chern-Simons (CS) 3-form
which we have discussed in section \ref{CStheory}.
As in that case, the level number $k$ should be quantized for a consistent
quantum theory.

The CS term is odd under parity and time-reversal. We have already mentioned its role as a mass term.
This is
made clear by considering the propagator for the theory. In a gauge where
$\del_\mu A_\mu = 0$, we find
\beq
\bra{0} \T A_i^a(x) A_j^b (y) \ket{0}
= \delta^{ab} \int {d^3p \over (2\pi)^3} e^{- i p(x-y)} {i \over p^2 - \mu^2 +i \e} \left( \delta_{ij}
+ i \mu { \e_{ijk} p^k\over p^2} \right)
\label{Ext1b}
\eeq
where $\mu = (e^2 k /4\pi)$.
This shows that perturbatively $\mu$ is the mass of the gauge particle.

As in the case of the pure Yang-Mills theory, we shall now use
the $A_0 =0$ gauge to set up the Hamiltonian formalism
\cite{KKN-CS}.
The canonical momenta can be easily identified from the action
(\ref{Ext1})  and are related to the electric fields
${\dot A}$ as
\beqar
E^a&=& {e^2\over 2} \Pi^a +{i e^2 k\over 8\pi} A^a= -{i e^2\over 2}{\delta \over \delta \bA^a}
+{i e^2 k\over 8\pi}A^a\nonumber\\
\bE^a&=& {e^2\over 2}{\bar \Pi}^a -{i e^2 k\over 8\pi} \bA^a= -{i e^2\over 2}{\delta \over
\delta A^a}
-{i e^2k\over 8\pi}\bA^a
\label{Ext2}
\eeqar
The commutation rule for $E^a,\bE^a$ is given by
\beq
[\bE^a(\vx) , E^b(\vy)]= {e^4 k\over 8\pi} \delta^{ab}\delta (\vx-\vy)
\label{Ext3}
\eeq 
The Gauss law operator $G_0(\theta)$ is given by
\beq
G_0(\theta) = \int \theta^a \left[ (D{\bar \Pi}+\bD \Pi )^a +{ik\over 4\pi} (\partial \bA^a -\bdel A^a)\right]
\label{Ext4}
\eeq

As before, we can parametrize the fields $A$, $\bA$ in terms of $M$ and $M^\dagger$, with wave functions taken as functionals
of $M$ and $M^\dagger$. 
The Gauss law operator generates gauge
transformations on the argument $(M, M^\dagger )$ of the
wave functions. The 
Gauss law condition (\ref{G23a}) for physical states is then
equivalent to
\beq
\Psi (hM,M^\dagger h^{-1})= \left[ 1+ {k\over 2\pi} \int \Tr \left( M^{\dagger
-1}\bdel
M^\dagger\, \partial \theta +\bdel \theta\, \partial M M^{-1}\right)\right]~\Psi
(M,M^\dagger)
\label{Ext5}
\eeq
where $h(x)\approx 1+\theta (x)$, $\theta =-it^a\theta^a$.
The general form of the wave function obeying this condition
can be written as
\beqar
\Psi (M,M^\dagger ) & =& \exp\left[ {k\over 2}\left( S_{\rm wzw} (M^\dagger )- S_{\rm wzw} (M)\right)\right]~
\chi (H) \nonumber\\
& \equiv& e^ {i \omega (M, M^{\dag})} \chi (H) 
\label{Ext6}
\eeqar
where $\chi$ is gauge-invariant, depending on $M,M^\dagger$ only via the
combination $H=M^\dagger M$. $S_{\rm wzw}(M)$ is the familiar
WZW action for $M$, see (\ref{vol21}).
The Chern-Simons term may be viewed as a ``velocity-dependent" potential, since it involves the time-derivative of the $A$'s.
The appearance of a phase factor, $e^ {i \omega (M, M^{\dag})} $
in (\ref{Ext6}) is in accordance with the fact that the wave functions must carry phase factors when we have
velocity-dependent potentials \cite{Asorey}.
The gauge-invariant volume element is still as given in 
section \ref{Vol}, so that the inner product is
\beq
\la 1\vert 2\ra = \int d\mu (H)\, e^{2c_A S_{\rm wzw} (H)}\, \chi_1^* \,\chi_2
\label{Ext7}
\eeq

We can formulate the Schr\"odinger equation in terms of
$\chi (H)$. But it will turn out to be simpler to
use another wave function $\Phi(H)=
e^{-\half k S_{\rm wzw} (H)}~\chi (H)$.
The original wave function $\Psi$ is related to $\Phi (H)$ as
\beqar
\Psi &=& e^ {i \omega (M, M^{\dag})} \exp\left[
{\half } k S_{\rm wzw}(H) \right] \, \Phi (H)\nonumber\\
&=& \exp\left[ {k\over 2}\left( S_{\rm wzw} (M^\dagger ) -S_{\rm wzw} (M) + S_{\rm wzw} (H)
\right)\right]
~\Phi (H)\nonumber\\
&=& \exp\left[ k S_{\rm wzw} (M^\dagger ) -{k\over 4\pi} \int A^a\bA^a \right] ~\Phi (H)
\label{Ext8}
\eeqar
In going to the last line of this equation from the second, we have used the Polyakov-Wiegmann identity (\ref{vol22}).
The inner product of the states, expressed
in terms of $\Phi$'s, is given by
\beq
\la 1\vert 2\ra = \int d\mu (H) e^{(k+2c_A) S_{\rm wzw} (H)}~\Phi_1^* \Phi_2
\label{Ext9}
\eeq
This inner product agrees with what is obtained for the pure Chern-Simons theory as well. Notice that, compared to the pure YM case, 
the key difference in the integration measure
is in the coefficient of the WZW action 
$S_{\rm wzw} (H)$, with $2 c_A \rightarrow k+2c_A$.
Also, the integration measure is identical to what is obtained for the pure
CS theory.

Since the Chern-Simons term is independent of the spacetime metric,
it does not contribute to the energy-momentum tensor and
the Hamiltonian. Thus $\H$ is still of the form
\beq
\H = {1\over 2 e^2} \int \bigl[ (E^a)^2 + (B^a)^2 \bigr]
\label{Ext10}
\eeq
However, the action of this on the wave functions is altered as the electric fields have
additional terms when expressed in terms of functional derivatives
as in (\ref{Ext2}).

There are several steps involved in working out the Hamiltonian
or the expression for the kinetic energy as a functional differential
operator. First of all, we express the derivatives with respect to
$A$, $\bA$ in terms of the translation operators
$p_a$, ${\bar p}_a$ on $M$, $M^\dagger$, as in 
(\ref{ham5}).
In $E$, $\bE$, we also have the additional terms proportional to
$A$, $\bA$. Finally, for the action on $\Phi (H)$, we
need $\H \rightarrow e^{ -{\half}k S_{\rm wzw}(H)  - i \omega}
\H e^{ i \omega + {\half}k S_{\rm wzw}(H) }$.
Since
\beq
\bp _a S_{\rm wzw} (M^{\dag})  = - {i \over 2\pi} \bdel(\del M^{\dag} M^{\dag -1})_a ,
\hskip .2in
p _a  S_{\rm wzw} (M) = - {i \over 2\pi} \del(M^{-1} \bdel M)_a ,
\label{Ext11}
\eeq
we find
\beqar
e^{- i \omega}\, T \,e^{i \omega}  &=&{e^2 \over 2} \int K_{ab} (\vx) \left[ \int_y {\bar{G}}(\vx,\vy) \bp (\vy) -
{ik \over 4\pi} (\del H H^{-1})(\vx) \right]^a \nonumber\\
&&\hskip .3in\times \left[ \int_u {G}(\vx,\vu) p(\vu) + {ik
\over
4\pi} (H^{-1}
\bdel H) (\vx)\right]^b
\label{Ext12}
\eeqar
Including the $e^{{\half}k S_{\rm wzw}(H)}$ factors, we then obtain
\beqar
{\tilde T}
&=& {e^2 \over 2} \int K_{ab} (\vx) \left( \int_y \bar{\cal{G}} (\vx,\vy)\, \bp
_a (\vy)
- {ik
\over 2\pi} (\del H H^{-1})_a \right) \int_u{\cal{G}}(\vx,\vu)\, p_b (\vu)
\nonumber\\
& =& {e^2 \over 2} \int K_{ab}\,  e^{- k S_{\rm wzw} (H)}\,  \bar{\cal{G}}  \bp _a 
\,e^{k S_{\rm wzw} (H)}\, {\cal{G}} p_b \label{Ext13}\\
{\tilde T} &\equiv& e^{-{\half}k S_{\rm wzw}(H)} e^{- i \omega}\, T\, e^{i \omega} 
e^{{\half}k S_{\rm wzw}(H)}
\nonumber
\eeqar
It is possible to write this in a more symmetric form, similar to
the expression (\ref{ham7}) for the pure Yang-Mills case, but
it will not be important for us at this stage.
(The relevant expressions are given in \cite{KKN-CS}.)

In the Yang-Mills theory, the observables are all obtained in terms of the current $J^a$. However, in the YMCS theory, there are additional
observables.
Notice that the inner product (\ref{Ext9})
expresses matrix elements of operators in
terms of a hermitian WZW model of level
$(k+2c_A)$. The correlators of the hermitian WZW model
with level number $(k+2c_A)$
 are the analytic continuation of the correlators of the
level $k$ $SU(N)$ WZW-model with
$\kappa = k+c_A$ replaced by $-\kappa = -(k+c_A)$. 
The level $k$ $SU(N)$ WZW model has integrable primary operators
(of finite norm) other than the identity.
These are the additional observables compared to the pure YM theory.
However, we expect that the vacuum wave function can still be expressed in terms of the currents. 
For a wave function $\Phi$ which depends on $J$, rather than $H$ in general, the
expression for the Hamiltonian can be written in terms of functional derivatives with respect to $J$.  
The result is
\beqar
\tilde{T}&=&T_{\rm YM} +{e^2k\over 4\pi} \int J^a {\delta \over \delta J^a}\nonumber\\
T_{\rm YM}&=& {e^2c_A\over 2\pi}\left[ \int_u J^a(\vu) {\d \over \d J^a(\vu)} ~+~
\int
\Omega_{ab} (\vu,\vv) 
{\d \over \d J^a(\vu) }{\d \over \d J^b(\vv) }\right]\nonumber\\
\Omega_{ab}(\vu,\vv)&=& {c_A\over \pi^2} {\d_{ab} \over (u-v)^2} ~-~ 
i {f_{abc} J^c (\vv)\over {\pi (u-v)}} +{\cal O} (\e )
\label{Ext14}
\eeqar
We see that the coefficient of the $\int J{\delta /\delta J}$-term
is $(k+2c_A)e^2/4\pi$, giving a mass of this value to every factor
of $J^a$. The perturbative mass $\mu$ gets corrected
by the addition of $(e^2 c_A/2\pi)$.
This is also consistent with the shift $2c_A\rightarrow 
k+2c_A$ in the integration measure in
(\ref{Ext9}), and also in accordance with the intuitive argument
for the mass gap given in subsection \ref{Int-mass}.

The expression (\ref{Ext14}) for ${\tilde T}$ can be rewritten as
\beqar
\tilde{T} &=& {\tilde m}\int_u J^a(\vu) {\d \over \d J^a(\vu)} ~+~ m \int
\Omega_{ab} (\vu,\vv) 
{\d \over \d J^a(\vu) }{\d \over \d J^b(\vv) } \nonumber\\
&=&{\tilde m} \left[ \int_u \xi^a(\vu) {\d \over \d \xi^a(\vu)} ~+~ \int
\Omega_{ab} (\vu,\vv) 
{\d \over \d \xi^a(\vu) }{\d \over \d \xi^b(\vv) }\right]
\label{Ext15}
\eeqar
where ${\tilde m}= (k+2c_A)e^2/4\pi$ and $\xi= \sqrt{{\tilde m}/m}~J$. The 
potential energy $\int B^2$ is as it was
in the pure YM case.

Given the similarity of these expressions to what we obtained for the YM theory, we can use the expansion scheme outlined in section
\ref{SchE} and work out the lowest order vacuum wave function as
\beqar
\Phi_0&=& \exp\left[ -{\pi\over {\tilde m} c_A}\int \bdel \xi\left( {1\over {\tilde m}+\sqrt{ {\tilde m}^2 -\nabla^2 }}\right) \bdel \xi \right]\nonumber\\
&=&\exp\left[ -{\pi\over mc_A}\int \bdel J\left( {1\over {\tilde m}+
\sqrt{ {\tilde m}^2 -\nabla^2 }}\right) \bdel J \right]
\label{Ext16}\\
&\approx& \exp \left( - {1\over 4 g^2} \int F^2\right)\nonumber
\eeqar
where $g^2 = {\tilde m} e^2 = e^4 {(k + 2 c_A)/ 4 \pi}$.
In the last line, we give the expression for the modes of low momentum or long wave length.
For an observable $\O$ (involving long wave length modes of the fields),
 the expectation value is thus
\beqar
\la {\cal O}\ra &\approx& \int d\mu (H) e^{(k+2c_A)S_{\rm wzw} (H)}~e^{-{1\over 4g^2}\int F^2} ~{\cal O}
\approx \int d\mu\left( {\cal{C}}\right)~e^{k S_{\rm wzw} (H)}e^{-{1\over 4g^2}\int F^2}~{\cal O}\nonumber\\
&\approx& \int [dQ]d\mu\left({\cal{C}}\right)~e^{-{\cal S}}~{\cal O}\nonumber\\
{\cal S}&=& \int d^2x~\left[ {1\over 4g^2}F^a_{\mu\nu}F^{a\mu\nu} +\sum_{i=1}^k
{\bar Q}^i\gamma
\cdot D Q^i \right]
\label{Ext17}
\eeqar
where we have used the fact that 
$e^{k S_{\rm wzw} (H)}$ can also be expressed in
terms of integration
over fermions (in two dimensions) as
\beqar
e^{S_{\rm wzw} (H)}&=& \det (- D\bD)=\int [dQ]e^{-\int ({\bar Q}_LD Q_L +{\bar Q}_R \bD Q_R )}\nonumber\\
&=& \int [dQ]~e^{\int {\bar Q}\gamma\cdot D Q}
\label{Ext18}
\eeqar
Here $Q$'s are fermions in the fundamental representation of $SU(N)$
and we use
$k$ flavors to get the factor $e^{k S_{\rm wzw} (H)}$. 

Effectively, we have the expectation value which can be
computed in a two dimensional 
YM theory coupled to $k$ flavors of fermions
in the fundamental representation.
(We are not saying that there are fermions in the spectrum of the theory; 
(\ref{Ext17}) is just a useful way of expressing expectation values.)
These fermions can screen charges in any representation and hence the
expectation value of the Wilson loop will not display an area law.
Already, at the level of perturbation theory, we have seen that the
Chern-Simons term in the action (\ref{Ext1}) acts as a mass term for the gluons. Therefore we should expect that the interaction energy
between charges cannot be of long range.
The fact that we do not obtain an area law for the Wilson loop is entirely in accordance with this expectation.
\subsection{Supersymmetric theories}\label{SUSY}
In this subsection we will make a few observations about supersymmetric Yang-Mills theories.
(This analysis is heavily based on
\cite{{Aga-N2},{Aga-N3}}.) In subsection \ref{Int-mass} we have seen that the mass gap is closely related to the integration measure used for the
inner product of the wave functions. 
We have also seen that this correlation holds also for the YMCS theory where the inner product involves the integral of
$e^{(k + 2 c_A) S_{\rm wzw}(H)}$, see (\ref{Ext9}), and the mass
${\tilde m} = e^2 (k + 2 c_A)/ 4\pi$.
Further, the measure (\ref{Ext9}) is the same for the pure CS theory and for the YMCS theory.
With these observations in mind, we see that we can make some statements
regarding supersymmetric theories via the following strategy.
We calculate the integration measure for the supersymmetric CS theory,
taking $k \rightarrow 0$ to obtain the YM case.
{\it The key point is that it
is possible to identify the integration measure without detailed calculations by using a set of indirect, although slightly intricate,
arguments.}
For this, we will consider the Hamiltonian analysis
of the level $k$ CS theory coupled to a set of point charges, the charge matrices being $T^a$ in some representation $r$ of the group $G$, 
as discussed in section \ref{CStheory}.

Again, as in that section, we will
consider two point-charges, conjugate to each other, at positions 
$\vx_1$ and $\vx_2$. 
Since we are now using $M$-dependent wave functions, the
solution to the Gauss law condition is
\beq
\Psi = N (z_1 , z_2 )\, M(x_1) \, M^{-1}(x_2) \, \Psi_0
\label{SUSY8}
\eeq
where $N (z_1, z_2)$, which depends on the positions of the charges, is
determined by
the requirement that $\Psi$ should obey the Schr\"odinger equation. 
The Hamiltonian for the CS theory with charges is given as
\beq
{\cal H} =  \sum_r (T^a)_r \left[ {2\pi \over k} {\dot\bz}_r {\delta \over \delta A^a(x_r)} + {\dot z}_r A^a(x_r) \right]
\label{SUSY9}
\eeq
When ${\cal H}$ acts on $\Psi$ we will encounter
singular terms due to terms like $\delta M(x_1) /\delta A(x_1)$, exactly as
explained in section \ref{CStheory}. 
The properly regularized version then leads again to the following
 two features.
First, $k$ in the expression (\ref{SUSY9}) is shifted to
$\kappa = k+ c_A$.
We have already obtained this shift entirely within
the Hamiltonian framework in section \ref{CStheory}.
(It is also derived in a different way in
in \cite{witten-jones}.) The Schr\"odinger equation is then
identical to the Knizhnik-Zamolodchikov (KZ) equation \cite{Kniz-Z} for
the chiral blocks of  the WZW theory with parameter $\kappa$.
Thus $N(z_1, z_2)$ becomes a chiral block of the level $k$
$SU(N)$-WZW theory.

Finally, we consider the normalization of the state
$\Psi$ in (\ref{SUSY8}). We can write the required integral as
\beq
\vert N (z_1, z_2) \vert^2 \, \int d \mu (H)\, e^{{\tilde k} \,S_{\rm wzw}(H)}~ 
H(x_1) \, H(x_2)^{-1} = 1
\label{SUSY10}
\eeq
We have put in the measure factor with an arbitrary coefficient ${\tilde k}$
for $S_{\rm wzw}(H)$ to show how we can determine it.
There are two points we can make about this normalization:
\begin{enumerate}
\item The integral in (\ref{SUSY10}) will yield the correlator
$\la H(x_1) \, H(x_2)^{-1}\ra$ of the hermitian WZW model (for $SL(N, \mathbb{C})/SU(N)$)
of level ${\tilde k}$. The $(z_1, z_2)$-dependence of this correlator
must exactly cancel the similar dependence of 
$\vert N (z_1, z_2) \vert^2$ to allow for a proper normalization
of (\ref{SUSY8}). So $\la H(1) \, H(2)^{-1}\ra$ should be given by
the solution of the KZ equation with $\kappa = (k+c_A)
\rightarrow - \kappa = - (k+c_A)$.
\item At the same time, we also know that the correlator
$\la H(x_1) \, H(x_2)^{-1}\ra$ of the hermitian theory (of level ${\tilde k}$)
is the same as the corresponding correlator
of the $SU(N)$ theory of level $-{\tilde k}$ \cite{Gaw-K}.
\end{enumerate}
These two statements together imply that
\beq
- {\tilde k} ~+~ c_A = - \left( k+ c_A \right)
\label{SUSY11}
\eeq
We see that this determines ${\tilde k}$ to be $k + 2 c_A$, as expected.

This illustrates how we can obtain the measure in more general cases.
First we calculate the shift in the level number $k$ to identify the
KZ parameter $\kappa$. This can be done via the Hamiltonian method, or 
in an even simpler way, by straightforward use of Feynman diagrams
\cite{Pis-R}.
Once this is done, the compatibility of the two requirements
given above for $\la H(x_1) \, H(x_2)^{-1}\ra$ will be
\beq
\boxed{
\left. \begin{array}{r}
{\text{KZ parameter of}}~ SU(N)\\
{\text{WZW theory of level}} - {\tilde k}\\
\end{array} \right\}
= - \left\{ \begin{array}{l}
{\text{KZ parameter of}}~ SU(N)\\
{\text{WZW theory of level}}~ k\\
\end{array} \right.}
 \label{SUSY12}
\eeq

We are now ready to look at supersymmetric theories. 
From diagrammatic calculation (for which the supersymmetric Yang-Mills term may be viewed as a regulator), the KZ parameters are given as
 \cite{lee-scs}
\beq
 k \rightarrow \left\{ \begin{array} {l c l}
 k+ c_A&~ \hskip .4in~& {\cal N} = 0\\
 k+ {\half} {c_A}& \hskip .3in& {\cal N} = 1\\
 k&\hskip .3in& {\cal N} \geq 2\\
 \end{array}\right.
 \label{SUSY13}
 \eeq
The normalization of the wave functions for the supersymmetric YMCS 
theories are thus given by
\beqar
\la 1\vert 2\ra &=& \int d\mu(H)\, \exp[ {\tilde k} \, S_{\rm wzw}(H)]~
d\mu[{\rm Fermions}]~ \Psi_1^* \, \Psi_2
\label{SUSY14}\\
{\tilde k} &=&  \left\{ \begin{array}{l c l}
 k+ 2\, c_A&~\hskip .4in& {\cal N} = 0\\
 k+ {c_A}& \hskip .2in& {\cal N} = 1\\
 k\hskip .2in& &{\cal N} \geq 2\\
 \end{array}\right.
 \label{SUSY15}
 \eeqar
 For ${\cal N} = 0$, we can take $k =0$ and obtain the result for the pure YM case.
 For ${\cal N} = 1$, we cannot take $k = 0$ since there is a
 parity anomaly, so we need $k =1$ as the minimal choice for a consistent theory \cite{seiberg-witten, n=1-lattice}. In this case, the value of ${\tilde k}$ suggests that there will be a mass gap, of a magnitude different from the case of ${\cal N } = 0$.
 For ${\cal N} \geq 2$, we can take $k = 0$. For these cases, we should expect that there will be no mass gap. 
 
 These statements are in accordance with expectations from other analyses. 
 For the ${\cal N} = 4$ case, constraints of unbroken
 supersymmetry prevent a mass term \cite{seiberg-witten},
 but a partial spontaneous breaking of the
 gauge symmetry is possible.
For ${\cal N} = 2$ theories, no mass gap is expected,
but there may be no stable supersymmetric vacuum \cite{seiberg-witten,
witten-index, N=2-vac}.
The absence of mass gap 
 for ${\cal N} = 2$ has also been analyzed by different methods in \cite{N=2-brane, unsal-bion}.

While the arguments presented above for the measure bypassed direct calculations, one can ask whether the same result is obtained in a straightforward Hamiltonian formulation of the supersymmetric theories.
This is indeed the case, as discussed in some detail in
\cite{Aga-N2}. 
Here we will briefly indicate the steps to highlight a subtle point
in obtaining the Hamiltonian.
The classical action for the ${\cal N} = 1$ theory is given by
\beq
S = -\frac{1}{4\,e^2}\int F^a_{\mu \nu}F^{a\mu \nu} - \frac{i}{2e^2} \int \bar{\psi}^a(\gamma ^\mu D_\mu \psi)^a\label{SUSY16}
\eeq
The supersymmetry transformation is given by
\beq
\delta_\epsilon A_\mu^a = -i \, {\bar\epsilon}\, \gamma_\mu \psi^a, \hspace{.3cm} \delta_\epsilon \psi^a =  \frac{1}{2}F^a_{\mu \nu }\gamma^{\mu \nu}\epsilon\label{SUSY17}
\eeq
The action is invariant under
this transformation with the
supercharges given by
\beq
Q^\dagger  =\int( i\psi^\dagger \gamma^i \frac{\delta}{\delta A^i} + \frac{1}{e^2}\psi^\dagger B), \hspace{.3cm} Q  =\int( i \gamma^i \psi \frac{\delta}{\delta A^i} + \frac{1}{e^2}\psi B)
\label{SUSY18}
\eeq
$Q$ is a two-component spinor, and we make the identification
$Q^1 = q$, $Q^2 = q^\dagger$.
As mentioned before, 
the parity anomaly will make the partition function of this
theory vanish, rendering it trivial or inconsistent
\cite{seiberg-witten,n=1-lattice}.
To get a consistent theory, we must include a supersymmetric
Chern-Simons term
\beq
S_{\rm SCS} = -\frac{k}{4\pi} \int d^3x ~\Tr \left[\left(A_\mu\partial_\nu A_\alpha - \frac{2}{3}A_\mu A_\nu A_\alpha\right)\epsilon^{\mu \nu \alpha} + ie^2\bar{\psi}\psi\right]\label{SUSY19}
\eeq
The full action is thus $S_{\rm SYM} = S + S_{\rm SCS}$.
Being a supersymmetric theory, the Hamiltonian can be obtained
as the anticommutator of supercharges.
Towards this, we first define the gauge-invariant wave function
$\Phi (H)$
as in (\ref{Ext8}),
\beq
\Psi = e^ {i \omega (M, M^{\dag})} \exp\left[
{\half } k S_{\rm wzw}(H) \right] \, \Phi (H)
\label{SUSY20}
\eeq
The supercharge in terms of its action on $\Phi$ is given by
 \beq
 q' = i\int \chi^{\dagger a}(\mathcal{G}p)^a - \frac{1}{e^2}\frac{2\pi}{c_A}\int \chi ^a \bar\partial J^a
 \label{SUSY21}
 \eeq
where ${\mathcal{G}}$ is the regularized version
of the Green's function $G = \del^{-1}$ and $\chi$ is the gauge-invariant version of the fermion field defined by
$\chi^b = (M^{-1})^{ab} \psi^a$,
$\chi^{b\dagger} = \psi^{a \dagger} M^{ab}$.

The integration measure for the inner product of the
$\Phi$'s is given as
\beq
d\mu = d\mu (H)\,  \exp \left[ (k + (2-n)c_A) \, S_{\rm wzw}(H)\right]
\label{SUSY22}
\eeq
For the present case, $n =1$, but we will leave it arbitrary for now.
The adjoint of the supercharge, which is consistently the adjoint with
(\ref{SUSY22}) defining the integration measure, is
 \beq
 q^{\prime \dagger} =  - i \int \chi ^a\left((\bar{\mathcal{G}}\bar{p})^a - i\frac{k}{2\pi}(\partial HH^{-1})^a + i \frac{nc_A}{2\pi}(\partial HH^{-1})^a\right) - \frac{1}{e^2}\frac{2\pi}{c_A}\int (\chi ^a \bar\partial J^a)^\dagger
 \label{SUSY23}
 \eeq 
Recall that, by virtue of the physical states being annihilated by the
Gauss law operator, we were able to eliminate $E^a$ in favor
of ${\bar E}^a$ and the currents, in the simplification of the
kinetic energy operator, see equations (\ref{ham12}) to
(\ref{ham21}). Equivalently, we can eliminate $\bp^a$
in favor of $p^a$. 
Effectively, this amounts to the statement
 \beq
 \bar p^a = (Kp)^a + \frac{1}{e^2}f^{alm}(K\chi^\dagger)^l\chi^m
 \label{SUSY24}
 \eeq
 When this is used in (\ref{SUSY23}), we have to move
 $\chi^a$ to the right end to obtain normal ordering. This results in
 a singular term ${\bar{\mathcal{G}}}(x,x)$, exactly the same kind of term
 we encountered in section \ref{Ham}. Evaluating it as before, we
 end up with
 \begin{align}
 -i\int \chi^a(x)(\bar{\mathcal{G}}\bar p) ^a(x) = &-i\int \chi^a(x)(\bar{\mathcal{G}}K p) ^a(x) - \int \chi^a(x)J^a(x) \nonumber\\
 &+ \frac{i}{e^2}\int \bar{\mathcal{G}}^{ab}(x,y) f^{blm}(K\chi^\dagger)^l(y)\chi^a(x)\chi^m(y)
 \label{SUSY24}
 \end{align}
The $\chi^a J^a$-term arises from the normal ordering mentioned above.
When this expression is used for $q^{\prime \dagger}$ in
(\ref{SUSY23}) and the anticommutator is taken, we find the
gauge-invariant form of the
Hamiltonian for the supersymmetric theory as
\begin{align}
\!\!\!\!\!\mathcal{H} = \frac{1}{2} \{q^\prime, q^{\prime \dagger}\} &= \frac{e^2c_A}{2\pi}\left(\int J^a\frac{\delta}{\delta J^a} + \int \Omega ^{ab}(x, y)\frac{\delta}{\delta J^a(x)}\frac{\delta}{\delta J^b(y)}\right)
+ \frac{2\pi^2}{e^2c_A^2}\int(\bar \partial J^a \bar \partial J^a)\nonumber\\
&+ \frac{e^2(k-nc_A)}{4\pi}\int J^a \frac{\delta}{\delta J^a} - \frac{ic_A}{2\pi}\int f^{abc}\bar{G}(x,y)K^{cs}(y)\chi^{s \dagger}(y)\chi^b(y) \frac{\delta}{\delta J^a(x)}\nonumber\\
&-\frac{1}{e^2}\int (\chi ^\dagger \bar{\mathcal{D}}_{\bar{J}}\chi ^\dagger - \chi \mathcal{D}_J\chi) + \left(\frac{e^2}{4\pi}(k + 
2 c_A - nc_A) \right)\int \frac{1}{e^2}\chi^{\dagger a} (K^{-1})^{ab}\chi^b
\label{SUSY25}
\end{align}
Notice the equality of the masses for the $J$'s and the $\chi$'s, as expected
for a supersymmetric theory.\footnote{It may be useful to keep in mind that the anticommutator is given as $\{ \chi^a, \chi^{b\dagger}\} = e^2 \delta^{ab} \delta^{(2)}
(x-y)$.}
Also the value of the mass, namely, $((k+ 2 c_A - n c_A) e^2/4\pi)$
is in agreement with the measure of integration for the inner product and
the intuitive argument given earlier. We see the concordance between the measure, the mass gap and the explicit quantization
using gauge-invariant variables for the
supersymmetric theories.
\section{Entanglement in Yang-Mills (2+1)}
\label{Ent}
 \setcounter{equation}{0}
\begin{quotation}
\fontfamily{bch}\fontsize{10pt}{16pt}\selectfont
\noindent 
The main difference in entanglement in the vacuum state between a matter field theory and a gauge theory is in the role of the edge states
mentioned in section \ref{gaugeprinciple}. The quantization of a gauge theory in a finite region in space can allow for edge states on its boundary.
But these are factored out in implementing the Gauss law over all of space.
This difference leads to the so-called contact term in the entanglement entropy. We analyze the Maxwell theory first to obtain this result and
then give the analogous result for the Yang-Mills theory. The contact term is also
shown to be related to the surface terms in the BFK gluing formula.
\end{quotation}
\fontfamily{bch}\fontsize{12pt}{16pt}\selectfont
Entanglement is a property of the state and can be characterized by
a reduced density matrix obtained by integrating  $\Psi^*[\vf]  \Psi[\vf']$
over fields in some subregion of space. So it would seem
 that if there is any feature of the quantum theory for which wave functions provide a better framework than manifestly covariant methods, it would be
entanglement. 
And this is indeed the case, although, for ease of calculation a path integral
with a cut on (the unintegrated) part of space is often used (with a replica
trick as well). In the case of gauge theories, this led to the identification
of an extra term in the entanglement entropy, known as the contact term
(or Kabat term) \cite{kabat}, compared to what is expected for matter fields.
Here we will consider the contact term for YM(2+1) in a Hamiltonian formulation and relate it to something familiar in mathematics literature, known as the BFK gluing formula \cite{BFK}.
(This analysis is basically taken from \cite{AKN-ent}; see also
\cite{Gomes-R}.)
\subsection{Entanglement in Maxwell theory}
It is simpler and conceptually more clarifying to consider the Maxwell theory first. The Gauss law condition takes the form $G_0 = \nabla \cdot E = 0$.
We will choose a conjugate constraint $\chi = \nabla\cdot A $.
In general, if we have constraints
$\zeta_i$ and conjugate constraints $\chi_j$, the Hamiltonian path integral 
is given by
\beq
Z = \int [d\mu] \,\delta(\zeta )\, \delta(\chi) \, \det[\{ \zeta_i, \chi_j\} ]
\, e^{i S}
\label{Ent1}
\eeq
Here $d\mu$ is the phase space measure of integration, 
the constraints are enforced by $\delta$-functions and
we also need the determinant of the Poisson brackets of the constraints.
Also $S$ in this formula is the action expressed in terms of the phase space variables.
For the Maxwell theory we thus get
\beq
Z = \int [d\mu] \,\delta (\nabla\cdot E) \, \delta (\nabla\cdot A) \,
\det[-\nabla^2]~ e^{i S}
\label{Ent2}
\eeq
We have already set $A_0 =0$, so that the phase space
variables are $E_i$ and $A_i$, $i = 1, 2$.
Consider the theory in some region of space ${\rm V}$
with a boundary $\del {\rm V}$. 
The fields can be parametrized as
\beq
A_i = \del_i \theta + \epsilon_{ij} \del_j \vf, \hskip .3in
E_i = {\dot A_i} = \del_i {\sigma} + \epsilon_{ij} \del_j {\Pi}
\label{Ent3}
\eeq
We separate the fields into a bulk part and a boundary part by writing
\begin{alignat}{2}
\theta (x) &= {\tilde \theta} (x) + \oint_{\del {\rm V}} \theta_{0} (y)\,
n \cdot \del \mathsf{G} (y,x), \hskip .2in
\vf (x) =& {\tilde \vf} (x) + \oint_{\del {\rm V}} \vf_{0} (y)\,
n \cdot \del \mathsf{G} (y,x)\nonumber\\
\sigma (x) &= {\tilde \sigma} (x) + \oint_{\del {\rm V}} \sigma_{0} (y)\,
n \cdot \del \mathsf{G} (y,x), \hskip .2in 
\Pi  (x) =& {\tilde \Pi} (x) + \oint_{\del {\rm V}} \Pi_{0} (y)\,
n \cdot \del \mathsf{G} (y,x)
\label{Ent4}
\end{alignat}
The tilde-fields all obey Dirichlet conditions, vanishing on
$\del{\rm V}$. The values of the fields on
the boundary are designated with a subscript $0$ and are continued into the interior of ${\rm V}$ via Laplace's equation,
i.e.,
\beq
\nabla_x^2 \oint_{\del {\rm V}} \theta_{0} (y)\,
n \cdot \del \mathsf{G} (y,x) = 0
\label{Ent5}
\eeq
The decomposition of fields as in (\ref{Ent4}) follows from
Green's theorem.
The Green's function $\mathsf{G}(y,x)$ for the Laplace operator,\footnote{$\mathsf{G}$ obeys i.e., $\nabla_y^2 \mathsf{G} (y,x) = \delta^{(2)}(y,x)$.}
also obeys Dirichlet conditions. 
If we consider a box of length $L$ along the $x_1$-axis, the
Green's function is given by
\begin{alignat}{2}
\mathsf{G} (y, x) =& \int {dp \over 2\pi} e^{i p(y_2 - x_2)} \mathsf{G}_p (y_1, x_1)\label{Ent5a}\\
\mathsf{G}_p (y_1, x_1) =&{1\over 2 \omega}
\left[  {e^{\omega (y_1 + x_1) }\over e^{2 L \omega} -1} +
{e^{-\omega (y_1 + x_1) }\over 1- e^{- 2 L \omega} } \right]
- \left\{ {e^{\omega (y_1 - x_1) }\over e^{2 L \omega} -1} + {e^{- \omega (y_1 - x_1) }\over 1- e^{- 2 L \omega} } \right\}, \hskip .2in &y_1> x_1\nonumber\\
=&{1\over 2 \omega}
\left[  {e^{\omega (y_1 + x_1) }\over e^{2 L \omega} -1} +
{e^{-\omega (y_1 + x_1) }\over 1- e^{- 2 L \omega} } \right]
-\left\{ {e^{-\omega (y_1 - x_1) }\over e^{2 L \omega} -1} + {e^{\omega (y_1 - x_1) }\over 1- e^{- 2 L \omega} } \right\}, \hskip .2in  &y_1< x_1\nonumber
\end{alignat}
where $\omega = \vert p\vert$.
It is easy to verify that this is zero at $x_1 = 0, L$ and at $y_1 = 0, L$.
By choosing $L$ appropriately, this can be used for the whole volume 
${\rm V}$ or for the subregions ${\rm V}_1$ and ${\rm V}_2$ considered later.

The canonical one-form is given by
$\A = \int E_i \delta A_i$ and by direct substitution  of
(\ref{Ent4}), we find
\begin{align}
\A&= \int E_i \, \delta A_i\nonumber\\
=&\int_{\rm V} \left[ (-\nabla^2 {\tilde \sigma})\, \delta {\tilde \theta}
+ {\tilde \Pi} \, \delta B \right]
+ \oint \E\,  \delta \theta_{0} (x) 
+ \oint Q\, \delta \vf_{0} (x) 
\label{Ent6}
\end{align}
where $B = -\nabla^2 {\tilde\vf}$ is the magnetic field. 
Also $\E$ and $Q$ are given by
\beqar
\E (x)&=& \oint_y \sigma_{0} (y) \mathfrak{M}(y,x)   + {\del_{\tau}} \Pi_{0}(x) \nonumber\\
Q (x) &=& \oint_y \Pi_{0 } (y) \mathfrak{M}(y,x)  - {\del_{\tau}} \sigma_{0} (x)\label{Ent7}\\
\mathfrak{M}(x, y)  &=&  n\cdot \del_x \,n\cdot \del_y G(x,y)\Big\vert_{ x,~y ~on~ \del {\rm V}} \label{Ent8}
\eeqar
$\mathfrak{M}(x,y)$ is what is usually referred to the Dirichlet-to-Neumann operator. This can be worked out explicitly using 
(\ref{Ent5a}).
For example, for the boundary at $x_1 = y_1 = L$, we find 
\beqar
\mathfrak{M}(x, y)  &=& \int {dp \over 2\pi} e^{i p(y_2 - x_2)}
\, {\omega } \, \coth (\omega L )\nonumber\\
&\rightarrow& 
 \int {dp \over 2\pi} e^{i p(y_2 - x_2)}
\, {\omega } \hskip .2in {\rm as~} L \rightarrow \infty
\label{Ent8a}
\eeqar
Also in (\ref{Ent7}), ${\del_{\tau}} = n_i \epsilon_{ij} \del_j$ denotes the tangential derivative on the boundary. 
$\E$ and $Q$ are not independent, but are related by
\beq
\C = \del_y \oint \E (x) \mathfrak{M}^{-1}(x,y) + Q (x)= 0
\label{Ent9}
\eeq
This can be verified directly using (\ref{Ent8a}).
In the sense of Dirac's theory of constraints, $\C$ is of the first class; one can choose a conjugate constraint $\vf_0 = 0$ and eliminate the pair, so that
\beq
\A = \int_{\rm V} \left[ (-\nabla^2 {\tilde \sigma})\, \delta {\tilde \theta}
+ {\tilde \Pi} \, \delta B \right]
+ \oint_{\del{\rm V}} \E\, \delta \theta_{0}  
\label{Ent10}
\eeq
The phase volume associated with this canonical structure is\footnote{The determinants of $-\nabla^2$ are calculated with Dirichlet boundary conditions.}
\beq
d\mu  = [d{\tilde \sigma} d {\tilde \theta}]\,
 [d \E\, d\theta_0] \, [ d{\tilde \Pi} d B] \,  \det (-\nabla^2)
 \label{Ent11}
 \eeq
 The constraints entering the path integral (\ref{Ent2}) can therefore
 be written out as
 \beq
 \delta(\nabla\cdot E) = (\det(-\nabla^2))^{-1}\, \delta ({\tilde\sigma}),\hskip .2in
\delta(\nabla\cdot A)  = (\det(-\nabla^2))^{-1}\, \delta ({\tilde\theta})
\label{Ent12}
\eeq
Since ${\tilde \theta}$ vanishes on $\del{\rm V}$, we are imposing the Gauss
law with test functions vanishing on $\del{\rm V}$. The value
of $\theta$ on the boundary and its conjugate $\E$ represent physical degrees of freedom.
Using (\ref{Ent11}) and (\ref{Ent12}), we see that we can set ${\tilde \sigma} = {\tilde \theta} = 0$, and all factors
of $\det (-\nabla^2)$ cancel out, so that $Z$ in (\ref{Ent2})
becomes
\beq
 Z = \int [d \E\, d\theta_0] \, [ d{\tilde \Pi} d B]~e^{i S}
 \label{Ent13}
 \eeq
 The action $S$ also involves only the fields $\Pi$, $B$, $\E$, $\theta_0$.
 We see that we have a theory of the bulk fields $\Pi$, $B$, which constitute a single bulk field, with ``edge modes"
 described by $\E$, $\theta_0$.
\begin{figure}[!t]
\begin{center}
\scalebox{.8}{\includegraphics{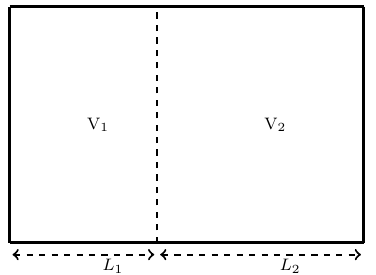}}
\caption{Showing division of space into two regions for entanglement considerations}
\label{entangle}
\end{center}
\end{figure}

Now consider going through this procedure for a region of space
divided into two with an interface (dashed line) as shown in Fig.\,\ref{entangle}. We construct the theory on the whole space
and then in  ${\rm V}_1$ and
${\rm V}_2$ separately, put them together and compare the results.
Eventually, we will take $L_1, L_2 \rightarrow \infty$, where
$L_1$, $L_2$ are the lengths along the $x_1$-direction for
${\rm V}_1$ and ${\rm V}_2$, respectively.
Thus $\mathfrak{M}(x,y)$ with $L \rightarrow \infty$ will be what is
relevant.
For the
theory on the whole space, the result is basically as we
have already discussed, with
\beqar
Z_{\rm whole} &=&\int [d\sigma d\theta] \, [d\Pi dB] \, [\det -\nabla^2]\,
\delta (\nabla\cdot E)\, \delta(\nabla\cdot A)~ \det[-\nabla^2]~ e^{i S}\nonumber\\
&=&\int [d\Pi d B ]~ e^{i S} 
\label{Ent14}
\eeqar
We do not consider any edge modes for the boundary of the whole space
since our focus will be on entanglement across the interface.
(They can be included without changing the essence of the argument.)

Now consider building the theory separately in ${\rm V}_1$ and
${\rm V}_2$.
The fields on the interface can be continued into ${\rm V}_1$ and
${\rm V}_2$ again using Laplace's equation, so that we have
\beq
\theta (x) =
\begin{cases} 
{\tilde \theta}_1 (x) + \oint_{\del {\rm V}_1} \theta_{0} (y)\,
\,n \cdot \del G_1(y,x) &~~~~{\rm in ~V}_1\\
{\tilde \theta}_2(x) + \oint_{\del {\rm V}_2} \theta_{0} (y)\,
\,n\cdot \del G_2 (y,x) &~~~~{\rm in~ V}_2\\
\end{cases}
\label{Ent15}
\eeq
 with similar expressions for the other fields. $G_1$ and $G_2$ are Green's functions for the Laplacian for regions ${\rm V}_1$ and ${\rm V}_2$, respectively, vanishing on the interface. The phase volume has the form
 \beq
d\mu_{\rm split}  = [d{\tilde \sigma} d{\tilde \theta}]_1 \,[d{\tilde \sigma} d{\tilde \theta}]_2 \,\det (-\nabla^2)_1 \, \det (-\nabla^2)_2\,
[d \E d\theta_0] \times [d\mu_{\Pi, B}\text{-part}]
\label{Ent16}
\eeq

The key issue is about the constraints.
Using $f, ~h$ for test functions,
with boundary values on the interface designated as $f_0, ~h_0$, respectively, 
the constraints are
\beqar
\int \del_i  f\, E_i &=& \int_{{\rm V}_1} \tilde{f }_1 (-\nabla^2 {\tilde \sigma}_1) + 
 \int_{{\rm V}_2} \tilde{f}_2 (-\nabla^2 {\tilde \sigma}_2)  + \oint
 f_0 \, \E \approx 0\nonumber\\
\int \del_i  h\, A_i &=& \int_{{\rm V}_1} \tilde{h}_1 (-\nabla^2 {\tilde \theta}_1
+  \int_{{\rm V}_2} \tilde{h}_2 (-\nabla^2 {\tilde \theta}_2)  + \oint
  h_0 \, (\mathfrak{M}_{\rm 1} + \mathfrak{M}_{\rm 2} )\,\theta_0  \approx 0
\label{Ent17}
\eeqar
$\mathfrak{M}_1$, $\mathfrak{M}_2$ are the Dirichlet-to-Neumann operators for
$G_1$ and $G_2$, respectively.
For the theory on the full space, $\theta$-dependence is eliminated everywhere
including the interface, so each term in (\ref{Ent17}) must vanish separately and the $\delta$-functions for the constraints
must be interpreted as
\beqar
\delta (\nabla \cdot E)~\delta(\nabla \cdot A) &=& \delta [ -\nabla^2{\tilde \sigma}_1]\,\delta [ -\nabla^2{\tilde \sigma}_2]
~ \delta [ -\nabla^2{\tilde \theta}_1]\,\delta [ -\nabla^2{\tilde \theta}_2]
\nonumber\\
&&\hskip .1in \times\delta [ \E ]\,\delta [(\mathfrak{M}_1 + \mathfrak{M}_2)  \theta_0 ]
\label{Ent18}
\eeqar
We also have the BFK gluing formula \cite{BFK}\footnote{This formula tells us that if a Riemannian manifold is separated into ${\rm V}_1$, ${\rm V}_2$, etc. by suitable hypersurfaces, then the determinant of the Laplacian for the full space can be obtained as the product of similar determinants
with Dirichlet boundary conditions in each of the regions ${\rm V}_1$, ${\rm V}_2$, etc. times a set of interface contributions which are the determinants of the Dirichlet-to-Neumann operators.}
\beq
\det (-\nabla^2) = \det (-\nabla^2)_1 \, \det (-\nabla^2)_2 \, 
\det (\mathfrak{M}_1 + \mathfrak{M}_2)
\label{Ent19}
\eeq
If we use results (\ref{Ent18}) and (\ref{Ent19}), we get back to
(\ref{Ent14}) as expected; splitting the fields is only a more involved way of
writing the path integral for the full space.

Consider now integrating out fields in ${\rm V}_2$.
Since the interface is a boundary to ${\rm V}_2$, from 
the point of view of the theory in
${\rm V}_2$, we can only impose the Gauss law with test functions which 
vanish on the interface. The
edge modes $\E$, $\theta_0$ are physical degrees of freedom.
Thus we must take $f_0 = h_0 = 0$, and the constraints become
\beq
\delta (\nabla \cdot E)~\delta(\nabla \cdot A) = \delta [ -\nabla^2{\tilde \sigma}_1]\,\delta [ -\nabla^2{\tilde \sigma}_2]\,
~ \delta [ -\nabla^2{\tilde \theta}_1]\,\delta [ -\nabla^2{\tilde \theta}_2]
\label{Ent20}
\eeq
The determinant $\det (\mathfrak{M}_1+ \mathfrak{M}_2)$ is not canceled out and the reduced theory in ${\rm V}_1$ takes the form
\beq
Z_{\rm red}  = \det (\mathfrak{M}_1 + \mathfrak{M}_2) \int  
[d\E d\theta_0] d\mu_{\Pi, B}\, e^{i S}
\label{Ent21}
\eeq
There is an extra factor $\det (\mathfrak{M}_1+ \mathfrak{M}_2)$; since this is part of the phase volume, it is to be considered as a degeneracy factor. Thus
if we define a reduced density matrix, it takes the form
 \beq
 \rho = {{\mathbb 1}  \over \det (\mathfrak{M}_1 + \mathfrak{M}_2)}\, ~(\rho_{\rm bulk})_{\rm red}
 \label{Ent22}
 \eeq
 where $(\rho_{\rm bulk})_{\rm red}$ refers to the reduced
 density matrix for all the remaining physical degrees of freedom and
 $\mathbb{1}$ is a matrix such that $\Tr\, \mathbb{1} = \det (\mathfrak{M}_1 + \mathfrak{M}_2)$.
 
 This determinant $\log \det (\mathfrak{M}_1+ \mathfrak{M}_2)$ is Kabat's contact term
 \cite{kabat}.
 Its origin is due to the simple fact that
in the full space Gauss law eliminates 
 $\theta$, $\E$, but for the theory in each region, these are not eliminated.
The contact term can also be identified as
the interface term in the BFK gluing formula \cite{BFK}.
\subsection{The case of YM(2+1)}
It is now straightforward to consider the situation for the Yang-Mills theory.
Since we phrased the discussion  given above in the language of gauge-fixing,
the simplest way for us is to eliminate $E$ from our considerations
using the Gauss law as we did in section \ref{Ham}, see equations
(\ref{ham12} to (\ref{ham21}).
This is like a complex gauge-fixing, since $M^\dagger $ gets set to $1$.
The canonical one-form is 
 $\A = \int E_i^a \delta A_i^a = - 4  \int \Tr ( \bE \, \delta A + E \, \delta \bA )$ and the Gauss law takes the form
 \beq
 \G^a = 2 (\bD E + D \bE )^a
\label{Ent23}
\eeq 
As the conjugate constraint, we take $\chi^a = (D \bA)^a$.
Eliminating $E$, the canonical one-form can then be written as
\beq
\A = - {4}\int \Tr \Bigl[ \bE \, \delta A + \G (x)\,  (- D \bD )^{-1}_{x, y}\,  \delta \chi (y) \Bigr]
\label{Ent24}
\eeq
The corresponding phase volume is
\beq
d \mu =   \det [ (- D \bD )^{-1}]~ [d \bE d A ] \, [ d\G d\chi ]
\label{Ent25}
\eeq
We see, in a way similar to what happens in the Maxwell case, that
we will get $\det[(- D \bD )]_1$ and $\det[(- D \bD )]_2$ for ${\rm V}_1$ and
${\rm V}_2$, and $\det[(- D \bD )]_{{\rm V}_1\cup {\rm V}_2}$ for the full space
${\rm V} = {\rm V}_1 \cup {\rm V}_2$.
The contact term is then given by
\beq
\boxed{
S_{\rm contact} = \log  \left[ { \det (- D \bD )_{{\rm V}_1 \cup {\rm V}_2 } \over \det (-D \bD )_{{\rm V}_1}
\, \det (- D \bD )_{{\rm V}_2} }\right] }
\label{Ent26}
\eeq
Unlike the Abelian case, this expression
depends on the fields.
So one has to carry out an averaging over the physical fields, i.e.,
do the integration over $H$, to calculate the entropy.
If we ignore the field dependence, the
contribution of (\ref{Ent26}) is the same as the result
for (dim$G$ copies of) the Abelian theory.

A noteworthy point is the following. The operator $(-D\bD)$ which comes into the contact term is independent of the mass, even though the theory does have a mass gap. In a massive theory, the entanglement 
tends to vanish as the mass becomes large. But in the gauge theory
the contact term will lead to nonzero
entanglement even in the large $m$ limit.
\section{Discussion and comments}\label{Concl}
\setcounter{equation}{0}
We have reviewed a number of results for gauge theories in 2+1 dimensions obtained using the Schr\"odinger representation.
As mentioned in the Introduction, the key results which could be compared to numerical analysis were the string tension and the Casimir energy.
We have also analyzed propagator masses, string breaking effects, 
supersymmetric extensions, entanglement, etc.
In this section, we will now make a number of comments
on the status and prospects of this type of analysis.

First and foremost, regarding the string tension,
as explained in Appendix \ref{AppD}, the first set of corrections to the lowest order
result within the expansion scheme presented in section \ref{SchE}
turn out to be very small. While this is encouraging, it is clear from the cancellations in the partial sums shown in (\ref{Corr11})
that there must be a better way to organize the corrections,
where corrections to the wave function and the renormalization implicit in the subsequent integration over $\Psi^* \Psi$ are combined.
It will be very illuminating to formulate such an expansion.
Also, in Appendix \ref{AppD} we have mentioned some of
the other corrections (higher order, representation-dependent, etc.)
which we have not calculated.
It would be very useful to have some way of estimating these 
to see if they are also small and do not vitiate the results for the string tension.

Regarding the Casimir energy, recall that one could effectively use a scalar field theory of mass $m = (e^2 c_A / 2 \pi)$ with Neumann boundary conditions which are equivalent to perfect conductivity.
It is practically trivial to calculate the Casimir energy with other
boundary conditions. Lattice simulations with different boundary conditions
would be welcome as they can provide additional checks on our analysis.

Going beyond the string tension and the Casimir energy,
clearly the next natural step would be about glueballs. In section
\ref{Alter} we have already referred to the work of LMY
\cite{Leigh-MY} on estimates of glueball masses using correlators calculated using their wave function.
The existence of a nonzero string tension tells us that some effective string description of glueball states should be possible, but it is important to describe such excitations directly in terms of the Schr\"odinger equation.
This is necessary to work towards a theory for glueball interactions,
 their possible decays, etc. Similarly, the inclusion of fermions
 or quarks is a natural and important next step. Some work on this, focusing on the fermionic contribution to the integration measure for the inner product, has been carried out \cite{Aga-N3}, but formulating meson and baryon states in terms of the Schr\"odinger equation will be a really useful advance.
 
For most of the analysis presented here we have taken the spatial manifold to be
$\mathbb{R}^2$, although we briefly mentioned the formalism for
$S^2$ at the end of section \ref{Par}.
The case of the torus $T^2 = S^1 \times S^1$ is particularly interesting.
The theory on $\mathbb{R} \times T^2$, with a suitable Wick rotation, can be interpreted as the theory on the spatial manifold $\mathbb{R} \times S^1$ with a compactified imaginary time direction $S^1$.
Thus it will describe the theory at finite temperature.
Since the torus has a complex
modular parameter $\tau$, one can tune the real and imaginary parts of
$\tau$ independently and, presumably, one can get information about the high temperature phase of the theory, including the question of 
the deconfinement transition. We may also note that a parametrization for the gauge fields similar to what we have been using exists for the torus;
it is of the form
\beq
A = - \del M M^{-1} + M \left[ {i \pi \, a\over {\mathfrak{Im}}\tau}\right] M^{-1}
\label{Concl1}
\eeq
where $a$ is a (complex) constant with certain periodic identifications.
It defines a point on the Jacobian of the torus
(which is itself a torus).
Therefore various Jacobi $\theta$-functions will naturally appear in the analysis.
It is then clear that there will be some number-theoretic dimensions
to the problem of deconfinement.
Any developments on this question will be interesting from a mathematical as well as physical point of view. 
In this context, we also note that there has been some interesting
work on the torus with twisted boundary conditions
\cite{Cham-G-A}. The authors considered possible tachyonic instabilities and
showed that the absence of instabilities is related to a problem in number theory, specifically, the approximation of irreducible fractions
by other fractions of smaller denominator.

Regarding supersymmetric theories, we note that there are expectations on the infrared limit of extended supersymmetric Yang-Mills
theories, namely, they flow to supersymmetric CS theories with
${\cal N} = 6, 8$, with gravity duals in terms of M2 branes \cite{ABJM}.
It will be interesting to see the role of the measure for the inner product
in the context of the gravity duals.
There are also a number of duality-related properties of
Chern-Simons theories with matter \cite{Aharony-etal}; it may be interesting to consider som eof these in relation to the inner product.

In the section on entanglement, in
(\ref{Ent26}), we have identified the contact term of the entanglement
entropy
in the 3d gauge theory in terms of the BFK formula.
Strictly speaking the BFK formula refers to the case of the Laplace operator,
but here we are dealing with $D\bD$.
It will be interesting to see the field dependence of
the entropy since the contact term is the only long range contribution
given the existence of the mass gap.
The determinants can be expressed in terms of the WZW action
for $H$, so the BFK formula should be expressible in terms of
the WZW action as well.

In the Introduction, we mentioned the Lichnerowicz bound on the eigenvalues of the Laplacian. While this is not directly applicable to a field theory,
a suitable generalization to the infinite dimensional context could lead to an argument for a nonzero mass gap. The volume element given 
in (\ref{vol31}) is a good start in this direction
since the total volume can then be viewed
as the partition function for the hermitian WZW theory.
(The question here is not whether the integration needed to get the total volume
 requires regularization, it does;
but the contrast to be drawn is with the Abelian case, where
even for a finite number of modes, the result is divergent
since there is no damping from the exponential factor, $c_A$ being zero.)
The natural next step will be to seek a properly regularized expression for the Ricci tensor and set up a Lichnerowicz-type argument for the eigenvalues.
(The vacuum state has zero energy, so the eigenvalue of interest is for the first excited state.)
One has to then show that the bound survives
when the regularization parameter $\e$ is taken to be zero.
We note that some progress along this line of reasoning has been made recently
\cite{Mondal}, although many fine points remain to be ironed out.

We may also note that our parametrization of the fields, and the gauge-invariant description
based on it, has also been useful for some problems
beyond the usual variants of the gauge theory.
In quantum Hall effect, the construction of states of fractional
filling fraction is done via a procedure known as flux attachment;
these are used to define the so-called composite fermions.
The definition of the gauge-invariant version for the matter fields
in our language,
as in \cite{Aga-N3}, shows that it is basically multiplication by
a phase factor. One may view this as a statement of
``flux attachment".
This has proven to be useful in sorting out how the fractional states can arise,
particularly with nonabelian fields \cite{Aga-flux}.

Finally we come to a point we have not touched upon at all.
Although we have done everything in the Hamiltonian approach and the
Schr\"odinger representation, we can ask, with the benefit of hindsight,
whether there is a more covariant description. Is there an effective action
$\Gamma$ which captures the essence of the wave function
(\ref{SchE9})? The vacuum wave function
contains enough information to 
construct all excited states, with some mild requirements on the Hamiltonian,
so one can indeed obtain an effective action.
This has been done in \cite{Nair-eff} and gives an effective action
\beq
\Gamma = \int {1\over 4} F^a_{\mu\nu} F^a_{\mu\nu} + S_{\rm mass}(A) 
+ (\sigma^\mu D_\mu \Phi_A)^{a\dagger} (\sigma^\nu D_\nu \Phi_A)^{a} ~+~ \cdots
\label{Concl2}
\eeq
Here $S_{\rm mass}(A)$ is a gauge-invariant nonlocal mass term for the gauge field.
We can take it to be of the form given in (\ref{Res25}),
\beq
S_{\rm mass} (A) = {m^2 \over 2} \int \left[ A^2 + \del\!\cdot \!\!A (x) \, \left( {1\over \del^2}\right)_{x,y} \, \del\!\cdot \!\!A (y) + \cdots \right]
\label{Concl3}
\eeq
where the ellipsis stands for terms with higher powers of
$A$ needed for a gauge-invariant completion of the quadratic terms.
Only the leading quadratic terms suffice to determine the leading 
expression (\ref{SchE9}) for the wave function, so the higher terms are not
important at this stage.
The field $\Phi^a_A$, $a = 1, 2, \cdots, (N^2 -1)$,  $A= 1, 2$, is complex
and transforms according to the adjoint representation of $SU(N)$;
it also transforms as a 2-component spinor under the Lorentz group.
Further, in (\ref{Concl2}),
$\sigma^\mu$, $\mu = 1, 2, 3$, are the Pauli matrices and $D_\mu$ denotes the gauge-covariant derivative. 
The field $\Phi^a_A$ is a Lorentz spinor but it is a bosonic field
with a term in the action which is quadratic in spacetime
derivatives. This is certainly unusual,
but $\Phi^a_A$ is not to be considered as an observable field but simply as a method of capturing the physics of the wave function (\ref{SchE9}).
 The action (\ref{Concl2}) also has an additional $U(1)$ symmetry $\Phi \rightarrow e^{i \theta} \, \Phi$, which the original Yang-Mills theory does not have. 
 This is to be eliminated
by requiring that all physical operators must have equal numbers of 
$\Phi$'s and $\Phi^*$'s.
It is possible to show that this action leads to the wave function
(\ref{SchE9}) \cite{Nair-eff}.

What is interesting is that because of the auxiliary field
$\Phi^a_A$, one can have $\mathbb{Z}_N$-vortex
solutions to the equations of motion.
(These are actually particle-like since we are in two spatial dimensions, but they are not point-like, they may have some extent, but have spatial codimension
equal to 2.)
However, one can show that, because of the spinorial nature
of $\Phi^a_A$, the total vortex number should vanish for reasons of Lorentz invariance \cite{Nair-eff}.
(The reduced action for the parameters of the Lorentz transformations
for a configuration of nonzero vortex number
shows that the corresponding moment of inertia
is infinite because of the asymptotic behavior of the configuration.
 Therefore Lorentz transformations cannot be unitarily implemented.
This is similar to how nonabelian magnetic monopoles break global
color symmetry
 \cite{mon-color}.)
Thus the theory will allow for configurations or states corresponding to
a gas of 
$\mathbb{Z}_N$-vortices but with overall vortex number equal to zero.
In our considerations so far, all the
results were due to the geometry and topology of the
space $\A/\G_*$; specific configurations, which are at best a
subspace of points in $\A/\G_*$ of measure zero, have not played any significant role.
Nevertheless, since $\mathbb{Z}_N$ vortices
may serve to illuminate many aspects of the theory, see for example
\cite{ZN-rev}, this observation about the possibility of
their existence may still be of interest.
Clearly there are issues to be clarified regarding any
possible link with confinement viewed in terms of
the $\mathbb{Z}_N$ vortices.

\bigskip
I thank Dimitra Karabali for a careful reading of the manuscript and for
a number of useful comments.
I also thank Abhishek Agarwal for several useful comments.
This work was supported in part by the U.S. National Science
Foundation Grant No. PHY-2112729.

\newpage
\setcounter{section}{0}
\renewcommand\thesection{\Alph{section}}
\section{Conventions and Notations}\label{AppA}
\def\theequation{A\arabic{equation}}
\setcounter{equation}{0}

Summation over repeated indices is assumed. Greek letters
$\mu$, $\nu$, etc. are used to denote spacetime components,
taking values $0$, $1$, $2$, $3$ in $(3+1)$-dimensional spacetime,
and $0$, $1$, $2$ in $(2+1)$-dimensional spacetime.
The metric for flat Minkowski space is denoted by
$\eta_{\mu\nu}$; the contravariant version is denoted by
$\eta^{\mu\nu}$.
The components of $\eta_{\mu\nu}$ are given by
$\eta_{00} = 1$, $\eta_{ij} = - \delta_{ij}$, and $\eta_{0i}= \eta_{i 0} = 0$.

We will also use the abbreviation $\del_\mu = {\del \over \del x^\mu}$.
The scalar product of two vectors with components $A_\mu$
and $B_\nu$ will be written as $A\cdot B = \eta^{\mu\nu} A_\mu B_\nu
= A_0B_0 - A_i B_i$. In some cases, such as in writing
$e^{i p\cdot x}$ we often abbreviate the scalar product as
just $px \equiv p_0 x_0 - p_i x_i$.

The Levi-Civita symbol in three dimensions is
$\e^{ijk}$ which is totally antisymmetric under
exchange of any two indices and is normalized as
$\e^{123} = 1$. In 2+1 dimensions, we take
$\e^{012} = 1$.
$\e^{\mu\nu\alpha\beta}$ is defined in a
similar way, with $\e^{0123} = 1$.

The symbol $\del$ is also used to denote the boundary of a
spatial or spacetime region. Thus $\del V$ denotes the
boundary of the region $V$. Differential forms will be
used for certain discussion and have the usual expression
in terms of a coordinate basis. Thus if $B$ denotes a
differential $p$-form, it has the local coordinate expression
\beq
B = {1\over p!}B_{\mu_1 \mu_2 \cdots \mu_p} \, dx^{\mu_1}
\wedge dx^{\mu_2}\cdots \wedge dx^{\mu_p}
\label{A1}
\eeq
with the wedge symbol, as usual signifying the antisymmetrization of the
coordinate differentials.
 The symbol $d$ will be used for the exterior derivative of a differential form,
 \beq
 d B = {1\over p!} (\del_\mu B_{\mu_1 \mu_2 \cdots \mu_p} ) \, dx^\mu \wedge dx^{\mu_1}
\wedge dx^{\mu_2}\cdots \wedge dx^{\mu_p}
\label{A2}
\eeq

While a large part of the discussions will use flat space, there will
be occasions to discuss some curved manifolds. The appropriate
metric will be given as the occasion arises.
In contexts where there is no chance of confusion, we will omit
writing the wedge symbol. This is implied whenever
we are using differential forms.
Thus, for example, $\Tr (A d A)$ will stand for
\beq
\Tr (A dA ) = \Tr ( A_\mu \del_\nu A_\alpha) \,dx^\mu\wedge dx^\nu\wedge dx^\alpha
\label{A2a}
\eeq

For spinors, we will need the Dirac $\gamma$-matrices; these are defined by
\beq
\gamma^\mu \, \gamma^\nu + \gamma^\nu \, \gamma^\mu = 
2\, \eta^{\mu\nu} \, {\mathbb{1}}
\label{A3}
\eeq
In the case of nonchiral spinors in
four dimensions, $\gamma$'s can be realized as 
$4\times 4$ matrices. Thus
the $\mathbb{1}$ on the right hand side
of (\ref{A3}) denotes the 
$4\times 4$ identity matrix.
A specific choice for the $\gamma$'s is
\beq
\gamma^0 = \left( \begin{matrix}
1&0\\ 0&-1\\ \end{matrix} \right),
\hskip .2in
\gamma^i = \left( \begin{matrix} 
0& \sigma^i\\ -\sigma^i&0\\ \end{matrix} \right)
\label{A4}
\eeq
Each entry in the matrices in (\ref{A4}) is a $2\times 2$ matrix.
$\sigma^i$ are the Pauli matrices given as
\beq
\sigma^1 = \left( \begin{matrix}
0&1\\ 1&0\\ \end{matrix}\right), \hskip .2in
\sigma^2 = \left( \begin{matrix}
0&-i\\ i&0\\ \end{matrix}\right), \hskip .2in
\sigma^3 = \left( \begin{matrix}
1&0\\ 0&-1\\ \end{matrix}\right)
\label{A5}
\eeq

In three dimensions (or $2+1$ dimensions), the spinors are nonchiral
and the $\gamma$'s can be realized as
$2\times 2$ matrices. A specific choice is
$\gamma^0 = \sigma^3$, $\gamma^1 = i \sigma^1$, $\gamma^2
= i \sigma^2$.

For the group $SU(N)$, the generators of the Lie algebra in the fundamental
($N$-dimensional) representation are denoted by $t_a$, $a = 1, 2, \cdots, {\rm dim}G = N^2-1$.
They are taken to be normalized as $\Tr (t_a t_b ) = {\half} \delta_{ab}$.
The commutation rules are $[ t_a, t_b ] = i f_{abc} t_c$.
The corresponding generators in other representations are denoted by
$T_a$.
The quadratic Casimir operator has the value
$c_F = (N^2-1)/2 N$ for the fundamental representation and
$c_A = N$ for the adjoint representation.
For the case of $SU(2)$, the generators are given by
$t_a = {\half} \sigma_a$, with the Pauli matrices $\sigma_a$ as in
(\ref{A5}).

 \section{The topology and geometry of $\C$}\label{TopC}
 \def\theequation{B\arabic{equation}}
\setcounter{equation}{0}
 \setcounter{equation}{0}
 
 The space of gauge-invariant field configurations $\C$ can be identified 
 as $\A/\G_*$, where $\A$ is the space of gauge potentials (which are Lie-algebra valued 1-forms) and $\G_*$ is the set of gauge transformations, i.e.,
 group elements $g: \mathbb{R}^2 \rightarrow G$, with the condition that
 $g (x) \rightarrow \mathbb{1}$ as $\vert \vx \vert \rightarrow \infty$.
 One can think of $\A$ as a fiber bundle with $\G_*$ as
 the structure group and $\C = \A /\G_*$ as the base manifold,
\beq
\begin{array}{c c l l}
\G_*&\longrightarrow&\A&\\
&&\big\downarrow&\\
&&\C~=&\!\!\A/\G_*\\
\end{array}
\label{gorb1}
\eeq
 These are all infinite dimensional spaces. The topology and geometry of these
 spaces are clearly important for the study of gauge theories.
 The bundle structure (\ref{gorb1}) shows that, locally on a patch $U$
 of $\C$, we have the product structure $\A_U \sim \C_U \times \G_*$.
 On the patch $U$ we have a set of gauge potentials
 (corresponding to points in $\C$) with a fiber corresponding to
 the orbit of each such configuration by gauge transformations. One can specify the gauge-invariant degrees of freedom by choosing a representative configuration for each orbit; this is the process of gauge-fixing and
 is equivalent to specifying a section of the bundle. While this can be done on
 a local patch on $\C$, $\A$ as a $\G_*$-bundle is nontrivial and does not admit a global section. Thus there is no gauge fixing which is valid for all gauge potentials. This is the Gribov problem \cite{Grib}; for a more general discussion, see also
 \cite{Singer}.
 
 The nontriviality of the bundle can be seen by a slight variant of the {\it reductio ad absurdum} argument due to Singer \cite{Singer}.
 Assume that we can write $\A = \C \times \G_*$ globally, i.e., for all gauge potentials. As mentioned in section \ref{gaugeprinciple}, the space $\A$ is an affine space and all homotopy groups of $\A$ are trivial.
 If the condition $\A = \C \times \G_*$  is correct, then we must have trivial homotopy groups for $\C$ and for $\G_*$.
 Consider now $\Pi_1(\G_*)$. A typical element of this would be a sequence of group elements $g(x_1, x_2, \sigma )$ where $\sigma$ is a parameter 
(with values in $[0,1]$) along the loop of $\G_*$ elements.
 Specifically, we consider a loop starting and ending at the identity
 element, which implies that $g(x_1, x_2, \sigma ) \rightarrow
1$ at $\sigma = 0, 1$. From the definition of
$\G_*$ we also have
$g(x_1, x_2, \sigma ) \rightarrow 1$ as $\vert \vx\vert \rightarrow \infty$.
 Thus $g$ is a map from a cylinder (coordinatized by $x_1$, $x_2$, $\sigma$) to $G$, with
$g = 1$ on the boundary. Topologically, this is equivalent to
maps from a sphere to $G$,
 \beq
 g(x_1, x_2, \sigma ) : S^3 \rightarrow G
 \label{gorb2}
 \eeq
 The homotopy classes of such maps are classified by
 $\Pi_3 (G)$, implying $\Pi_1 (\G_*) = \Pi_3 (G)$.
This is nontrivial for all nonabelian Lie groups; for simple groups, we have
\beq
\Pi_3 (G ) = \begin{cases}
\mathbb{Z}\hskip .4in & {\text{ Any~simple}}~G,~{\text{except}}~ SO(4)\\
\mathbb{Z} \times \mathbb{Z} &~SO(4)\\
\end{cases}
\label{gorb3}
\eeq
The nontriviality of $\Pi_1(\G_*)$ shows that the initial assumption that
$\A = \C \times \G_*$ cannot be valid. This establishes the nontriviality of
the bundle (\ref{gorb1}). There is a Gribov problem for any nonabelian group.

Consider now a two-parameter family of gauge potentials of the form
\beq
A (x,1, x_2, \sigma , \tau ) =
\tau  \, A (x_1, x_2) +  (1- \tau) A^{g_1} (x_1, x_2, \sigma )
\label{gorb4}
\eeq
The $\sigma$-dependence of the potentials is due to
the $\sigma$-dependence of $g_1$ which we take
to be a nontrivial element of $\G_*$.
Taking $\sigma$, $\tau$ as coordinates in $\A$, this defines potentials over a
disc in $\A$. The potentials on the boundary of the disc are
$A$ at $\sigma = 0, 1$ and at $\tau = 1$, and
$A^{g_1}$ at $\tau = 0$.
Since these boundary values are all gauge-equivalent, they correspond to a single point in $\C$, so that the disc is a closed 2-surface in $\C$.
If this surface is contractible to a point in $\C$, the pre-image of that point
is a disc in $\A$ where all potentials inside are also gauge-equivalent to
$A$, of the form
$A^g$ with $g (x_1, x_2, \sigma, \tau)$ such that 
$g (x_1, x_2, \sigma, 0) = g_1 (x_1, x_2, \sigma)$
and $g (x_1, x_2, \sigma, 1) = 1$. If such a
$g (x_1, x_2, \sigma, \tau)$ were possible, it would give
a homotopy between
$g_1 (x_1, x_2, \sigma )$ and the identity. We know this is impossible
since $g_1$ is a nontrivial element of
$\Pi_3 (G)$. Therefore the conclusion is that the closed  2-surface in
$\C$ is not contractible. Rather than this long argument, we could also
have used exact homotopy sequence
\beq
\begin{array}{c c c c c c l}
\Pi_2 (\A) & \rightarrow&\Pi_2 (\C) &\rightarrow& \Pi_1(\G_*)&\rightarrow
&\Pi_1(\A)\\
0~& \rightarrow&~\Pi_2 (\C) &=& \Pi_1(\G_*)&\rightarrow
&0\\
\end{array}
\label{gorb5}
\eeq
to arrive at the same conclusion.
The implication of  the nontrivial nature of the bundle at the level
of using $\A$ is the Gribov problem and the impossibility of a global
section. At the level of directly using $\C$, it is manifest in the nontrivial topology of $\C$, the lowest dimensional such feature being
$\Pi_2 (\C) \neq 0$.

Our aim now is to construct an example of the
 set of configurations which form a noncontractible
two-surface, i.e., a nontrivial element of
$\Pi_2(\C)$. (This discussion follows \cite{KKN1}.)
The winding number, which we may take as
characterizing the element of $\Pi_2(\C)$
can be related to the instanton number of
a four-dimensional gauge theory. This can be seen as follows.
In addition to the homotopy group
$\Pi_2({\cal C})$ being nontrivial,
the second cohomology group of ${\cal C}$ is nontrivial as well. 
Thus there is 
a closed but not exact two-form on ${\cal C}$. 
In terms of the
potentials, the generating element of this
cohomology can be written as
\beq
\Omega ={1\over 4\pi} \int \tr (\delta A ~ \delta A)
\label{gorb6}
\eeq
Here $A$ is a one-form on the spatial manifold, $\delta$ denotes
the exterior derivative on $\A$. 
If we use $w$, $\bw$ to denote the coordinates along the
two-surface in $\C$, $\delta$ is given by
$\delta = dw \partial_w + d{\bar w}\partial_{\bar w}$. 
The integration in (\ref{gorb6}) is over the spatial manifold,
making $\Omega$ a two-form on $\C$.
The integral of $\Omega$ over the closed noncontractible two-surface in 
${\cal C}$ will give
a winding number $\nu$ by
$\int \Omega = 2\pi \nu $. 

The two-surface in
${\cal C}$
(with the coordinates  $w$, $\bw$) and the two-dimensional
spatial manifold can be considered together as
a four-dimensional space. 
The instanton number on this 4d-space
is given by
\beq
\nu = {1\over 8\pi ^2} \int \Tr ({\tilde F} ~{\tilde F}) 
\label{gorb7}
\eeq
where ${\tilde F} = (d+\delta ) {\tilde A} +{\tilde A}{\tilde A}$. 
The operator $(d + \delta)$ denotes the full exterior derivative on the four-dimensional space
and ${\tilde A}$ is the
four-dimensional gauge potential.
The 4d-potential can be constructed from the two-dimensional potential $A$
as ${\tilde A} = A+ A'$, where we take $A'$ to be given in terms
of $M$, $M^\dag$ by
\beq
A'=  - \partial_w M ~M^{-1} \, dw + (M^\dag)^{-1}\partial_{\bar w}M^\dag d{\bar w} 
\label{gorb8}
\eeq
While $A$ transforms as a connection under gauge transformations $g( x)$,
$A'$ is gauge-covariant since $g(x)$ does not depend on $w$, $\bw$.
In other words, $\delta g = 0$ for gauge transformations.
The field strength can be written out as
\beq
{\tilde F}  = F + F' + \delta A + D A', \hskip .2in
D A' = d A' + A A' + A' A
\label{gorb9}
\eeq
It is then easy to see that
\beq
\tr ({\tilde F} {\tilde F} ) =  \tr ( \delta A \,\delta A ) 
+ d \left[ \tr \left( A' \,D A' + 2\, \delta A \, A' \right) \right] + \delta\left[ \tr ( F \, A' )
\right]
\label{gorb10}
\eeq
In integrating this expression over the spatial manifold and the internal
closed two-surface, the terms which are total derivatives give zero.
(Notice that the integrands are gauge-invariant, so there is not
problem of the potentials being patchwise defined with
transitional gauge transformations on the overlap regions.
Therefore the total derivatives indeed integrate to zero.)
From the integral of (\ref{gorb10}), we see that
\beq
\nu = \int {\Omega \over 2 \pi} 
\label{gorb11}
\eeq
where $\Omega$ is as given in (\ref{gorb6}).
Using the expression for $A'$ from (\ref{gorb8}) and the parametrization
(\ref{par6}) for the spatial components, $\Omega$ takes the form
\beq
\Omega ={1\over 2\pi} \int \Tr \left[ \partial (H^{-1}\bdel H)\delta (H^{-1} {\bar \delta}
H)~+\partial (H^{-1}{\bar \delta} H) \delta (H^{-1}\bdel H)\right]
\label{gorb12}
\eeq

We can exploit this connection between the two-form
$\Omega$ on ${\cal C}$ and the instanton number to construct an example
of the noncontractible
two-surface of configurations.
Towards this, we write
the standard instanton in ${\mathbb{R}}^4$
using complex coordinates
and interpret one pair of complex coordinates as internal coordinates
parametrizing the two-surface in ${\cal C}$.
Explicitly this gives the expression
\beq
H= \exp (2f J^3) ~=\cosh 2 f ~+J^3 \sinh 2f 
\label{gorb13}
\eeq
Here $J^3 =\sigma \cdot n $ with $\sigma^a$, $a=1, 2, 3$, being the Pauli matrices and the unit
vector $n^a$ is
given by
\beq
n^a={1\over ({\bar z}z+{\bar w}w)}
\left ( {\bar z} w+{\bar w}z, ~i({\bar w}z -{\bar z}w),~ {\bar z}z-{\bar w}w \right)
\label{gorb14}
\eeq
The function $f$ is given by
\beq
f = {1\over 2} \log \left( {{\bar z}z+{\bar w}w +\mu^2 \over {{\bar z}z+{\bar w}w}}\right)
\label{gorb15}
\eeq
$\mu$ is a scale parameter and $(w, {\bar w})$ parametrize the two-surface
in ${\cal C}$. Using the formula
(\ref{gorb12}), it is easy to verify that
\beq
\nu =\int {\Omega \over 2\pi } = 3
\label{gorb16}
\eeq
for this set of configurations (\ref{gorb13}).
Therefore (\ref{gorb13}) does correspond to a noncontractible two-surface
in ${\cal C}$, although not the minimal one.

We have specified the configurations (\ref{gorb13}) directly in terms
of the gauge-invariant variable $H$, so there is no Gribov
problem {\it per se}. However, the existence of nontrivial elements in
$\Pi_2 (\C)$ means that we have to choose coordinate patches (in $\C$)
to specify the whole set of configurations in a nonsingular way.
This will be related to the freedom of the holomorphic transformations
mentioned earlier. We will illustrate this in our explicit example now.

Notice that, as ${\bar z}z \rightarrow \infty$,
$H\rightarrow 1$. Further, for almost all 
$w, {\bar w}$, $H$ is nonsingular; however, the particular
configuration at $w=0$ has a singularity
at the spatial point $z=0$. We can change the position of this singularity
by transformations of the type $H\rightarrow V H{\bar V}$, where $V$ is holomorphic in
$z$. Consider the configuration
for which $w = \bw = 0$; it is given by
\beqar
f &=& {1\over 2} \log ( \bz z + \mu^2 /\bz z)\nonumber\\
H &=& \exp ( 2 f \sigma^3 ) =
 \exp \left( \sigma^3 \left[ \log (\bz z+ \mu^2) - \log z -\log \bz \right]\right)
 \nonumber\\
 &=&\exp\left( {- \sigma^3 \log z}\right) \, \exp\left({ \sigma^3 \log (\bz z + \mu^2)}\right) 
 \, \exp\left({- \sigma^3 \log \bz }\right)
\label{gorb17}
\eeqar
Using $V = e^{\sigma^3 \log (z/ z- a) }$, we find
\beq
V \, H \, {\bar V} = \exp\left( \sigma^3 [ \log ( \bz z+ \mu^2 )
- \log (z-a) - \log (\bz - \ba ) ]\right)
\label{gorb18}
\eeq
We see that the singularity has been shifted from $z = 0$ to
$z = a$.

This tells us that, at least for configurations of the type given here, we can specify field configurations
by nonsingular formulae for 
$H$ in different coordinate patches with transition relations given by
transformations of the form
$H \rightarrow V (z) \, H\, {\bar V}(\bz )$. This shows the importance of the holomorphic invariance.

Since the singularity
in our example is at a point, namely at $w=0$, for this specific case, even if we simply use the formulae
(\ref{gorb13}-\ref{gorb15}) with the coordinate singularity, the effect on the quantum wave functions
is minimal. This is something that can be verified in terms
of the wave functions given later.
Also, as remarked earlier, the
WZW-action $S_{\rm wzw}(H)$ is invariant under transformations of the 
type $H\rightarrow V H{\bar V}$ and therefore
we do not expect any pathology for the wave function.
Explicitly, for the set of configurations (\ref{gorb13}), the WZW-action is given by
\beqar
{S}_{\rm wzw} (H) &=& {{ 5 \mu^2 +4w{\bar w}}\over {w{\bar w}+\mu^2}} ~-
{{ 3 \mu^2 +4 w{\bar w}}\over{\mu^2}} \log \left[ {{ \mu^2 +w{\bar w}} \over
{w {\bar w}}}\right] \nonumber\\
&=& 5 + 3 \log (w\bw )  + {\cal O}(w\bw )
\label{gorb19}
\eeqar
When $w\rightarrow 0$, 
$\exp ( 2 c_A S_{\rm wzw})$ vanishes as $(w{\bar w})^{6 c_A}$. 
The coordinate singularity does not
lead to difficulties, at least for this case.

Another interesting feature which we alluded to in the Introduction is about the compactness of $\C$. One can find configurations, i.e., points of
$\C$, which are separated by an arbitrarily large distance.
This can be illustrated by a simple example. Consider, in an
$SU(2)$ gauge theory, the configuration
\beq
A = (-it^3) ~in (z\bz )^{n-1} {(z d\bz - \bz dz) \over [1+(z\bz )^n]}
\label{gorb20}
\eeq
where $n$ is an integer.
This corresponds to the field strength
\beq
F = (-it^3) ~ (-4 n^2) {(z\bz )^{n-1}  \over [1+ ( z\bz )^n]^2} ~dx\wedge dy
\label{gorb21}
\eeq
We can estimate the distance of this configuration from another one, say
$A'$ as follows. First consider the Euclidean
distance two points on the orbits corresponding to $A$ and $A'$.
This is given by
\beqar
s^2 &=& - 2 \int  d^2x\,\Tr \left[ (A^{g_1} - A'^{g_2} )^2\right]\nonumber\\
&=& \int d^2x\, (A - A'^g )^a (A- A'^g)^a 
\label{gorb22}
\eeqar
where $g = g_2 g_1^{-1}$. The minimum distance between the two orbits
 is thus given by
minimizing this with respect to $g$. We can thus take
\beq
s^2_{\cal C} (A, A') = {\rm Inf}_g \int d^2x~ (A- A'^g )^a (A- A'^g)^a
\label{gorb23}
\eeq
The minimal distance of the orbit $A$ from the orbit of $A = 0$ is thus
\beqar
s^2_{\cal C} (A, 0) &=& {\rm Inf}_g \int d^2x~ (A- g^{-1} \del g )^a (A- g^{-1} \del g)^a\nonumber\\
&=&{\rm Inf}_g \int d^2x~ \left[
\left( A - {i (f d{\bar f} - {\bar f} df) \over 1+f {\bar f} } + d\vf \right)^2 +
4 {\del_i {\bar f} \del_i f \over (1+ f {\bar f})^2} \right]
\label{gorb24}
\eeqar
where we have parametrized $g$ as
\beq
g = {1\over \sqrt{1+ f {\bar f}}} \left[ \begin{matrix} 
1&f \\
-{\bar f} &1\\
\end{matrix}\right] 
~\left( \begin{matrix} 
e^{-i\vf /2} &0\\
0&e^{i\vf/2}\\
\end{matrix}
\right)
\label{gorb25}
\eeq
The last term in the brackets in (\ref{gorb24}) has a minimum given by
$8\pi Q[f]$, where $Q[f]$, which is an integer, is the topological charge of $f$, given by
\beq
Q[f] = {i\over 2\pi} \int d^2x~ \epsilon^{ij} { \del_i {\bar f} \,\del_j f \over (1+ f {\bar f})^2}
\label{gorb26}
\eeq
We see that the first term in brackets in (\ref{gorb24}) is minimized by the choice $f = z^n$, $\vf = 0$. Calculating $Q[f]$ for this case, we find
$s^2_\C (A, 0) \geq 8 \pi n$.
By taking $n$ large enough, we can get arbitrarily large distances.
This shows that the space $\C$ could have arbitrarily long ``spikes" like these;
compactness for $\C$ is not obtained.
One could then envisage constructing a wave function of arbitrarily long wavelength along such a spike
and this could lead to an infinitesimally small
eigenvalue for the kinetic term of the Hamiltonian. This is essentially
the counterargument to what Feynman was attempting to show.
However, the transverse measure of such spikes is important and the
total volume of $\C$ being the partition function for the hermitian
WZW theory shows that the transverse measure is almost zero,
in a regularized sense.
The zero-point fluctuations can then potentially lift the energy.
Presumably this is how the theory still leads to a
mass gap.\footnote{We may note that the measure calculation did not exist at the time Feynman worked on this, so the needed mathematics for him to complete the argument was not available.}
\section{Regularization}\label{AppC}
\def\theequation{C\arabic{equation}}
\def\thesubsection{C.\arabic{subsection}}
\setcounter{equation}{0}
\subsection{The regularized form of the operators}
We will now go over some of the issues related to defining the regularized form of the operators for the kinetic and potential energies, and the Hamiltonian. A good regularization procedure must preserve gauge invariance. With our choice of variables, it is also important to preserve the holomorphic invariance. This property was discussed at the end of 
section \ref{Par}. The matrices 
$(M, M^\dagger )$ and $(M', M'^\dagger )$ where
$M' = M {\bar V}$, $M'^\dagger = V M^\dagger$
will give the same potentials $(A, \bA)$. Here $V$ is a holomoprhic function of the coordinates, ${\bar V}$ is an antiholomoprhic function.
Generally, we need this freedom in how we define $M$ and
$M^\dagger$, so that configurations can be represented
on various coordinate patches without singularities.
The calculations we do will involve the Green's functions
for $D = \del +A$ and $\bD = \bdel + \bA$. These were introduced in section
\ref{Vol} in the form
\beq
D^{-1}(x, y) = M(x) G(x, y) M^{-1}(y), \hskip .2in
\bD^{-1} (x, y) = M^{\dagger -1} (x) \bG (x, y) M^\dagger (y)
\label{B1}
\eeq
where $G$ and $\bG$ are the Green's functions for
$\del$ and $\bdel$, respectively. 
For a particular coordinate patch, we can take these to be
\beq
G(x, y) = {1\over \pi (\bx - \by )}, \hskip .2in
\bG(x, y) = {1\over \pi (x-y) }
\label{B2}
\eeq

Consider the construction of $D^{-1}(x, y)$ and $\bD^{-1} (x, y)$
using $M'$ and $M'^\dagger$. These Green's functions are unchanged
if we define
\beq
G'(x, y) = {\bar V}^{-1} (x) \, G(x, y) \, {\bar V} (y),
\hskip .2in
\bG' (x, y) = V(x) \, \bG(x, y) V^{-1} (y)
\label{B3}
\eeq
Notice that these will still satisfy the required equations
\beq
\del_x G (x, y) = \bdel_x  \bG(x, y) = \delta^{(2)}(x-y)
\label{B4}
\eeq
We see that the use of different forms for the matrices $M$, $M^\dagger$ must be
accompanied by the use of different definitions for
the Green's functions $G$ and $\bG$.

Our aim is to use a point-splitting regularization
for the Green's functions $G(x, y)$ and $\bG (x, y)$
which preserves the transformation property
(\ref{B3}). This can be done by use of the Gaussian approximation to the 
Dirac $\delta$-function given in (\ref{ham10a}),
namely,
\beq
\sigma (\vx,\vy;\e) ={{e^{-|\vx-\vy|^2/\e}} \over {\pi \e}}
\label{B5}
\eeq
Based on this we define
\beqar
\G (\vx,\vy) &=& \int_u G (\vx,\vu) \s (\vu,\vy;\e) K^{-1}(y,\bu) K (y,\by) \nonumber\\ 
\bar{\G} (\vx,\vy) &=& \int_u \bG (\vx,\vu) \s (\vu,\vy;\e) K(u,\by) K^{-1} (y,\by) 
\label{B6}
\eeqar
Since we will be using the matrices $M$, $M^\dagger$, $H$ in the adjoint representation for most of the calculations, we have given these expressions in the appropriate form. 
Here $K_{ab} = 2 \Tr (t_a H t_b H^{-1} )$ is the same
as the matrix $H$ but in the adjoint representation.
It is easy enough to verify that $\G$ and $\bar{\G}$ have the same
transformation as $G$ and $\bG$.
By expanding the $K$'s in (\ref{B6}),
it is possible to carry out the integration and reduce these to the form
\beqar
\G _{ma} (\vx,\vy)  &=&  G (\vx,\vy) [ \d _{ma} - e^{-|\vx-\vy|^2/\e} \bigl(
K^{-1}(y,\bx) K (y, \by) \bigr) _{ma}]\nonumber\\
\bar{\G} _{ma} (\vx,\vy)  &=&  \bG (\vx,\vy) [ \d _{ma} - e^{-|\vx-\vy|^2/\e} \bigl(
K(x,\by) K^{-1} (y, \by) \bigr) _{ma}] 
\label{B7}
\eeqar
The coincident point limit of $\bar\G$ can be read off from these
as
\beq
{\bar \G} (x, x) = - {\del K K^{-1} \over \pi} 
\label{B8}
\eeq
Correspondingly, we have
\beq
\bD ^{-1} (\vx,\vx) _{\rm reg}  =  -{1 \over \pi} M^{\dag -1} (\vx) (\del K K^{-1})
M^{\dag} (\vx) 
= {1 \over \pi} (A- M^{\dag -1} \del M^{\dag}) (\vx) 
\label{B9}
\eeq
This reproduces the results in equations
(\ref{vol19}) and (\ref{vol20}) used in the calculation of the volume element
for $\C$.

Turning to the kinetic energy operator, we start with the form given in
(\ref{ham7}), 
\beq
T = {e^2 \over 4}  \int_x e^{-2c_A S_{\rm wzw}  (H)} \Bigl[ {\bar
\G}\bp_a(\vx) K_{ab}(\vx) \,e^{2c_A S_{\rm wzw}  (H)} \G p_b(\vx) +
 \G p_a(\vx) K_{ba}(\vx)\,
e^{2c_A S_{\rm wzw} (H)} {\bar \G}\bp_b(\vx)\Bigr]
\label{B10}
\eeq
We have put in the regularized form of the Green's functions.
In this expression, we use the abbreviation
\beq
\G p_b(\vx) =
\int_u \G_{bc}(\vx, \vu) p_c(\vu), \hskip .2in
{\bar \G}\bp_a (\vx) = \int_u {\bar \G}_{ac}(\vx, \vu) \bp_c (\vu), ~~{\rm etc.}
\label{B11}
\eeq
In (\ref{B10}), in moving $p_a$ and $\bp_a$ to the right,
we encounter the commutators
\beq
{\bar\G} [\bp_a(\vx) , K_{ab}(\vx) \,e^{2c_A S_{\rm wzw}  (H)}],
\hskip .2in
\G [p_a(\vx), K_{ba}(\vx)\,
e^{2c_A S_{\rm wzw} (H)}]
\label{B12}
\eeq
These involve the coincident point limit of the Green's functions.
Using (\ref{B7}) we can calculate the commutators and see that they are zero
as $\e \rightarrow 0$; see 
\cite{KKN3} for more details. The expression for $T$ can then be brought to the form
\beqar
T&=& {e^2 \over 2} \int \Pi_{rs} (\vu,\vv) \bp _r (\vu) p_s (\vv)
\label{B13}\\ 
\Pi_{rs} (\vu,\vv) &=& \int_x \bar{\G} _{ar} (\vx,\vu) K_{ab}(\vx) \G _{bs} (\vx,\vv) \nonumber
\eeqar
This is the regularized version of (\ref{ham6}), thus establishing its equivalence with (\ref{ham7}) as well.

The action of $p$ and $\bp$ on the current $J$ is given by
\beqar
[p_s (\vv),~J_a (\vz)] & =& -i {c_A \over \pi} K_{as} (\vz) \del _z \d
(\vz,\vv) \nonumber\\
{} [\bp _r (\vu),~ J_b (\vw)] & = & -i ({\cal{D}} _w) _{br} \d (\vw-\vu),
\hskip .2in ({\cal{D}}_{w})_{ab}  = {c_A\over \pi}\partial_w \delta_{ab} +if_{abc}J_c (\vw)
\label{B14}
\eeqar
We can then use the form of $T$ from
(\ref{B13}) and work out its action on a functional of the currents;
basically this involves using the chain rule and the commutators
(\ref{B14}). The result is
\beq
 T~\Psi (J) = m \left[ \int_z \omega_a(\vz){\delta \over \delta J_a(\vz)}~+\int_{z,w}
\Omega_{ab}(\vz,\vw) {\delta \over \delta J_a(\vz) }{\delta \over \delta J_b(\vw)}\right]
\Psi(J)\label{B15}
\eeq
where
\beqar
\omega_a(\vz)&=& -i f_{arm} \bigl[ \del _z \Pi_{rs} (\vu,\vz) \bigr]_{\vu \rightarrow \vz}
~K^{-1}_{sm} (\vz) = i f_{arm} \Lambda _{rm} (\vu,\vz)  {\big |}_{\vu
\rightarrow \vz}\nonumber\\
\Omega_{ab}(\vz,\vw)&=& -\left[ \left[{c_A\over \pi}\partial_w \delta_{br}  +if_{brm}J_m
(\vw)\right] ~\partial_z \Pi_{rs} (\vw,\vz) \right] K^{-1}_{sa} (\vz)\nonumber\\
&=&
{\cal{D}} _{w~br} \Lambda_{ra} (\vw,\vz) 
\label{B16}\\
\Lambda _{ra} (\vw,\vz) & =& -(\del _z \Pi_{rs} (\vw,\vz)) K^{-1}_{sa} (\vz) 
\label{B17}
\eeqar
Using $\Pi_{rs} (\vu,\vv)$ from (\ref{B13}), the
expression for $\Lambda _{ra}$ can be written out as
\beqar
\Lambda _{ra} (\vw,\vz) &=&\int _x \bar{\G} _{mr} (\vx,\vw) G(\vx,\vz)
e^{-|\vx-\vz|^2/\e} \Bigl[  {{\bx - \bz} \over \e} K (x, \bx) K^{-1} (z,\bx)\nonumber\\
&& \hskip .2in+
K(x,\bx) \del_z ( K^{-1} (z,\bx)K(z,\bz)) K^{-1}(z,\bz) \Bigr] _{ma} 
\label{B18}
\eeqar
Because of the exponential $e^{-|\vx-\vz|^2/\e}$, the region
$|\vx-\vz|^2 {\lesssim} ~\e$ is what is relevant for
$\omega_a (\vz)$. Expanding around $z$, we get
\beq
\omega_a(\vz) = J_a (\vz) + \O (\e) 
\label{B19}
\eeq
If we use the expression (\ref{B7}) for $\bar{\G} _{mr} (\vx,\vw)$ in (\ref{B18}), the expression for $\Lambda _{ra}$ will split into four terms.
One can expand the integrands in powers of $x-w$, $\bx - \bw$ and carry out the $x$-integration to generate an expansion in powers of $\e$.
We then find
\beqar
\Lambda _{ra} (\vw,\vz) &=& {1 \over {\pi (z-w)}} \bigl[ \d _{ra} - \bigl( K(\vw) K^{-1}
( z, \bw )\bigr) _{ra} e^{- \vert z-w\vert^2 /2\e}  \bigr] \nonumber\\
&&+ ( {\text{terms of higher order in}~\e~\text{or}~(z-w),~(\bz - \bw)})
\nonumber\\
&\equiv & ~\bar{\G}' _{ra} (\vz,\vw)+ \cdots
\label{B20}
\eeqar
Here $\bar{\G} '$ is the transpose of $\bar{\G}$
 with $\e$ replaced by $2\e$.  
We can use this expression in (\ref{B16}) for
$\Omega_{ab}(\vz, \vw )$
and write the kinetic energy operator as
 \beqar
  T \Psi (J) &=& m \left[ \int J_a (\vz) {\d \over {\d J_a (\vz)}} + \int \bigl(
{\cal{D}} _w \bar{\G}' (\vz,\vw) \bigr) _{ab} {\d \over {\d J_a (\vw)}} {\d \over {\d
J_b (\vz)}} \right]  \Psi (J) + \O (\e) \nonumber\\
&=& m \int _{z,w}\left[  \bdel J_a (\vw) \bar{G} (\vz,\vw) {\d \over {\d
J_a (\vz)}} +  \bigl( {\cal{D}} _w \bar{\G}' (\vz,\vw) \bigr) _{ab} {\d \over {\d J_a
(\vw)}} {\d \over {\d J_b (\vz)}} \right]  \Psi (J) + \O (\e) \nonumber\\
\label{B21}
\eeqar
  It is easy to see that as
$\e \rightarrow 0$, $ \Lambda (\vw,\vz) \rightarrow \bar{G} (\vz,\vw)$.
The first line of (\ref{B21}) then reproduces the expression
(\ref{ham10}) in text.
In simplifying (\ref{B16}) for $\omega_a$ to get
(\ref{B19}), we have cancelled powers of
$(z-w)$ against $\bar{G} (\vz,\vw)$.
This can lead to expressions which seemingly do not have the
holomorphic invariance. The second line of
(\ref{B21}) shows $T$ in a manifestly holomorphic invariant form.

As for the potential energy, we can do a point-splitting
and write
\beqar
V_{(\lambda')} &=& {\pi \over {m c_A}} \int_x : \bdel J_a (\vx) \bdel J_a  (\vx) : \nonumber\\
&=& {\pi \over {m c_A}} \Bigl[ \int_{x,y} \s (\vx,\vy;\lambda' ) \bdel J_a (\vx) (K(x,\by)
K^{-1} (y,\by))_{ab} \bdel J_b (\vy) - {{c_A {\rm dim} G} \over {\pi^2
\lambda'^{2}}} \Bigr]
\label{B22}
\eeqar
{\it A priori} we have the freedom to choose a different value $\lambda'$, rather than $\e$, for the width
of $\s (\vx,\vy;\lambda' )$, so we have displayed the expression
for such a choice. The action of $T$ on $V$ is important for
solving the Schr\"odinger equation. Since we have regularized 
all operators, it is straightforward to work this out
and obtain
\beq
T_{(\e )} ~V_{(\lambda' )} = 2m \left[ 1 + \half \log (\lambda' /2\e )\right] ~V_{(\lambda' )}~+\cdots
\label{B23}
\eeq
Since it is different from $\lambda'$,
we display the regularization parameter $\e$ for
$T$ as a subscript.

To understand how $\e$ and $\lambda'$ may be related, we
first note that since
we are using the $A_0 = 0$ gauge, the Coulomb potential at short distances will be obtained from the action of the kinetic term on wave functions. In 2+1 dimensions, the Coulomb potential is logarithmic
and so a subtraction point  needs to be chosen to define the
zero of the potential. The freedom of choosing this point
is also a reflection of the fact that the
kinetic operator is scale invariant as $\e \rightarrow 0$.
In terms of the regularized version, this means that we can define
\beqar
T_{(\lambda )}&=& T_{(\e )} + { e^2 \over 2} \log ({2\e / \lambda })~ {\cal Q}\nonumber\\
 {\cal Q}&=& \e \int \s (\vu,\vv;\e) K_{rs} (u, \bv)~\Bigl( \bp_r (\vu) -i \bdel J_r(\vu )\Bigr)~ p_s(\vv)
 \label{B24}
\eeqar
Acting on $V$, we now get
\beq
T_{(\lambda )} V_{(\lambda' )} = 2m \left[ 1 + \half \log (\lambda' /\lambda )\right] ~V_{(\lambda')} ~+\cdots
\label{B25}
\eeq
The addition of ${\cal Q}$ to $T$ can be interpreted as follows.
Recall that, in covariant perturbation theory, the addition of local counterterms is equivalent to a change of regularization.
The addition of ${\cal Q}$ in (\ref{B24}) may be viewed as
the analogous procedure for the Hamiltonian formulation.
\subsection{A Lorentz-invariance argument}
We have regularized $T$ and $V$ with different parameters,
$\lambda$ (or $\e$) and $\lambda'$. 
Each expression will thus involve fields with modes of
momenta larger than $1/\sqrt{\lambda}$ and $1/\sqrt{\lambda'}$,
respectively.
Being short-distance regularization parameters, we need 
$\lambda$, $\lambda'$ to be much smaller than any physical scales
such as $1/(e^2)^2$.

The key missing ingredient in treating $T$ and $V$ separately is
Lorentz invariance. Recall that under a Lorentz transformation
corresponding to velocity $v_i$, the electric and magnetic fields
transform as
\beq
\delta E_i \approx -\epsilon_{ij} v_j B, 
\hskip .2in \delta B \approx \epsilon_{ij} v_i  E_j, \hskip .3in
\vert v\vert \ll 1
\label{B26}
\eeq
The Hamiltonian is the integral of the energy density $T_{00}$.
The momentum and stress densities are given by
\beq
T_{0i} = \e_{ij} E_j B, \hskip .2in
T_{ij} = - E_i E_j + \delta_{ij} T_{00}
\label{B27}
\eeq
Under a Lorentz transformation, we have
\beq
\delta T_{0i} = v_i T_{00} + v_j T_{ij} 
\label{B28}
\eeq
If we use 
 the transformation (\ref{B26}) for the fields,
\beq
\delta T_{0i} = \delta (\e_{ij} E_j B) = v_i (E^2 + B^2) - v_k E_k \, E_i
\label{B29}
\eeq
This is in agreement with (\ref{B28}). But it also
shows that if we regularize the momentum
$P_i = \int \e_{ij} E_j B$ with a parameter
$\lambda$, Lorentz invariance will require that
both terms in the energy (on the right hand side
of (\ref{B29}) should be regularized with the
same parameter.
(A variant of this argument was given in
 \cite{KN-robust}.)
Thus, although {\it a priori} we could use different
regularizations for $T$ and $V$,
consistency with Lorentz symmetry requires
$\lambda = \lambda'$ (with $1/\sqrt{\lambda} \gg e^2$).
In this case, (\ref{B25}) simplifies to
\beq
T_{(\lambda )} V_{(\lambda)} = 2m ~V_{(\lambda)} ~+\cdots
\label{B30}
\eeq
At this point, we can also go back and use $\e$ in place of $\lambda$ as
we did for the potential energy in (\ref{ham25}).

Another version of this argument, where the action of $T$ on $V$ and the
vacuum wave function are considered, is given in \cite{KN-robust}.
We may note that the issue of regularization and how it relates to
Lorentz invariance is somewhat tricky. For a discussion with
different points of view, see \cite{mansfield, krug}.

\section{Corrections to string tension}\label{AppD}
\def\theequation{D\arabic{equation}}
\setcounter{equation}{0}
In this Appendix, we will calculate the first set of corrections
to the formula (\ref{Res10}) for the string tension using the first order corrections to the vacuum wave function obtained in section \ref{SchE},
equations (\ref{SchE10})-(\ref{SchE16}).
From the recursive solution of the Schr\"odinger equation, 
the correction to the quadratic kernel in the wave function is
\beqar
e^2~f_2^{(2)}(q) &=& \frac{m}{E_q}\int \frac{d^2 k}{32\pi}\ \left(\frac{1}{\bar k}\ g^{(3)}(q,k,-k-q)\ + \frac{k}{2 \bar k}\ g^{(4)}(q,k;-q,-k)\right)\nonumber\\
&\approx&  \frac{\bar q^2}{2 m}\,(1.1308) +\ldots
\label{Corr1}
\eeqar
In the second line we give the lowest order (quadratic in $q^2$) term, as this is what is relevant for the calculation of the string tension.
Seemingly, this is a $113\%$ correction, but
there are important additional terms which should be included.
In calculating the vacuum expectation value of an operator
as $\la {\cal O}\ra = \int \Psi^* \Psi \, {\cal O}$, we have to do a functional integration over $H$, so this can be viewed as a two-dimensional
field theory. It is then clear that there are loop corrections, in the 2d field theory sense, to the quadratic kernel.
So we start with a procedure for simplifying and systematizing
these contributions. (The interpretation of $\la {\cal O}\ra = \int \Psi^* \Psi \, {\cal O}$ as equivalent to calculating the functional integral of a 2d field theory was also used in the calculation of the string tension
in section \ref{Res}.)

Since the measure of integration has the WZW action, our first step will be
to transform the functional integration over $\Psi_0^* \Psi_0 = e^\F$ into the integration over a two-dimensional chiral boson field $\vf, ~{\bar \vf}$.
(Although we use the same letter, this is not the $\vf$ we used in parametrizing $H$ as  $e^{t_a\vf^a}$.) 
The key point is that $\F$ is given in terms of currents, so
consider the calculation of the current correlators in just the hermitian WZW theory. We can write
\beqar
\frac{1}{{\cal Z}}\! \int d\mu (H) e^{2c_A S_{\rm wzw}(H) } e^{-{c_A\over \pi}
\int {\bar C}^a (\del H H^{-1})^a}
&=&\frac{1}{{\cal Z}}\! \int d\mu (H) e^{2c_A S_{\rm wzw}(UH) -2 c_A S_{\rm wzw}(U)}
\nonumber\\
&=&\displaystyle e^{-2 c_A S_{\rm wzw}(U)}
\label{Corr2}
\eeqar
where ${\bar C} = U^{-1}\bdel U$ and
we have used the Polyakov-Wiegmann identity 
\beq
S_{\rm wzw} (H) - {1\over \pi} \int \Tr ({\bar C} \del H H^{-1})
= S_{\rm wzw}(UH) ~-~ S_{\rm wzw} (U)
\label{Corr3}
\eeq
(Here ${\cal Z}$ is just a normalization factor; it is the partition function of the
hermitian WZW theory.)
Since $\exp (-2c_A S_{\rm wzw}(U))$ is the inverse of the chiral Dirac determinant in two dimensions, we can use the formula
\beq
\exp (-2c_A S_{\rm wzw}(U) ) =
\int [d\vf d{\bar \vf}]~ e^{-\int {\bar \vf} ( \bdel +{\bar C} )\vf}
\label{Corr4}
\eeq
The complex boson field $\vf$ transforms in the adjoint representation of $SU(N)$.
For the YM case, including $\Psi_0^* \Psi_0$, we have
\beqar
\la {\cal O}\ra &=& \int d\mu (H) \, e^{2 c_A S_{\rm wzw}(H)} 
e^{\F(J)} \, {\cal O}(J)\nonumber\\
&=&
\left[ {\cal O}({\hat J}) ~ e^{F({\hat J})}\right]~ \int d\mu (H)\, e^{2c_A S_{\rm wzw}(H) }\, e^{-{c_A\over \pi}
\int {\bar C}^a (\del H H^{-1})^a}\biggr]_{{\bar C} =0}\nonumber\\
&=& \int [d\vf d{\bar \vf}]~  e^{-S(\vf )}~{\cal O}(\sqrt{2\pi /mc_A} ~\bvf t^a \vf )
\label{Corr5}
\eeqar
where ${\hat J}^a = -\sqrt{2\pi /mc_A} ~{\delta \over \delta {\bar C}^a}$
and, after introducing the representation (\ref{Corr4}), 
we have evaluated the action of the ${\hat J}^a$'s and set ${\bar C}$ to zero.
The action $S(\vf)$ is given by
\beq
S(\vf ) = \int \bvf \bdel \vf - \F(\sqrt{2\pi /mc_A} ~\bvf t^a \vf )
\label{Corr6}
\eeq
There is a correction to be made to this formula once we have
$\F$ which will introduce additional interactions 
for the $\vf$-field. If we think of this as a 2d field theory, it is easy to see
that we will need renormalization constants
($Z$-factors) for the chiral boson action. The representation of the
determinant which is applicable in the presence of interactions is
thus
\beq
\exp (-2c_A S_{\rm wzw}(U) ) =
\int [d\vf d{\bar \vf}]~ \exp\left[ {-\int {\bar \vf} (Z_2 \bdel + Z_1 {\bar C} )\vf}
\right]
\label{Corr7}
\eeq
For the expectation values, we still get the formula
(\ref{Corr5}), but now with the action
\beq
S(\vf ) = \int ~(Z_2 \bvf \bdel \vf + Z_1 \bvf {\bar C}\vf )~-~ \F(Z_1\sqrt{2\pi /mc_A} ~\bvf t^a \vf )\label{Corr8}
\eeq
(We can set ${\bar C}$ to zero at the end, once the renormalization constants
$Z_1, Z_2$ have been calculated; see \cite{KNY} for more details.)
In this representation in terms of $\vf$, effectively, the current
$J^a$ is replaced by $Z_1\sqrt{2\pi /mc_A} ~\bvf t^a \vf$.
The function $\F(Z_1\sqrt{2\pi /mc_A} ~\bvf t^a \vf )$ contains vertices, $\F^{(2)}$ with two currents
(quartic in $\vf$, ${\bar\vf}$), $\F^{(3)}$ with three currents, etc.
 For example, we may diagrammatically represent $\F^{(2)}$, with two $\vf$'s and two ${\bar\vf}$'s, as
\begin{center}
\begin{minipage}{8.5cm}
{\Large$\F^{(2)}= {2\pi \over mc_A}\int ({\bar\vf}t^a\vf)_x ~f^{(2)} (x, y)({\bar\vf}t^a\vf)_y ~= $}\\
\end{minipage}
\hskip .2in
\begin{minipage}{5cm}
\includegraphics[height = .3\textwidth, width=.6\textwidth]{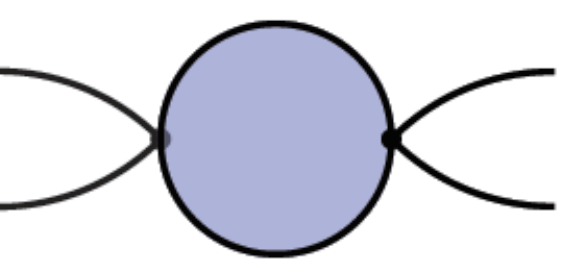}\\
\end{minipage}
\end{center}
The corrections to $\F^{(2)}$, which is what we are interested in, may be viewed as loop corrections to the quartic vertex in this two-dimensional field theory.

In calculating the corrections to $\F^{(2)}$, the vertices
$\F^{(3)}$, $\F^{(4)}$, etc., can be included perturbatively since they carry powers of $e$. 
However, the lowest term in the vertex $\F^{(2)}$, corresponding to
$f^{(2)}_{0a_1a_2}(x_1, x_2)$, has no powers of $e$ and hence its contributions will have to be included to all orders and
summed up. 
The result for the current-current correlator is
\beq
\la \bvf t^a \vf (x)~ \bvf t^b \vf (y)\ra = 
\delta^{ab} {c_A \over \pi} \int {d^2k \over (2\pi )^2} e^{ik(x-y)}~
 {k\over {\bar k}}~ {\left( {m\over E_k}\right)}
 \label{Corr9}
 \eeq
Here $E_k = \sqrt{k^2 +m^2}$; the $(m/E_k)$ factor arises from the summation of corrections due to $\F^{(2)}_0$, 
shown diagrammatically in Fig.\,\ref{Corr-graph1}.
\begin{figure}[!t]
\begin{center}
\scalebox{.95}{\includegraphics[height = .23\textwidth, width=.9\textwidth]{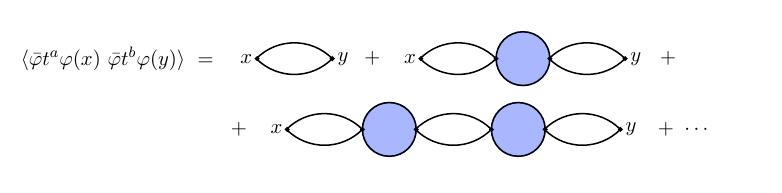}}
\caption{The current-current correlator including all contributions from $\F_0^{(2)}$}
\label{Corr-graph1}
\end{center}
\end{figure}
Any vertex can acquire a series of corrections from $\F_0^{(2)}$, so that we may consider an effective vertex
\beq
\bvf t^a \vf (x)\Bigr]_{\rm eff} =
 \int {d^2k\over (2\pi )^2} 
 e^{ik(x-z)} ~{m\over E_k} ~ (\bvf t^a \vf )(z)
 \label{Corr10}
 \eeq
 This is shown diagrammatically in Fig.\,\ref{Corr-graph2}.
 As a result, all corrections acquire powers of $(m/E_k)$ in the integrands
and in fact, we can classify contributions in powers of $(m/E_k)$.
Since $(m/E_k) \leq 1$, the numerical values will decrease as we
go down the series.
 \vskip .1in
\begin{figure}[!b]
\begin{center}
\scalebox{.95}{\includegraphics{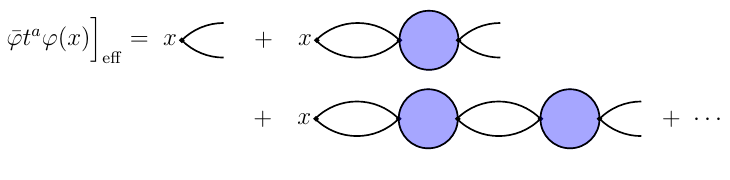}}
\vskip .1in
\caption{The effective current vertex}
\label{Corr-graph2}
\end{center}
\end{figure}

The basic strategy for calculating corrections
may then be summarized as follows:
\begin{enumerate}
\item Construct loop diagrams  generated by $\F^{(3)}$ (3 factors of ${\bar\vf}t^a \vf$) and $\F^{(4)}$
(4 factors of ${\bar\vf}t^a \vf$).
\item They can have arbitrary insertions of $\F_0^{(2)}$'s, leading to a factor of $(m/E_k)$, as in Fig.\,\ref{Corr-graph3}.
\item Sum up $\F_0^{(2)}$ insertions in all diagrams (of order $e^2$) generated by $\F^{(3)}$ and $\F^{(4)}$.
\item Classify and group these by the number of factors of $(m/E_k)$.
\end{enumerate}
\begin{figure}[!t]
\begin{center}
\includegraphics[height = .13\textwidth, width=.8\textwidth]{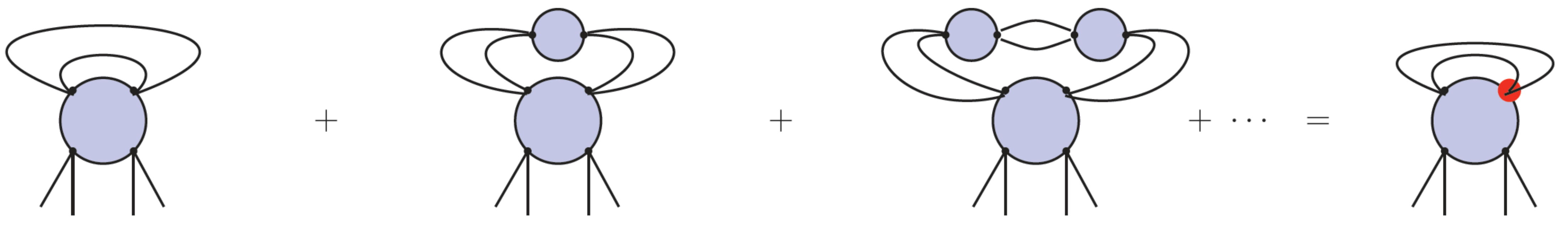}\\
\caption{Corrections from $F_0^{(2)}$ summed up as a factor of $m/E_k$ (shaded circle at vertex) and sample renormalization diagrams}
\label{Corr-graph3}
\end{center}
\end{figure}
There will be corrections generated to the terms
$\bvf \bdel \vf$ and $\bvf {\bar C}\vf $ in the action; these are 
renormalization effects due to
$\F_0^{(2)}$. These have to be cancelled by $Z_1$ $Z_2$ factors.
They are discussed in more detail in \cite{KNY}.

We have calculated corrections to {order $e^2$} and up to $4$ 
powers of $(m/E_k)$.
Denoting the factors of $(m/E_k)$ by shaded circles at the vertices, the corrections
to the low momentum limit of $f^{(2)}$ may summarized as in Fig.\,\ref{Corr-graph4}.
We show the coefficients of ${\bar q}^2/2m$, for small $q, ~{\bar q}$
for each diagram. 
\begin{figure}[!t]
\begin{center}
\begin{tabular}{c c c c}
\scalebox{.78}{\includegraphics[height = .13\textwidth, width=.13\textwidth]{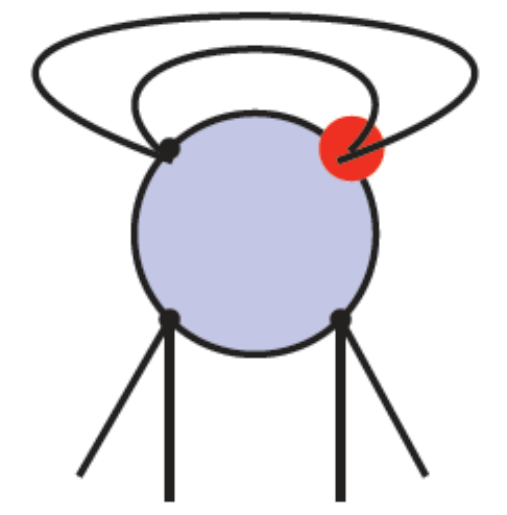}}&\scalebox{.85}{\includegraphics[height = .13\textwidth, width=.13\textwidth]{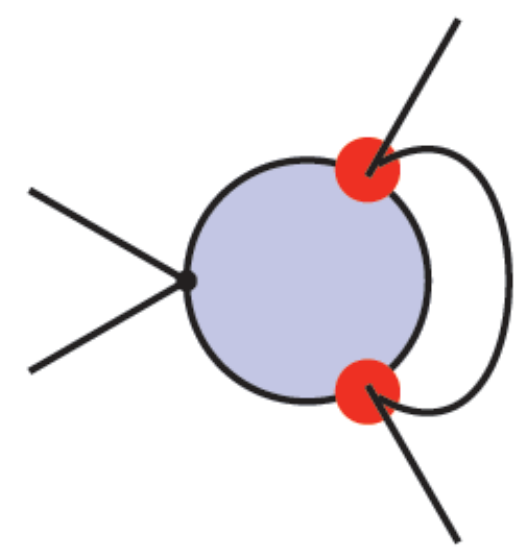}}&
\scalebox{.75}{\includegraphics[height = .11\textwidth, width=.24\textwidth]{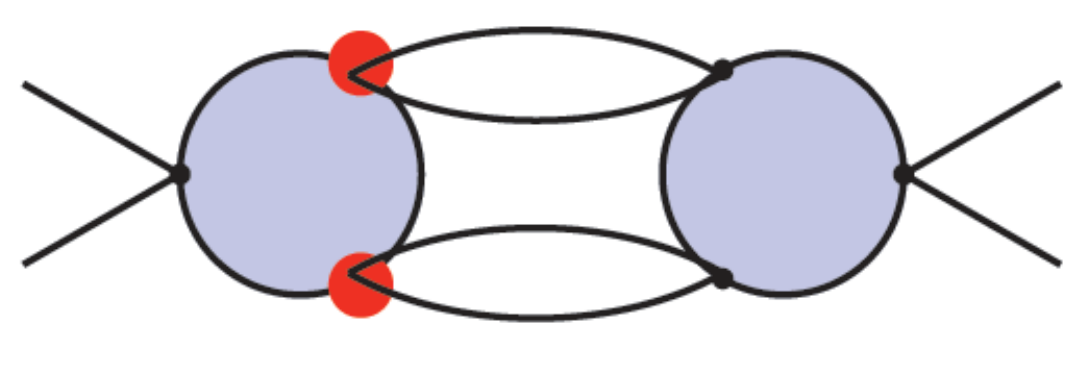}}&\scalebox{.75}{\includegraphics[height = .13\textwidth, width=.13\textwidth]{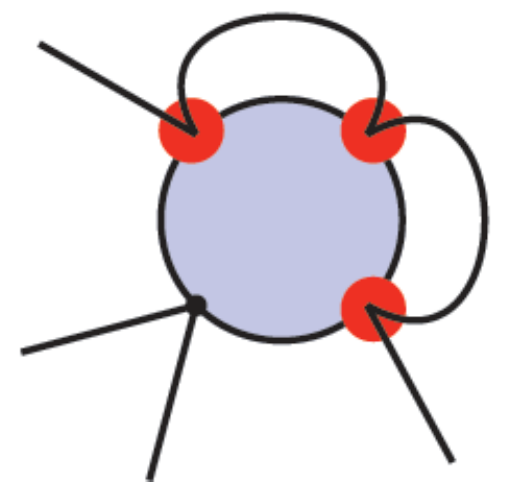}}\\
~~~~~~~$-0.58118 $~~~~~~~&
~~~~~~~$-0.47835$~~~~~~~&
~~~~~~~$0.20169$~~~~~~~&~~~~~~~$-0.23569$~~~~~~~\\
\end{tabular}\\
\vskip .1in
\begin{tabular}{c c c}
\scalebox{1.1}{\includegraphics[height = .06\textwidth, width=.18\textwidth]{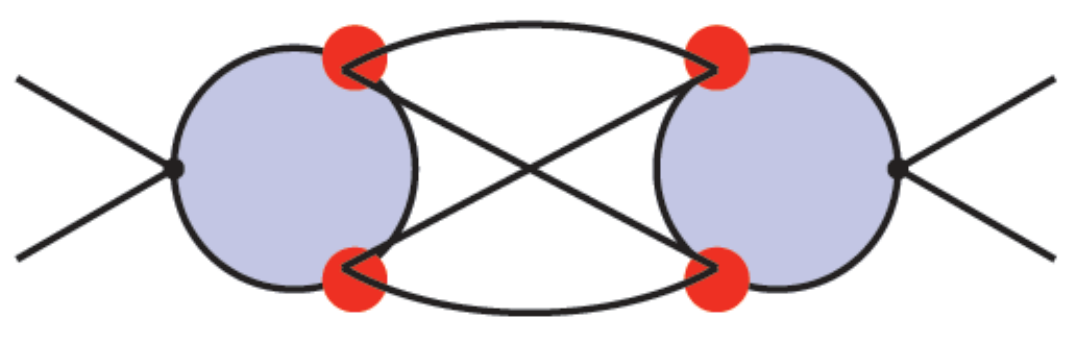}}~~~~~&
\scalebox{1.1}{\includegraphics[height = .07\textwidth, width=.2\textwidth]{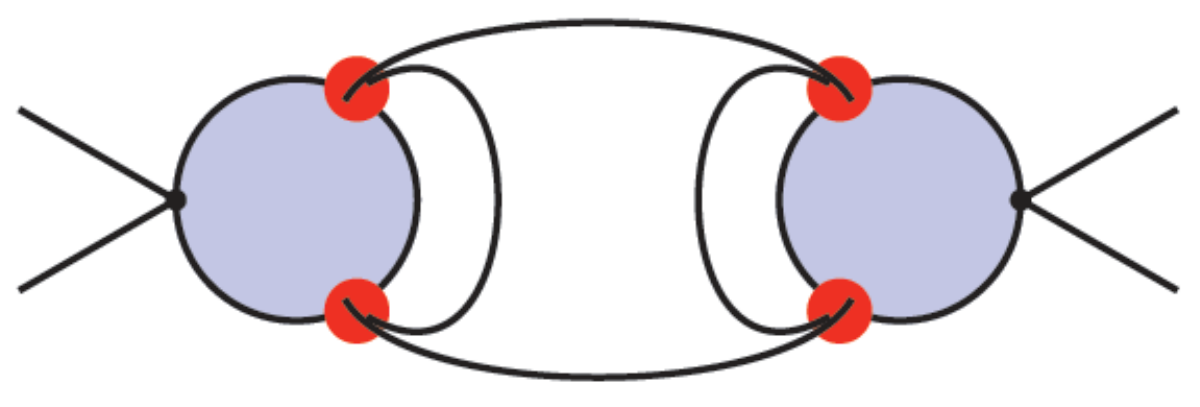}}~~~~~&
\scalebox{1.1}{\includegraphics[height = .08\textwidth, width=.1\textwidth]{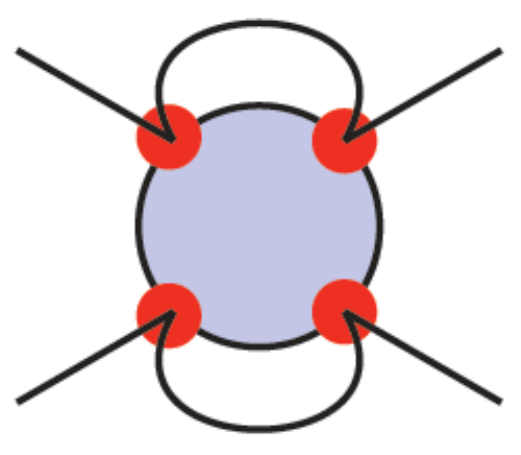}}\\
$0$&$0.02083$&$-0.06893$\\
&&\\
\scalebox{1.1}{\includegraphics[height = .09\textwidth, width=.08\textwidth]{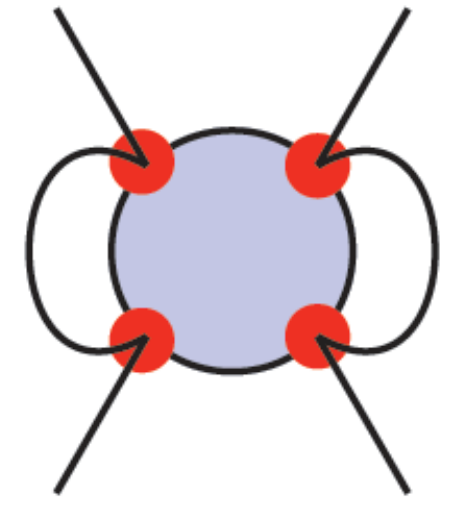}}~~~~~&
\scalebox{1.1}{\includegraphics[height = .09\textwidth, width=.22\textwidth]{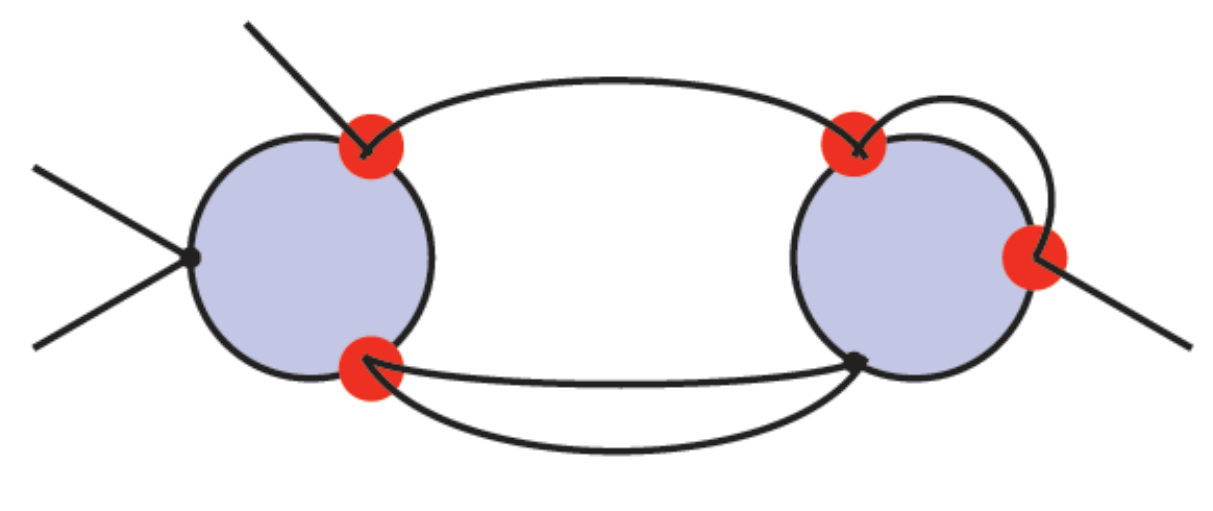}}~~~~~&
\scalebox{1.1}{\includegraphics[height = .085\textwidth, width=.2\textwidth]{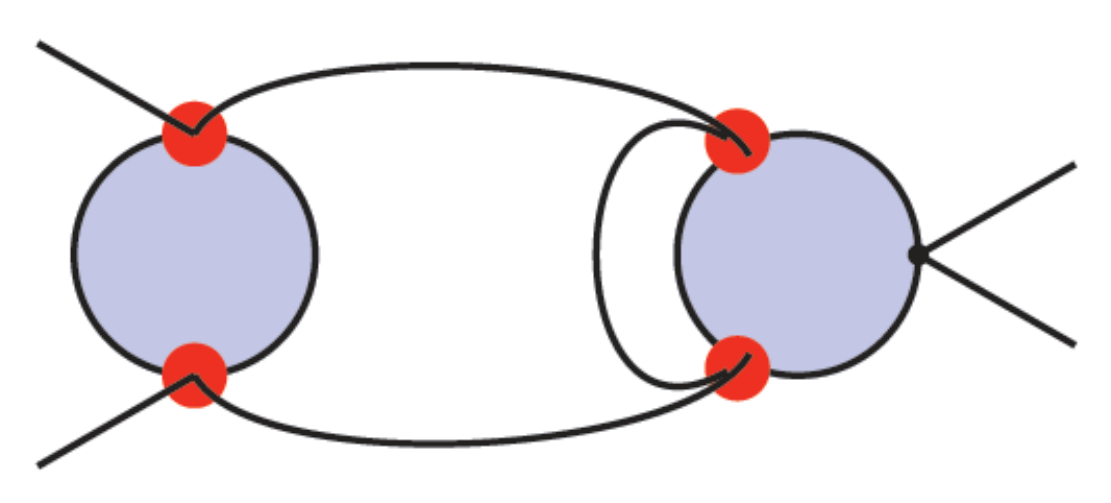}}\\
$-0.01216$&$0.06824$&$ ( - 0.1037)~{\rm to}~(-0.166)$\\
\end{tabular}
\vskip .05in
\caption{Corrections to the low momentum limit of the $\F^{(2)}$ vertex}
\label{Corr-graph4}
\end{center}
\end{figure}
The low momentum limit of $f^{(2)}$, including the renormalization corrections, can be written as
\beq
e^2 f^{(2)}(q) = {{\bar q}^2 \over 2 m} \left(
C_0 + d_1 + d_2 + \cdots\right)
\label{Corr10a}
\eeq
where $C_0 = 1.1308$ from the recursive procedure as in
(\ref{SchE16} or as quoted in
(\ref{Corr1}). $d_1$ corresponds to the contribution
from the term with one power of $(m/E_k)$ in the integrand corresponding
to the diagrams in Fig.\,\ref{Corr-graph4}, $d_2$
is the contribution from all terms with two powers of
$(m/E_k)$ in the integrand and so on.
Let $C_n$ denote the partial sum of corrections up to terms with
$(m/E_k)^n$, i.e., $C_1 = C_0 + d_1$, $C_2 = C_0 + d_1 + d_2$, etc.
Then we find
\beqar
C_1 &=& 0.5496\nonumber\\
C_2 &=& 0.2730\nonumber\\
C_3 &=& 0.0373\nonumber\\
C_4&=&
-0.05843~{\rm to}~ -0.00583
\label{Corr11}
\eeqar
Many of the integrals have to be evaluated numerically.
There is a small ambiguity in one of the integrals for the last diagram in 
Fig.\,\ref{Corr-graph4} \cite{KNY}, which is why a range of values is
indicated for $C_4$.

Notice that the partial sums are systematically decreasing in value, showing that the ordering of diagrams by powers of $m/E_k$ does constitute a viable expansion. 
The cumulative value
of the corrections to the order we have calculated is indeed small. For the string tension, we then find
\beq
\sqrt{\sigma_R} = 
 e^2 \sqrt{ c_A c_R\over
4\pi }~\left\{ 
\begin{matrix}
\bigl( 1 - 0.02799 +\cdots \bigr)\\
\!\!\!\bigl( 1 - 0.0029 +\cdots \bigr)\\
\end{matrix}\right.
\label{Corr12}
 \eeq
This correction, of the order of $-2.8\%$ to $-0.03\%$, is entirely consistent with lattice calculations. Terms of order $(m/E_k)^5$ are expected to contribute at the level of a fraction of $1\%$.
\begin{figure}[!t]
\begin{center}
\scalebox{.8}{\includegraphics{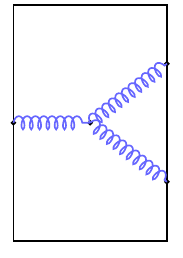}}
\caption{Example of a correction from $\F^{(3)}$ term to the Wilson loop
expectation value. Each``gluon" stands for a $\la J^a J^b\ra$ propagator.}
\label{Corr-graph5}
\end{center}
\end{figure}

The corrections seem to play out in an intriguing way here.
The first term $C_0$ is rather large, but is then mostly cancelled out
by the ``loop corrections". This suggests that there must be a 
different way to organize the corrections so that many of the 
cancellations are packaged together.

There are also several types of corrections we have not calculated.
First of all, there is the issue of corrections to $\Psi_0$
due to the next set of terms (of order $e^4$) in the solution of the Schr\"odinger equation, as in section \ref{SchE}.
Secondly, even to the order we are calculating here, there are
diagrams with two or more current loops in the effective
2d theory of the $\vf$-fields. Finally, there could also be
corrections which do not appear as loop corrections to the quadratic terms
in $\Psi_0$ but 
have to be included in the computation of the
expectation value of the Wilson loop operator. 
An example of such a diagram is shown in Fig.\,\ref{Corr-graph5}.
These types of corrections can also be representation-dependent
 in general and they can be important for string-breaking effects as well.

\newpage

\end{document}